\documentclass[aip,jcp,amsmath,amssymb,reprint,]{revtex4-1}
\pdfoutput=1

\usepackage{graphicx}
\usepackage{dcolumn}
\usepackage{bm}
\usepackage{natbib}
\usepackage{gensymb}

\begin{document}

\title{Ice Formation on Kaolinite: Insights from Molecular Dynamics Simulations}

\author{Gabriele C. Sosso}
\email{g.sosso@ucl.ac.uk}
\affiliation{Thomas Young Centre, London Centre for Nanotechnology and Department of Physics and Astronomy, University College London, Gower Street London WC1E 6BT, United Kingdom}
\author{Gareth A. Tribello}
\affiliation{Atomistic Simulation Centre, Department of Physics and Astronomy, Queen's University Belfast, University Road Belfast BT7 1NN, United Kingdom}
\author{Andrea Zen}
\affiliation{Thomas Young Centre, London Centre for Nanotechnology and Department of Physics and Astronomy, University College London, Gower Street London WC1E 6BT, United Kingdom}
\author{Philipp Pedevilla}
\affiliation{Thomas Young Centre, London Centre for Nanotechnology and Department of Physics and Astronomy, University College London, Gower Street London WC1E 6BT, United Kingdom}
\author{Angelos Michaelides}
\email{angelos.michaelides@ucl.ac.uk}
\affiliation{Thomas Young Centre, London Centre for Nanotechnology and Department of Physics and Astronomy, University College London, Gower Street London WC1E 6BT, United Kingdom}

\date{\today}

\begin{abstract}

The formation of ice affects many aspects of our everyday life as well as important technologies such as cryotherapy and
cryopreservation.  Foreign substances almost always aid water freezing through heterogeneous ice nucleation, but the
molecular details of this process remain largely unknown.  In fact, insight into the microscopic mechanism of ice
formation on different substrates is difficult to obtain even if state-of-the-art experimental techniques are used.  At
the same time, atomistic simulations of heterogeneous ice nucleation frequently face extraordinary challenges due to the
complexity of the water-substrate interaction and the long timescales that characterize nucleation events.  Here, we
have investigated several aspects of molecular dynamics simulations of heterogeneous ice nucleation considering as a
prototypical ice nucleating material the clay mineral kaolinite, which is of relevance in atmospheric science.  We show
via seeded molecular dynamics simulations that ice nucleation on the hydroxylated (001) face of kaolinite proceeds
exclusively via the formation of the hexagonal ice polytype. The critical nucleus size is
two times smaller than that obtained for homogeneous nucleation at the same supercooling.
Previous findings suggested that the flexibility of the kaolinite surface can alter the time scale for ice nucleation within molecular
dynamics simulations. However, we here demonstrate that equally flexible (or non flexible) kaolinite surfaces 
can lead to very different outcomes in terms of ice formation, according to whether or not the surface
relaxation of the clay is taken into account.  We show that very small structural changes upon relaxation
dramatically alter the ability of kaolinite to provide a template for the formation of a hexagonal overlayer of water
molecules at the water-kaolinite interface, and that this relaxation therefore determines the nucleation ability of this
mineral. 

\end{abstract}

\maketitle

\section{\label{SEC_1} Introduction}

From the immense extent of glaciers to the microscopic length scale of living cells, ice shapes life as we know
it~\cite{bartels-rausch_chemistry:_2013}. For instance, the formation of clouds~\cite{tang_interactions_2016} and the
weathering of rocks~\cite{bartels-rausch_ice_2012} originate from  water freezing in the atmosphere and on earth
respectively.  Moreover, technologies such as cryotherapy and
cryopreservation~\cite{tam_solution_2009} are greatly influenced by the microscopic details of ice formation.  It is 
surprising to discover that pure water freezes only at very strong supercooling, i.e. when it is brought to temperatures
lower than -30$\degree$C~\cite{sellberg_ultrafast_2014}. Water must, therefore, be freezing heterogeneously, as in a
world where water would only freeze homogeneously, the Arctic Ocean would hardly turn into the icy playground of polar
bears~\cite{martin_frazil_1981,lepparanta_2015}.  Furthermore, if water only froze homogeneously, too much solar
radiation would reach us as there would be no screening by ice-rich clouds~\cite{welch_1980,yang_radiative_2014}.
Luckily, ice can form at mild supercooling (i.e. at few degrees only below 0$\degree$C)
heterogeneously~\cite{murray_ice_2012}, with the aid of foreign substances which lower the free energy cost needed to
form a sufficiently large (or \textit{critical}) nucleus of crystalline ice within supercooled liquid water.  The nature
of these impurities is astonishingly diverse~\cite{murray_ice_2012}: for example, bacterial fragments, soot, pollen,
volcanic ashes and mineral dust have all been shown to boost the rate of ice nucleation. Hence the question:
\textit{what makes these very different substrates so effective in promoting ice formation?}  Surprisingly, a conclusive
answer is yet to be found, mainly because the microscopic details of heterogeneous ice nucleation are still largely
unresolved~\cite{slater_blue-sky_2015}.

Experiments can quantify the ability of a given substance to promote the formation of ice: for instance, a common
approach consists of comparing the fraction of water droplets that freeze in a given time interval at a certain
temperature with or without the presence of foreign
particles~\cite{murray_heterogeneous_2010,campbell_is_2015,whale_ice_2015}.  However, the early stages and the atomistic
mechanism of nucleation remain exceedingly challenging to probe experimentally. This is in large part because once the
critical size (typically of the order of nanometers) of the ice nucleus has been reached, nucleation proceeds on very
fast time scales (pico- or nanoseconds).  This is why atomistic simulations can provide valuable insight, and complement
experimental evidence by unraveling the details of the nucleation mechanism on different substrates at the molecular
level.  Having said that, however, nucleation is a \textit{rare event} and seconds can pass before the spontaneous
formation of a nucleus of critical size occurs. Running molecular simulations of these lengths is simply not tractable.
This is why in the last few years substantial effort has been devoted to developing enhanced sampling techniques capable
of tackling this time scale problem~\cite{fecc,sosso_crystal_2016}. Moreover, equally serious issues often go unacknowledged, such as the ability of
the interatomic potentials of choice to represent the water, the substrate, and the interaction between the two, or the
extent to which simulations of ice nucleation are affected by specific computational details, such as the flexibility of
the substrate.

Probing the importance of such aspects would require to investigate heterogeneous ice formation across a collection of
different substrates: this is currently possible only by taking advantage of the computational speed granted by the
coarse grained mW model~\cite{molinero_water_2009} for water. This approach led to many important findings, such as the
influence of the hydrophobicity and/or lattice mismatch of different crystalline surfaces in promoting ice
formation~\cite{fitzner_many_2015} and the characterization of ice formation on carbonaceous
particles~\cite{lupi_does_2014,lupi_heterogeneous_2014,cabriolu_ice_2015,lupi_pre-ordering_2016,bi_heterogeneous_2016}. However, fully atomistic models are needed to
deal with water at complex interfaces, such as crystalline surfaces of organic crystals or mineral dust particles. In
these cases, it remains to be seen whether the description of the surface and most importantly of the water-surface
interaction is accurate enough to allow for reliable results to be obtained. 

Only a few works have probed ice formation on crystalline substrates by means of all-atom
models~\cite{yan_heterogeneous_2011,yan_ice_2013,zielke_molecular_2014,cox_microscopic_2014,sosso_microscopic_2016}.
Here, we study heterogeneous ice nucleation at strong supercooling ($\Delta T=T_m-T$=42K, where $T_m$ is the melting
temperature) on the (001) surface of kaolinite
using molecular dynamics (MD).  Kaolinite is of great relevance in atmospheric science, and its surface structure and
ice nucleation ability have been extensively investigated in both
experiments~\cite{murray_heterogeneous_2011,murray_ice_2012,tobo_impacts_2012,welti_exploring_2013,wex_kaolinite_2014}
and simulations~\cite{hu_kaolinite_2010,tunega_ab_2004,cox_microscopic_2014,zielke_simulations_2016}.  In particular,
Cox and coworkers~\cite{cox_microscopic_2014} performed brute-force MD simulations of ice nucleation on kaolinite using
small (10$^2$ molecules) models of the clay-water interface. Despite the substantial finite-size effects affecting these
simulations, the results suggested that non-basal faces of ice, specifically the primary prism face of the hexagonal ice
polytype, can nucleate on the hydroxylated basal face of kaolinite. This evidence has also been observed in brute-force
MD simulations~\cite{zielke_simulations_2016} employing much larger (10$^4$ molecules) models. However, in that case nucleation was observed 
when the kaolinite surface was almost entirely kept \textit{frozen}, i.e. atoms have been kept fixed at the experimental atomic
positions of the bulk phase.  In fact, it has been suggested~\cite{zielke_simulations_2016} that the flexibility of the
kaolinite surface can substantially affect the time scale over which heterogeneous ice nucleation occurs.  Moreover, we
have recently succeeded in elucidating the kinetics of ice formation on this mineral~\cite{sosso_microscopic_2016} by
means of forward flux sampling (FFS) simulations~\cite{allen_forward_2009}, an accurate path sampling technique which
has been successfully employed to investigate crystal nucleation and growth in different
systems~\cite{li_ice_2013,bi_probing_2014,gianetti_computational_2016}.  

In this work, we investigate: (i.) the type of ice (cubic, $I_c$, or hexagonal, $I_h$) that forms on the hydroxylated basal face
of kaolinite; (ii.) how the surface relaxation of the clay alters ice formation at the water-kaolinite interface; (iii.) some aspects
of how the force fields used perform for this system.

We demonstrate by seeded MD simulations that the hexagonal polytype of ice is likely to be the only one involved in
the nucleation process, due to the favorable interaction between the hydroxylated (001) surface of kaolinite and the
prism face of hexagonal ice. In fact, while large ($\sim$ 450 molecules) $I_c$ seeds dissolve into liquid water, $I_h$
nuclei of the same size or smaller ($\sim$ 250 molecules) grow within a wide temperature range.  In addition, we show by
means of very long (up to 2 $\mu$s) unbiased MD simulations that nuclei of hexagonal ice exposing the prism face to the
clay surface spontaneously occur as natural fluctuations of the water network.  These findings are consistent with
previous computational studies~\cite{cox_microscopic_2014,zielke_simulations_2016,sosso_microscopic_2016}, and provide
conclusive evidence of the dominant role of the hexagonal polytype in the early stages of ice nucleation on the
hydroxylated (001) surface of kaolinite.

We have also addressed whether the flexibility of the substrate, in this case the kaolinite surface, can affect
the kinetics of ice formation within MD simulations.  We find that equally flexible (or non-flexible) kaolinite surfaces 
can lead to very different outcomes in terms of ice formation. In particular, it seems that small
structural changes can significantly alter the nucleation ability of this mineral.  Thus, we assess the sensitivity of
the nucleation mechanism to the microscopic structural features of the water-kaolinite interface.  We find that surface
relaxation, however small, can substantially alter the templating effect of kaolinite and that these effects can
facilitate the heterogeneous formation of ice.  Specifically, small structural changes in the arrangement of the
hydroxyl groups at the surface affect the free energy cost needed to form a hexagonal motif of water molecules within
the first overlayer: this templating structure is in turn very effective in promoting the formation of ice.

Finally, we briefly investigate whether the TIP4P/Ice and CLAY\_FF force fields are capable to describe supercooled
water and kaolinite respectively.  We find that the CLAY\_FF force field seems to provide a reliable description of ice
nucleation at the water-KAO$_{OH}$ interface.  However, the surface relaxation of the siloxane (001) surface of
kaolinite predicted by the CLAY\_FF is not consistent with first principles calculations results, thus putting into
question the ability of this force field to deal with ice nucleation and growth on the siloxane face of kaolinite.

\section{\label{SEC_2} Computational Methods}

\begin{figure}[t!]
\begin{centering}
\centerline{\includegraphics[width=7cm]{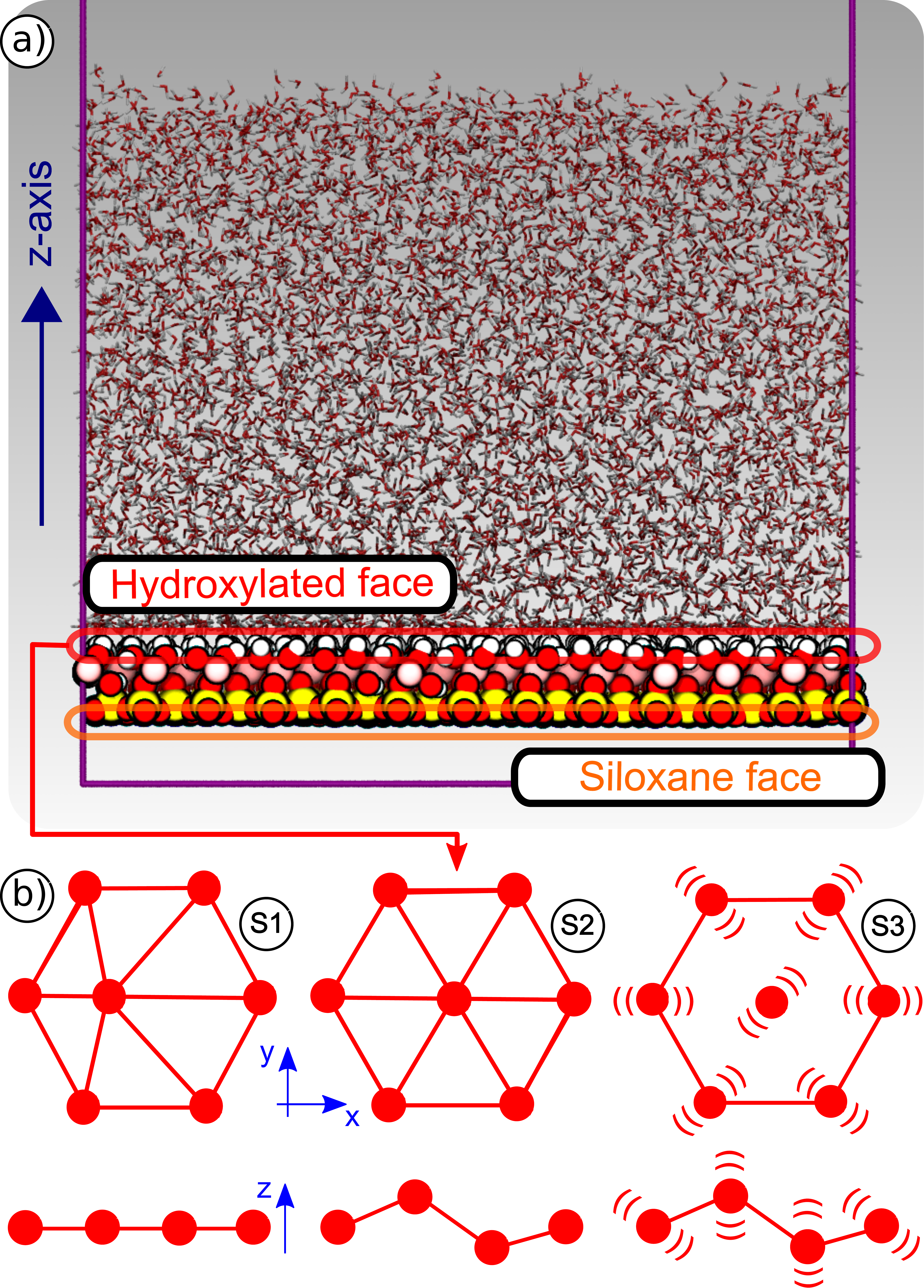}}
\par\end{centering}
\protect\caption{a) A kaolinite slab (spheres), as cleaved along the (001) basal plane normal to the z-axis, is
in contact with a water film containing $\sim$6000 water molecules (sticks). This particular computational geometry
corresponds to interface S3 (see text).  Oxygen, silicon, aluminum and hydrogen atoms are colored in red, yellow, pink
and white respectively.
b) Schematic representation of the three water-kaolinite interfaces S1, S2 and S3
considered in this work (top and side views).  For the sake of simplicity, just the oxygen atoms in the outer layer of the KAO$_{OH}$ face are shown.
S1: atoms are kept frozen in the unrelaxed experimental positions of bulk kaolinite.  S2:  atoms are kept frozen in a
configuration obtained upon surface relaxation. S3: atoms are unconstrained and thus the surface is flexible.  The extent of surface relaxation has been
deliberately exaggerated in these cartoons.} 
\label{FIG_1} 
\end{figure}

MD simulations have been performed using the GROMACS simulation package~\cite{van_der_spoel_gromacs:_2005}.  The CLAY\_FF~\cite{cygan_molecular_2004} and
the TIP4P/Ice~\cite{abascal_potential_2005} force fields were used to model kaolinite and water respectively. We have not
included the optional angular term (see Ref.~\citenum{cygan_molecular_2004}) in the CLAY\_FF force field, as we have
verified that it does not affect the structure of the surface. So as to address the question of surface flexibility and relaxation, 
we have considered three different water-kaolinite
interfaces, where water molecules are in contact with the hydroxylated (001) face of kaolinite (KAO$_{OH}$): 

\begin{itemize}
\item[\textbf{S1 - }] {\textbf{Frozen Surface, Unrelaxed:} 
all the kaolinite atoms are kept fixed during the MD simulations at the experimental positions of the bulk system, except for
the hydrogen atoms of the hydroxyl groups on the outer layer of the slab. These hydrogen atoms are bonded to the
corresponding oxygen atoms via
a harmonic constraint characterized by a spring constant of 2.3185$\cdot 10^3$ kJ/mol \AA$^{-2}$ acting on the O-H bond
length (1.0 \AA), as required by the CLAY\_FF force field~\cite{cygan_molecular_2004}. 
About 6000 water molecules have been placed between two kaolinite slabs mirroring
each other. This interface is identical to that reported in Ref.~\citenum{zielke_simulations_2016}.}
\item[\textbf{S2 - }] {\textbf{Frozen Surface, Relaxed:} all the kaolinite atoms are kept fixed during the MD simulations at 
the average atomic positions of the system previously equilibrated in the absence of restraints at 230 K,
except for the hydrogen atoms of the hydroxyl groups on the outer layer of the slab.
The O-H bonds are treated with the same harmonic constraint of S1.
Again in this system about 6000 water molecules have been placed between two kaolinite slabs mirroring each other.}
\item[\textbf{S3 - }] {\textbf{Flexible Surface:} the positions of the silicon 
atoms within the kaolinite slab are restrained during the MD simulations by means of a harmonic potential
characterized by a spring constant of 1.0$\cdot 10^3$ kJ/mol \AA$^{-2}$.  The O-H bonds are treated with the same
harmonic constraint of S1 and S2.
All the other kaolinite atoms are unrestrained.
About 6000 water molecules have been placed on top of a single kaolinite slab, as depicted in Fig.~\ref{FIG_1}a.}
\end{itemize}

\noindent Schematic representations of S1, S2 and S3 are shown in Fig.~\ref{FIG_1}b. Note that upon surface relaxation the arrangement of the
hydroxyl groups becomes more symmetric in the xy plane and more corrugated with respect to the z axis (normal to the 001 plane).
Additional details about S1, S2
and S3 as well as about additional models for the water-kaolinite interfaces are discussed in the Supplementary Material
(SM).  The interaction parameters between the clay and the water were obtained using the standard Lorentz-Berthelot
mixing rules~\cite{lorentz_ueber_1881,ahthefrench}, which yields water-surface interaction energies in good agreement
with high quality reference data from quantum Monte Carlo calculations~\footnote{Specifically, the adsorption energy $E_{ads}$
of a single water molecule on top of the hydroxylated (001) surface of kaolinite (calculated as $E_{ads}=E_{KAO+W}-E_{KAO}-E_{W}$ where $E_{KAO+W}$, 
$E_{KAO}$ and $E_{W}$ are the total energy of a kaolinite slab with a water molecule on top, the energy of the slab alone and the energy of the water molecule
in vacuum respectively) on its most favorable adsorption site is -646 $\pm$ 60 meV, in excellent agreement with the quantum Monte Carlo result of
-648 $\pm$ 18 meV~\cite{ANDREA_PREPRINT}}. The equations of motion were integrated using a leap-frog
integrator with a timestep of 2 fs. The van der Waals (non bonded) interactions were considered up to 10 \AA, where a
switching function was used to bring them to zero at 12 \AA.  Electrostatic interactions have been dealt with by means
of an Ewald summation with a real space cutoff at 14 \AA. The NVT ensemble was sampled at 230 K using a stochastic velocity rescaling
thermostat~\cite{bussi_canonical_2007} with a very weak coupling constant of 4 ps in order to avoid temperature
inhomogeneities throughout the system. The geometry of the water molecules (TIP4P/Ice being a rigid model) was
constrained using the SETTLE algorithm~\cite{miyamoto_settle:_1992} while the P\_LINCS algorithm~\cite{hess_lincs:_1997}
was used to constrain the O-H bonds within the clay.  The system was equilibrated at 300 K for 10 ns, before being
quenched to 230 K over 50 ns. For each interface S1, S2 and S3 (plus the other interfaces discussed in the SM), 10
independent MD simulations have been performed to look for nucleation events and in order to extract the equilibrium
canonical averages used to compute free energies. Details concerning the order
parameter used to identify ice-like water molecules are reported in the SM.

We have also performed density functional theory (DFT) calculations as part of this study: periodic kaolinite slab models were
used within the plane wave pseudopotential approach, using both GGA (Generalized Gradient Approximation) 
and dispersion inclusive GGA exchange-correlation functionals. Full details are included in the SM.

\section{\label{SEC_3} Results}

\subsection{\label{SSEC_3.1} Ice Nucleation on Kaolinite: Hexagonal vs Cubic Ice}

\begin{figure*}[t!]
\begin{centering}
\centerline{\includegraphics[width=\textwidth]{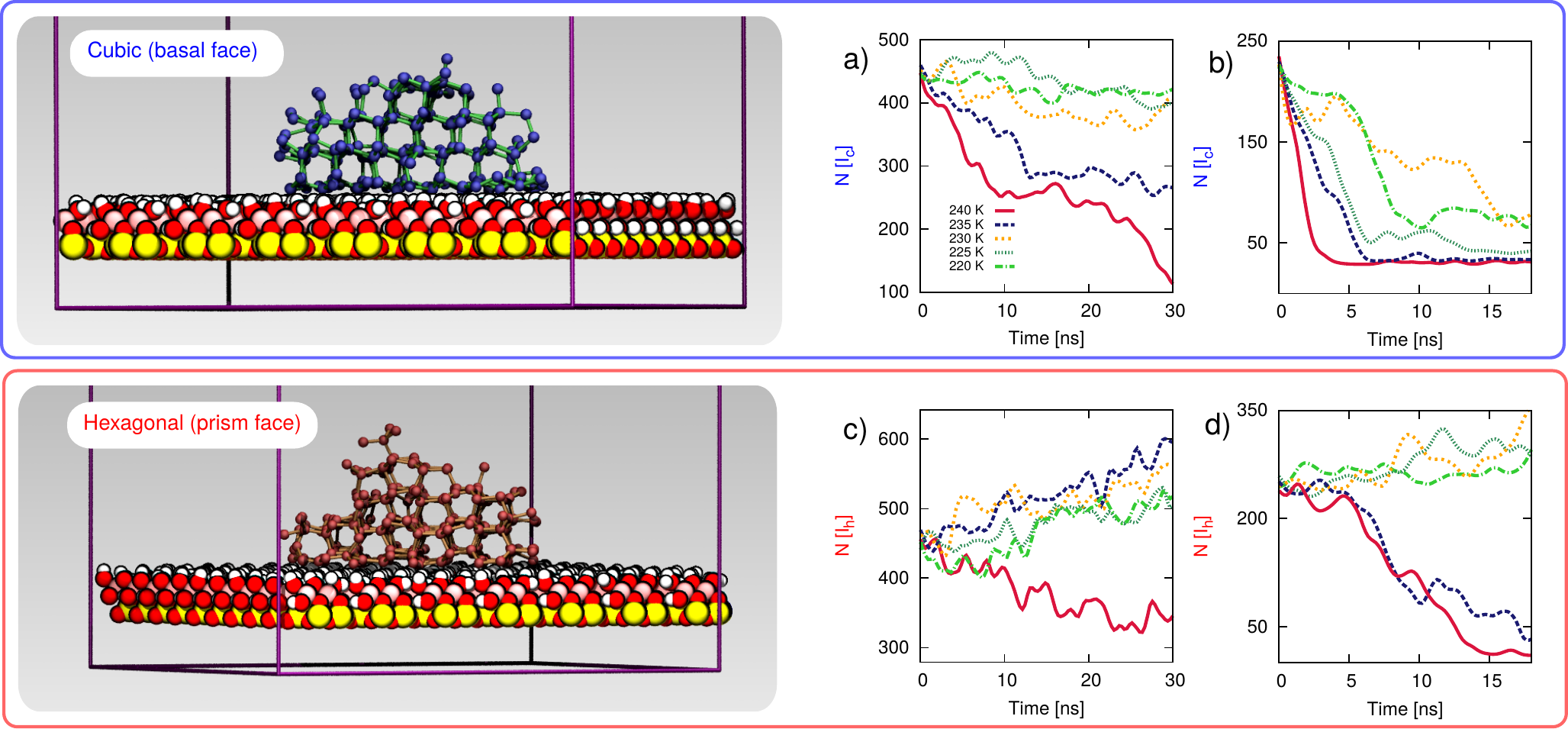}}
\par\end{centering}
\protect\caption{Seeded MD simulations of heterogeneous ice nucleation on the KAO$_{OH}$ surface.  Ice nuclei of $I_c$
(top, blue) and $I_h$ (bottom, red), as obtained from metadynamics simulations provided high-quality starting points in
terms of the hydrogen bond network between ice and the kaolinite surface.  The number $\lambda$ of ice-like molecules
within the largest connected cluster is reported as a function of time for seeds with initial sizes of about 250 and 450
ice molecules of $I_c$ [a) and b)] and $I_h$ [c) and d)]. For each seed five different temperatures (220, 225, 230, 235
and 240 K) have been considered.} \label{FIG_2} \end{figure*}

\begin{figure}[t!]
\begin{centering}
\centerline{\includegraphics[width=8cm]{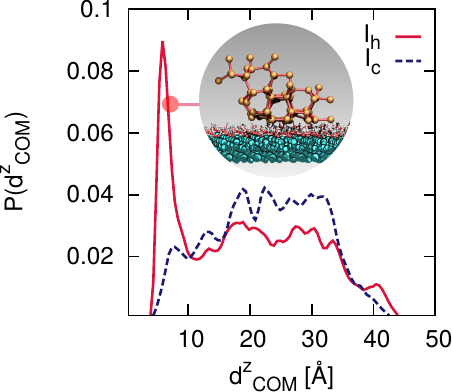}}
\par\end{centering}
\protect\caption{Natural fluctuations of the TIP4P/Ice water network on top of the KAO$_{OH}$ surface at 230 K, as
obtained from a 2 $\mu$s long unbiased MD simulation. The probability density $P(d^z_{COM})$ of the distance $d^z_{COM}$ of the ice nuclei center of mass from the kaolinite
surface (along the z direction normal to the surface) is reported for cubic ($I_c$) and hexagonal ($I_h$) nuclei.
These pre-critical nuclei can contain up to 80 molecules (according to the order parameter detailed in the SM).
A typical example of an I$_h$ (orange spheres, red sticks) cluster
exposing the prism face to the KAO$_{OH}$ surface is shown in the inset. Most of the KAO$_{OH}$ surface is depicted
using light blue spheres irrespective of the atomic species, although the oxygen and hydrogen atoms of the surface 
hydroxyl groups are shown in red and white respectively).} 
\label{FIG_3} 
\end{figure}

At strong supercooling ($\Delta T < \sim $ 40 K ), homogeneous water freezing results in a mixture of $I_h$ and $I_c$
that is known as stacking disordered ice~\cite{moore_is_2011,hansen_formation_2008,malkin_stacking_2014,slater_crystal_2014,quigley_communication:_2014} ($I_{sd}$). However, previous
computational studies~\cite{cox_microscopic_2014,zielke_simulations_2016} suggest that the formation of ice at the
water-kaolinite interface proceeds via the nucleation of $I_h$ only, possibly due to the favorable interaction between
its prism face and the hydroxyl groups on the clay surface.

Here, we have explicitly compared the preference of the hydroxylated (001) surface of kaolinite for nucleating either
$I_h$ or $I_c$ by means of seeded MD simulations. Specifically, we have inserted several crystalline nuclei of either
cubic or hexagonal ice into the system, and then subsequently observed at which temperature they shrink into the liquid
phase and at which temperatures they proceed toward full crystallization.  Seeded MD simulations are an efficient way
of obtaining a qualitative picture of crystal nucleation and growth, having been successfully used recently to explore
homogeneous water freezing~\cite{sanz_homogeneous_2013,espinosa_homogeneous_2014,espinosa_seeding_2016}.  In the case of
heterogeneous ice nucleation, however, one serious issue with seeded MD simulations is the choice/construction of the
crystalline seeds.  In fact, it is: (i.) rather difficult to guess \textit{a priori} which crystallographic face - if any
- of a certain ice polytype will form at the water-kaolinite interface and; (ii.) it is even more challenging to construct
  a feasible hydrogen bond network between the ice seed and the surface. 
In this work it is clear how to resolve problem (i.) as we already know that the prism face of $I_h$ and the basal face
of $I_c$ bind to kaolinite most strongly~\cite{cox_microscopic_2014}.
As for the hydrogen bond network, we have employed metadynamics
simulations~\cite{laio_escaping_2002,laio_metadynamics:_2008} to generate reasonable models of $I_h$ and $I_c$ spherical
caps in contact with the kaolinite surface, as depicted in Fig.~\ref{FIG_2}.  The ice nuclei obtained in this way
(details are discussed in the SM) have a very good match between the crystalline ice seeds and the
kaolinite surface, that would be very difficult to obtain otherwise. We have considered ice nuclei containing $\sim$ 250 or
450 water molecules at a flexible interface (model S3, as discussed below). A criterion based on the $q_3$
Steinhardt order parameter~\cite{steinhardt_bond-orientational_1983} has been used to label each water molecule in the
system as liquid, ice-like, and/or belonging to the cubic or hexagonal polytype, as described in the SM. The starting
configurations that were obtained from the metadynamics simulations have been equilibrated at 220 K for 200 ps, before collecting a
series of unbiased MD runs at different temperatures (220, 225, 230, 235 and 240 K), that were initiated by randomizing the
initial atomic velocities in a manner consistent with the corresponding Maxwell-Boltzmann distribution at the temperature of
interest.

The results are summarized in Fig.~\ref{FIG_2}, where we report
the number of $I_c$ and $I_h$ molecules within the ice seeds as a function of simulation time.  We find that seeds
containing as many as 450 $I_c$ molecules are not stabilized by the presence of the surface and dissolve at any
temperature we have probed (see Fig.~\ref{FIG_2}a and Fig.~\ref{FIG_2}b). In contrast,
450-molecule $I_h$ seeds clearly grow up to 235 K (see Fig.~\ref{FIG_2}c), and even small 250-molecule $I_h$ nuclei proceed
toward crystallization below 235 K (see Fig.~\ref{FIG_2}d).  These findings indicate that the basal face of cubic ice is
exceedingly unlikely to form on the KAO$_{OH}$ surface, and that the heterogeneous critical
nucleus size $N^*$ for ice on top of this kaolinite surface is of the order of 250 molecules at 230 K. This is consistent with
the estimate of $N^*=$225$\pm$25 molecules obtained via the FFS simulations reported in
Ref.~\citenum{sosso_microscopic_2016}. Note that the homogeneous critical nucleus size at the same supercooling is more
than two times larger: $\sim$ 540 molecules~\cite{haji-akbari_direct_2015}, where the same order parameter
has to be used to compare our results with those of
Ref.~\citenum{haji-akbari_direct_2015}, as discussed in the SM.  Thus, our findings
suggest that ice nucleation on the KAO$_{OH}$ surface most likely proceeds via the formation of  the prism face of
$I_h$, in agreement with the simulations of Cox {\textit et al.}~\cite{cox_microscopic_2014} and Zielke {\textit et
al.}~\cite{zielke_simulations_2016} as well as with our FFS simulations~\cite{sosso_microscopic_2016}.

Before moving on we stress, however, one drawback of seeded MD simulation: they assume \textit{a priori} the composition, the structure, the
size and the shape of the crystalline seeds. Thus, even if the latter do grow, one has to verify that such seeds are
compatible with the spontaneous, fluctuations of the system, in this case of the water network.  To explore this, we
have performed a very long (2 $\mu$s long) unbiased MD simulation, looking at the natural, pre-critical fluctuations of
the water network toward the ice phase.  In particular, we have determined for each configuration along the trajectory
whether the largest ice nucleus in the system is predominantly~\cite{mostly} made of either $I_h$ or $I_c$. The
probability density of the distance of the center of mass of the ice nuclei from the kaolinite surface (along the z
direction normal to the surface) is reported in Fig.~\ref{FIG_3} for both $I_h$ and $I_c$ nuclei.  While a similar
fraction of $I_h$ and $I_c$ nuclei can be observed within the bulk of the water slab, close to the water-kaolinite
interface there is a very clear preference for $I_h$ nuclei, while the probability for the $I_c$ polytype drops sharply.
Importantly, we find that of all the large (i.e. containing more than 60 water molecules) pre-critical ice nuclei, a
substantial fraction (25\%) still sit on top of the kaolinite surface (that is, at least one water molecule that belongs
to the ice nucleus sits within the first overlayer on the clay).  In addition, 98\% of this subset consists of
nuclei of $I_h$ that expose the prism face to the kaolinite surface (as illustrated in the inset of Fig.~\ref{FIG_3}).
We can thus safely assert that the $I_h$ nuclei we have investigated with seeded MD can indeed form on the $KAO_{OH}$
face within spontaneous, pre-critical fluctuations of the water network, and that $I_h$ nuclei exposing the prism face
to the kaolinite surface are the most likely to nucleate at this supercooling.

\subsection{\label{SSEC_3.2} Surface Relaxation and Ice Nucleating Ability}

\begin{figure}[t!]
\begin{centering}
\centerline{\includegraphics[width=6cm]{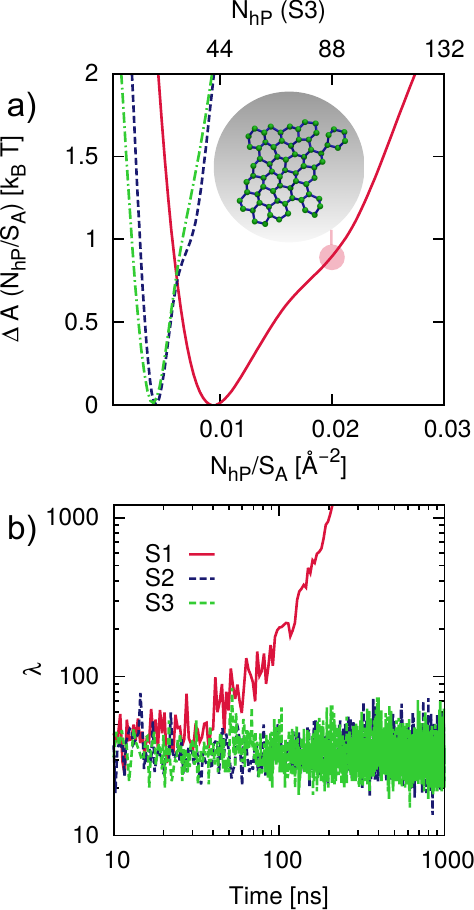}}
\par\end{centering}
\protect\caption{a)  Free energy relative to $k_B T$ as a function of the number $N_{hP}$ of water molecules involved in the biggest
hexagonal patch (or connected cluster of hexagons, see inset) within the first water overlayer, normalized by the
surface area $S_A$ of the kaolinite slab for S1, S2 and S3. The upper x-axis reports the $N_{hP}$ (not normalized by
surface area) for the S3 interface to convey the extent of the hexagonal motif.  b) Number of water molecules within the
largest ice cluster ($\lambda$) as a function of time for a typical MD trajectory obtained for S1, S2 and S3.}
\label{FIG_4}
\end{figure}

\begin{figure}[t!]
\begin{centering}
\centerline{\includegraphics[width=9cm]{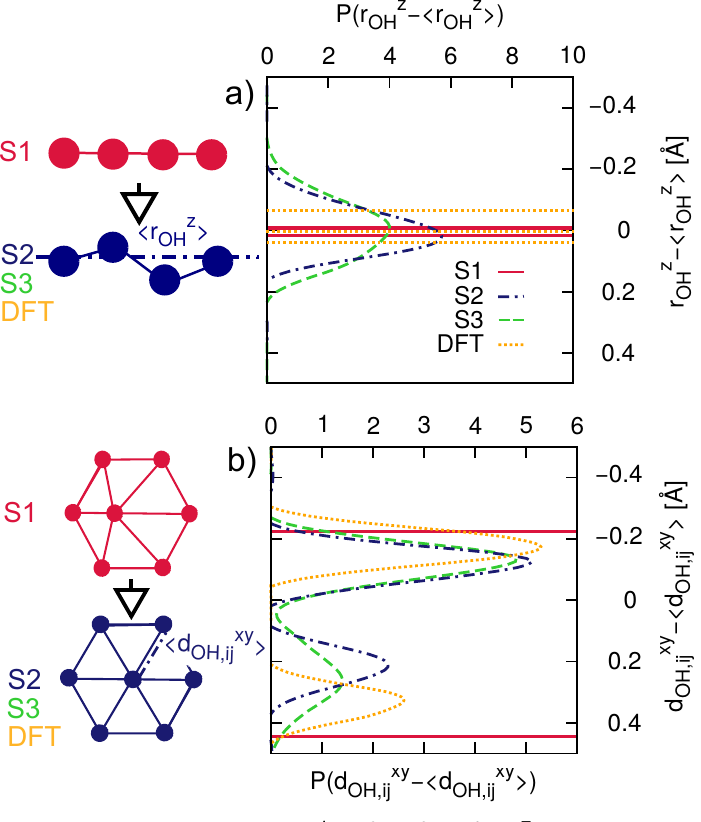}}
\par\end{centering}
\protect\caption{
Surface relaxation of kaolinite - particularly with respect to the hydroxyl groups at the water-kaolinite interface (see Fig.~\ref{FIG_1}).
a) Probability density distribution of the deviation of the height $r_{OH}^z$ (along the z direction normal to the slab)
of the oxygen atoms of the hydroxyl groups from their mean height $\langle r_{OH}^z \rangle$ for S1, S2 and S3. Results obtained from DFT
calculations with the optPBE-vdW exchange-correlation functional~\cite{klimes-vdW-DF} (see SM) are also shown.
A sketch of surface relaxation - deliberately exaggerated - along the z direction is shown on the left side. Note that for S1 and DFT the only three values
of $r_{OH}^z - \langle r_{OH}^z \rangle$ observed are reported as horizontal bars instead of continuous probability densities.
b) Probability density distribution of the deviation of the in-plane (xy plane parallel to the slab) nearest neighbor distance $d_{OH,ij}^{xy}$
of the oxygen atoms of the hydroxyl groups from their mean nearest neighbor distance $\langle d_{OH,ij}^{xy} \rangle$ for S1, S2, S3 and DFT. 
A sketch of surface relaxation - deliberately exaggerated -  in the xy plane is shown in the inset.  Note that for S1
the only three values of $d_{OH,ij}^{xy}-\langle d_{OH,ij}^{xy} \rangle$ are reported as horizontal bars instead of continuous probability densities.}
\label{FIG_5}
\end{figure}

It has been suggested that the ability of the KAO$_{OH}$ face to promote the formation of ice nuclei stems from the
templating effect of the hexagonal arrangement of hydroxyl groups~\cite{PK97,cox_microscopic_2014}, as depicted in
Fig.~\ref{FIG_1}. Moreover, recent MD simulations~\cite{zielke_simulations_2016} performed at strong supercooling
($\Delta T$=42 K) that employed the TIP4P/Ice water model indicate that the flexibility of these hydroxyl groups is
crucial when it comes to promoting heterogeneous ice nucleation. Specifically, the formation of ice on the KAO$_{OH}$
face happens spontaneously in unbiased MD simulations within $\sim$ 100 ns, provided that all the kaolinite atoms except
for the hydrogen atoms of the hydroxyl groups at the water-kaolinite interface are frozen at the experimental positions
of the bulk system.  We refer to this setup as interface S1 (see Sec.~\ref{SEC_2}). On the other hand, restraining the
dynamics of the oxygen atoms of the same hydroxyl groups using a harmonic potential, as opposed to completely freezing
the atomic positions, prevented the formation of ice within the $\mu$s time scale.

In order to investigate how exactly the flexibility of the substrate influences the formation of ice in MD simulations,
we have considered in addition to the S1 interface discussed in Ref.~\citenum{zielke_simulations_2016} two different
water-kaolinite interfaces S2 and S3 (see Sec.~\ref{SEC_2}).  In S2, kaolinite atoms are kept frozen during the MD
simulations exactly as for S1, but the starting configuration has been obtained from a kaolinite slab previously
equilibrated (without imposing any restraint) at 230 K. In S3, the kaolinite atoms are instead unconstrained and the
surface is flexible. Upon running unbiased MD simulations on these three substrates, we found that no nucleation events
occur for both S2 and S3 up to 2 $\mu$s, whereas on S1 ice formation occurs rapidly (within $\sim$ 100 ns).  We
expected this result for S3, as according to Ref.~\citenum{zielke_simulations_2016} the flexibility of the surface
should prevent ice formation on this timescale.  However, the fact that we do not observe ice nucleation on the S2
surface is surprising, as they are both frozen surfaces and the only difference between S1 and S2 is that the latter was
relaxed prior to the MD simulation of the frozen substrate.

The structure of the KAO$_{OH}$ face
facilitates the formation of a hexagonal motif of water molecules within the first overlayer (depicted in the inset of
Fig.~\ref{FIG_4}a). This structure is in turn compatible with the prism face of hexagonal ice or the basal face of cubic
ice~\cite{cox_microscopic_2014,zielke_simulations_2016}.  The likelihood for formation of ice on the KAO$_{OH}$ face is
related to the free energy cost needed to form a hexagonal patch (or connected cluster of water molecule hexagons)
containing $N_{hP}$ water molecules at the water-clay interface.  Here we have quantified this free energy cost for the
three interfaces S1, S2 and S3 as follows: we start by pinpointing six-membered rings of oxygen atoms within the first
water over layer on the surface. To do that, we took advantage of the R.I.N.G.S.~\cite{le_roux_ring_2010} code, 
and analyzed our simulations with a nearest-neighbors distance set to 3.2 \AA$\ $ and King's shortest path
criterion~\cite{king_ring_1967,franzblau_computation_1991}. We then selected only those rings for which
each triplet $i,j,k$ of adjacent oxygen atoms is characterized by an in plane angle $\alpha_h$ = 120 $\pm$ 20$\degree$,
where

\begin{equation}
\alpha_h=\arccos \left ( \frac{r_x(ij)\cdot r_x(kj) + r_y(ij)\cdot r_y(kj)}{|r(ij)|\cdot|r(kj)|} \right ).
\end{equation}

\noindent Here, $r_x(ij)$ is the $x$ component of the distance vector between oxygen atoms $i$ and $j$. We
then calculate the number of water molecules $N_{hP}$ within the largest connected clusters of hexagonal rings. We have
collected about 10$^4$ values of $N_{hP}$ from the first 10 ns of 10 independent MD runs for each interface.  As S1, S2
and S3 have slightly different surface area $S_A$, we have normalized by the latter the value of $N_{hP}$. We
subsequently evaluate the free energy profile as a function of $\frac{N_{hP}}{S_A}$ as:

\begin{equation}
\Delta A \left (\frac{N_{hP}}{S_A} \right )= -k_B T \log \left  [ P \left ( \frac{N_{hP}}{S_A} \right ) \right ]
\end{equation}

\noindent where $k_B$ is the Boltzmann constant and $P(\frac{N_{hP}}{S_A})$ is the equilibrium~\cite{equill} probability density
distribution for $\frac{N_{hP}}{S_A}$.  The results, obtained by taking into account the first 10 ns of 10 independent
simulations for each interface (in order to avoid the onset of ice formation for S1) are reported in Fig.~\ref{FIG_4}a.
In the case of S1, the free energy profile is rather shallow: for instance, $\sim$ 1 $k_BT$ is sufficient to produce
a rather large hexagonal patch containing $\sim$ 90 water molecules. As a result, the whole first overlayer of water molecules
relaxes into this hexagonal pattern within 50-100 ns, quickly triggering crystallization (as shown in
Fig.~\ref{FIG_4}b). This is why at this interface the formation of ice does not proceed via nucleation events, but
instead through a relaxation process.  In fact, the onset of crystallization is determined by the time needed for the
first water overlayer to relax into the hexagonal template. We have verified that for ten independent simulations the
induction times for "nucleation" at the S1 surface are all very similar, thus resulting in a survival probability for
the liquid phase (reported in the SM) which is typical of a relaxation process, as opposed to the stochastic nature of
nucleation events. We also note that the kinetics of ice formation on S1 is nonphysical, being about six orders of magnitude faster 
than the nucleation rate we have obtained for S3 via FFS calculations~\cite{sosso_microscopic_2016}. 

On the other hand, the occurrence of large ($N_{hP}>$$\sim$50) hexagonal patches for S2 and S3 is exceedingly rare 
compared to S1: for instance, the same free energy cost needed for S1 to form a hexagonal patch containing $\sim$ 90
water molecules results for both S2 and S3 in a patch about two times smaller. Hence, despite the fact that S2 is kept
frozen while S3 is fully flexible, the two interfaces show very similar free energy profiles. Indeed the free
energy cost needed for the interface to form a templating water overlayer is the same for S2 and S3, and as a
consequence, no ice nucleation is observed for either interface on the $\mu$s timescale (as illustrated in
Fig.~\ref{FIG_4}b).  We note that we observe a very similar scenario for S1 when we restrain the oxygen atoms of the
hydroxyl groups using a harmonic potential instead of freezing them completely: consistent with
Ref.~\citenum{zielke_simulations_2016}, we do not observe ice formation.  This happens because the harmonic restraints
do not prevent surface relaxation from taking place. Consequently, this constrained S1 surface resembles S2.  This
relaxation, however small, is enough to alter the ability of KAO$_{OH}$ to produce a large enough hexagonal patch.  In
fact, we have also verified that by just relaxing the S1 interface at zero K - without equilibrating the kaolinite
surface at 230 K, and subsequently freezing the atomic position exactly as we did for S1, we obtain results very similar
to what we observe for S2. In particular, ice does not form within the $\mu$s timescale, and the extent of surface
relaxation and the free energy cost to create the templating hexagonal overlayer are comparable to the outcomes of the
S2 scenario.

To understand these results we have examined the structures of the various slabs with the CLAY\_FF force field and also
DFT.  The changes in the structure of the KAO$_{OH}$ (001) surface upon relaxation, particularly the arrangement of the
hydroxyl groups, are summarized in Fig.~\ref{FIG_5}. We have quantified the corrugation (or surface
roughness) of the surface in terms of the deviation of the height $r_{OH}^z$ (along the z direction normal to the slab)
of the oxygen atoms of the hydroxyl groups from their mean height $\langle r_{OH}^z \rangle$, as shown in
Fig.~\ref{FIG_5}a. While S1 is basically flat, we observe a small degree of corrugation, up to 0.2 \AA, for S2 and S3.
Relaxing S1 at the DFT level leads to a very similar degree of corrugation. The in-plane arrangement of the hydroxyl groups
at the surface is also affected by surface relaxation. Fig.~\ref{FIG_5}b shows the probability density distribution for
the deviation of the in-plane (xy plane parallel to the slab) nearest neighbor distance $d_{OH,ij}^{xy}$ of the oxygen
atoms of the hydroxyl groups from their mean nearest neighbor distance $\langle d_{OH,ij}^{xy} \rangle$ for S1, S2 and
S3. Both S2 and S3 are characterized by a non negligible spread of $d_{OH,ij}^{xy}$, which in turn leads to a more
symmetric arrangement (see Fig.~\ref{FIG_5}b). The same conclusion holds for the structure obtained upon DFT relaxation of
S1, albeit the extent of surface relaxation appears to be less pronounced.  Despite the fact that the overall extent of
surface relaxation for the KAO$_{OH}$ surface appears to be quite small (ca. 0.1 \AA), these marginal structural changes
can play a significant role when it comes to the formation of ice on this clay.  This is not entirely unexpected,
as we have recently shown~\cite{fitzner_many_2015} that the ice nucleation rate for the coarse grained mW model of
water~\cite{molinero_water_2009} on Lennard-Jones crystals (at $\Delta T$ = 70 K) can change by several orders of
magnitude just because of deviations of ca. 0.2 \AA$\ $ in the lattice parameter of the crystalline surface.

It is important to note that S2 differs from S3 in terms of dynamical properties.  For instance, the structural relaxation
time (defined and discussed in the SM) of the water network within the first overlayer for S2 is two times larger than
that obtained for S3. This means that water dynamics at the interface with the KAO$_{OH}$ face is slower for S2
(and qualitatively for S1 as well, as discussed in the SM), as the frozen surface interacts more strongly with the water
molecules, in a manner which is consistent with early findings for Lennard-Jones interfaces~\cite{haji-akbari_effect_2014}. The absence of
nucleation for S2 and S3 on the same $\mu$s timescale indicates that the dynamics of water at the kaolinite-water interface has a lesser impact
than the structure of the clay on the tendency for ice to form.

This observation that the degree of surface relaxation can strongly affect the nucleation dynamics leads one to ask
whether other technicalities play a role. For instance, kaolinite slabs have a non-zero dipole moment, which in
principle can affect the nucleation process: in fact, it has been reported that electric fields can affect the freezing
of water~\cite{patey2011,patey2013}. Moreover, water-surface and/or water-vacuum interfaces introduce structural and
dynamical fluctuations into the water network, which must decay within the thickness of the water film on top of the
mineral. However, it seems that these issues do not affect the outcome of MD simulations
of ice formation in the case of kaolinite. In fact, we have been able to reproduce the results of the MD simulations of ice formation on
kaolinite reported in Ref.~\citenum{zielke_simulations_2016} (at the same supercooling and employing the same water
model) using a number of different computational setups, as discussed in the SM.  In contrast, we argue that the tiniest
structural details of the surface are crucial in determining the kinetics of ice formation on crystalline surfaces
within atomistic simulations, and that surface relaxation can play an even more relevant role than the flexibility of
the surface in promoting the formation of ice on this particular kaolinite surface. At this stage, it is reasonable to speculate that
both flexibility and surface relaxation are likely to be a general issue when dealing with MD simulations of heterogeneous ice nucleation.
It could be that the structural details of a particular substrate will determinate which one of the two would be the dominant factor in ruling the
kinetics of ice formation.

\subsection{\label{SSEC_3.3} The Impact of the Force Field}

\begin{figure*}[t]
\begin{centering}
\centerline{\includegraphics[width=10cm]{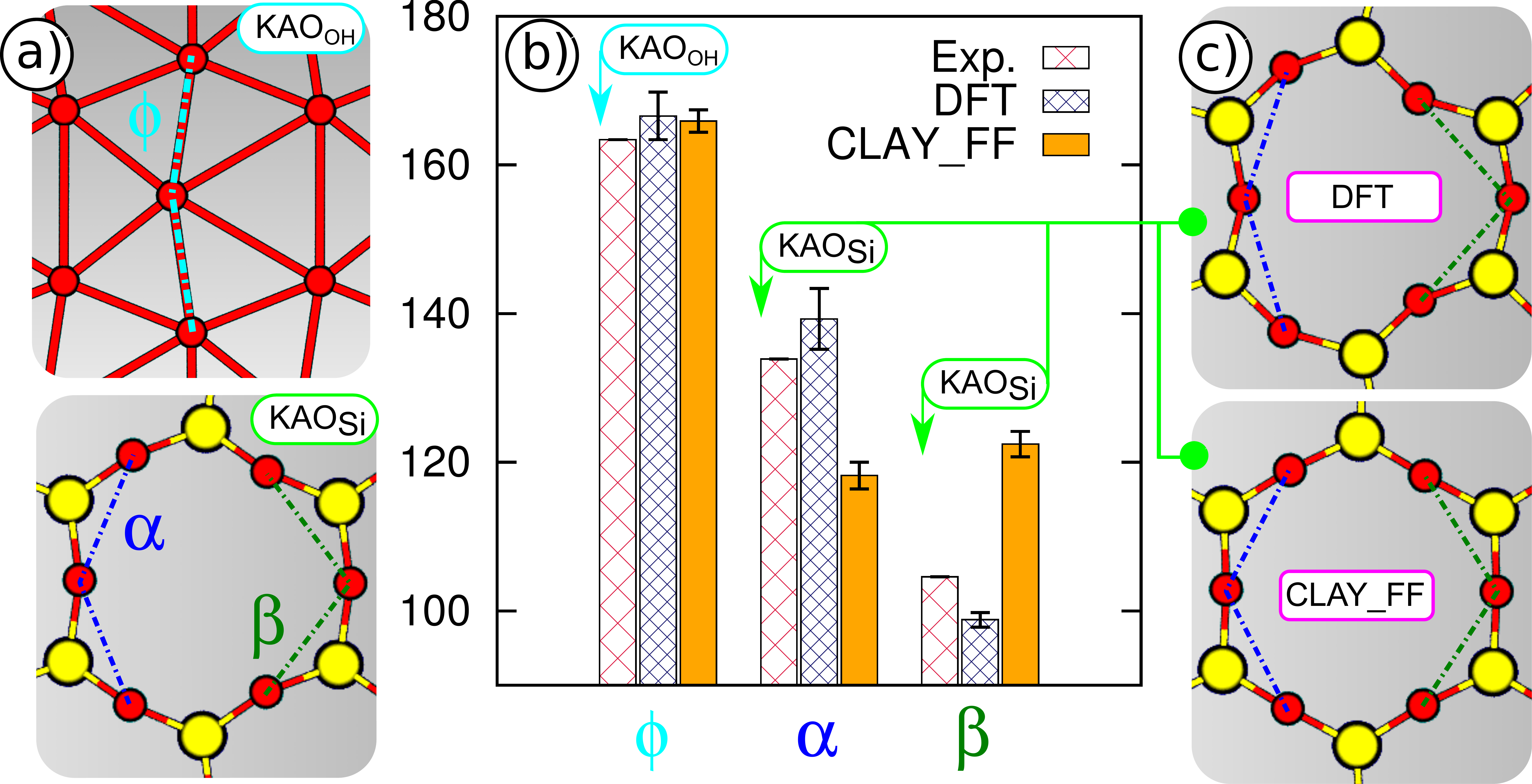}}
\par\end{centering}
\protect\caption{a) Arrangement of oxygen and silicon atoms in the outer layer of the
hydroxylated (KAO$_{OH}$, top) and the siloxane face (KAO$_{Si}$, bottom) of kaolinite.  Hydrogen atoms are not shown. The O-O-O
angles $\phi$ (for the KAO$_{OH}$ face), $\alpha$ and $\beta$ (for the KAO$_{Si}$ face), are highlighted. 
Oxygen, silicon, aluminum and hydrogen atoms are colored in red, yellow, pink and white, respectively.  
b) Surface relaxation of a single kaolinite slab. The experimental values (Exp.) of $\phi$,
$\alpha$ and $\beta$ for the bulk system are compared to the results obtained upon relaxation of a single kaolinite slab
via DFT and CLAY\_FF simulations. The outcomes in terms of surface structure predicted by DFT and CLAY\_FF for
KAO$_{Si}$ are shown in the top and bottom insets of panel c) respectively.} 
\label{FIG_6}
\end{figure*}

Before ending we briefly comment on the force field models employed.
Many options are available to simulate water~\cite{vega_melting_2005,molinero_water_2009,orsi_comparative_2014,sosso_crystal_2016}. The
coarse grained mW model~\cite{molinero_water_2009} is computationally very fast and as such it has been extensively used
to model heterogeneous ice nucleation on e.g. carbonaceous
particles~\cite{lupi_does_2014,lupi_heterogeneous_2014,cabriolu_ice_2015} and Lennard-Jones
crystals~\cite{fitzner_many_2015}.  However, with coarse grained approaches a truly atomistic description of the
nucleation mechanism cannot be achieved, and more importantly it is difficult to describe the interaction between water and a
complex material.  This is why here we have employed the atomistic TIP4P/Ice rigid model~\cite{abascal_potential_2005} for
water in this work. This model reproduces many structural and dynamical properties of liquid water as well as of different
ice phases correctly~\cite{abascal_potential_2005}, and it has been recently used to obtain an accurate reference for the
thermodynamics and kinetics of homogeneous freezing~\cite{haji-akbari_suppression_2014} at 230 K, corresponding to a
supercooling $\Delta T$=42 K (TIP4P/Ice water melts at 272 K). At this supercooling, the
dynamics of the water network is far from being homogeneous~\cite{mazza_relation_2006,kumar_relation_2007} and special
care has to be taken to correctly reproduce quantities like the self-diffusion coefficient and the structural
relaxation time, as detailed in the SM.

The CLAY\_FF force field~\cite{cygan_molecular_2004} is widely used to model clays as well as the interaction between clays and
water, including swelling properties~\cite{liu_thermodynamic_2006} and confinement effects~\cite{haria_viscosity_2013}.
As we are interested in having a reliable description of the water-surface interface, we have investigated the extent
of surface relaxation for two kaolinite surfaces customarily investigated in the context of heterogeneous ice
nucleation: the hydroxylated (KAO$_{OH}$) and siloxane (KAO$_{Si}$) (001) faces (see Fig.~\ref{FIG_1}a). As shown in
Fig.~\ref{FIG_6}a, the arrangement of oxygen atoms at the surface, which is critical in templating ice
formation~\cite{hu_ice_2007,cox_microscopic_2014,zielke_simulations_2016}, can be described by the O-O-O angles $\phi$
(for KAO$_{OH}$), $\alpha$ and $\beta$ (for KAO$_{Si}$).  We compare in Fig.~\ref{FIG_6}b the values of $\phi$, $\alpha$
and $\beta$ at the experimental atomic positions (Exp., as obtained upon cleavage of the bulk crystal) with those
obtained for the relaxed configurations of KAO$_{OH}$ and KAO$_{Si}$, calculated by DFT 
and CLAY\_FF. In the case of the KAO$_{OH}$ face, $\phi$ changes by $\sim$ 3 $\degree$ for both DFT
and CLAY\_FF. However, DFT and CLAY\_FF simulations give substantially different results for the KAO$_{Si}$ face: 
DFT predicts a marginal increase of the asymmetric buckling between oxygen and silicon atoms ($\alpha\simeq$140 and
$\beta\simeq$100), as depicted in the inset
of Fig.~\ref{FIG_6}b; in contrast, the CLAY\_FF relaxed structure is almost perfectly symmetric ($\alpha \simeq
\beta\simeq$120),
the buckling is absent and the atoms at the surface form regular hexagonal patterns.  Note that this 
symmetric arrangement of oxygen atoms at the KAO$_{Si}$ surface has also been predicted by the CLAY\_FF force field for
materials such as mica~\cite{wang_structure_2005}, while DFT calculations on the same system~\cite{odelius_two_1997}
resulted in buckled arrangements, in line with what we have observed in here. We remark that different starting
structures and/or different DFT exchange-correlation functionals do not affect these findings, as reported in the SM.
Whether the actual structure of the KAO$_{Si}$ face upon surface relaxation is closer to the CLAY\_FF or the DFT
prediction remains to be seen. This is one reason why here we have limited the discussion to the KAO$_{OH}$ (001) surface.

\section{\label{SEC_4} Discussion}

We have investigated several aspects of atomistic simulations of heterogeneous ice nucleation on a realistic surface,
choosing as an example the well-characterized (001) surface of kaolinite, a prototypical clay mineral of
relevance to ice formation in the atmosphere. 

Previous MD simulations~\cite{cox_microscopic_2014,zielke_simulations_2016} suggest that ice nucleation
occurs on the hydroxylated (001) kaolinite surface via the formation of hexagonal ice, due to the favorable interaction between its prism face
and the hexagonal arrangement of the hydroxyl groups at the surface of the clay. Here, we have established the preference of the
surface for hexagonal ice over the cubic polytype by means of seeded MD simulations using a fully flexible model
of kaolinite. We find that nuclei of cubic ice exposing the basal face
to the clay are not stabilized by the presence of the surface and that these nuclei therefore tend to shrink back into the liquid phase. On the other hand,
nuclei of hexagonal ice substantially wet the kaolinite surface and proceed to grow. We have estimated the critical nucleus size for these
nuclei of hexagonal ice to be roughly two times smaller than what has been reported for homogeneous water freezing at the same supercooling.

We have also verified by looking at the natural fluctuations of the water network at a flexible water-kaolinite
interface using a very long ($\mu$s)
unbiased MD simulation that the overwhelming majority of pre-critical ice nuclei that form on top of the clay are indeed made of hexagonal ice
exposing the prism face to the surface. This demonstrates that such nuclei spontaneously form at the surface, and that indeed the hexagonal
polytype is the only one involved in the early stages of heterogeneous ice nucleation on this particular kaolinite surface.

As discussed, it has recently been reported~\cite{zielke_simulations_2016} that the flexibility of the clay surface influences the
rate of ice formation. In this work, we find that surface relaxation can be equally important. In particular, small changes in the
structure of the hydroxylated (001) kaolinite face drastically alter the free energy cost needed to form an extended hexagonal
motif of ice-like molecules at the water-kaolinite interface. The occurrence of this templating layer leads to the formation of ice
within $\sim$ 100 ns if the atoms of the kaolinite surface are frozen at the experimental positions of the bulk phase.
However, upon surface relaxation the free energy cost for creating such a template is much higher and nucleation does not take place on
the $\mu$s timescale. Note that of all the water-kaolinite interfaces considered in this work, S3 is arguably the best
representation of the system, as actual surfaces are not frozen and/or unrelaxed. 

In addition, we note that the CLAY\_FF force field, customarily used to model clays as well as water interacting with clays, seems
to provide a reliable description of ice nucleation at the water-KAO$_{OH}$ interface. However, this classical force field predicts, in
the case of the siloxane (001) surface of kaolinite, a surface relaxation which is not consistent with the outcome of DFT calculations.
Thus, we argue that at this stage is not clear whether the formation of ice on this particular surface can be safely modeled using the CLAY\_FF force field.


We also remark that the fact that the nucleation process is so sensitive to the structure of the interface 
strongly suggests that future efforts should be
devoted to produce more accurate interatomic force fields for water at complex interfaces. In fact,
we have seen that subtle effects such as surface relaxation can truly affect the nucleation kinetics to an
point where it becomes rather difficult to benchmark computational results and most importantly to compare them with
experimental data.

\begin{acknowledgments}
This work was supported by the European Research Council under the European Union's Seventh Framework Programme
(FP/2007-2013)/ERC Grant Agreement number 616121 (HeteroIce project). A.M. is also supported by
a Royal Society Wolfson Research Merit Award. 
AM and AZ work has also been sponsored by the Air Force Office of Scientific Research, Air Force Material Command, USAF, under
grant number FA8655-12-1-2099.
The authors acknowledge the use of the UCL Grace and Legion High Performance Computing Facilities, the use of 
Emerald, a GPU-accelerated High Performance Computer,
made available by the Science \& Engineering South Consortium operated in partnership with the STFC Rutherford-Appleton Laboratory and the use of
ARCHER UK National Supercomputing Service (http://www.archer.ac.uk) through the Materials Chemistry Consortium through
the EPSRC grant number EP/L000202.
\end{acknowledgments}

\clearpage

\setcounter{section}{0}

\section*{SUPPLEMENTARY MATERIAL}

\begin{center}
\begin{table*}[t]
\label{TAB:CG}
   \caption{Details of the different computational geometries used to model the water-kaolinite interface. The last
column indicates whether ice formation has been observed (Y) or not (N) within $\sim$100 ns for each one of the ten
independent MD simulations performed for each interface.}
   \centering
   \resizebox{\textwidth}{!} {
   {\renewcommand{\arraystretch}{1.2}
   \begin{tabular}{ c c c c c c c c c }
   \multicolumn{1}{c|}{Interface} & \multicolumn{1}{|c|}{N. of water mol.} & \multicolumn{3}{|c|}{Cell vectors [\AA]} & \multicolumn{1}{|c|}{Slab geometry} & \multicolumn{1}{|c|}{Restraints} & \multicolumn{1}{|c|}{Starting conf.} & \multicolumn{1}{|c}{Ice nucl. within $\sim$ 100 ns} \\
   \hline
             &                  & A & B & C &               &            &                &                                \\
   \hline
   \hline
   S1    & 6464 & 51.5 & 62.6 & 84.2 & 2SM  & F  & Exp.  & Y \\
   S1R   & 6464 & 51.5 & 62.6 & 84.2 & 2SM  & F  & Relax & N \\
   S2    & 6464 & 51.8 & 62.9 & 84.2 & 2SM  & F  & 230K  & N \\
   S3    & 6144 & 61.8 & 71.5 & 150  & 1S   & N  & Exp.  & N \\
   SE\_2 & 6464 & 51.5 & 62.6 & 84.2 & 2SM  & FH & Exp.  & N \\
   SE\_3 & 6464 & 51.5 & 62.6 & 84.2 & 1S   & F  & Exp.  & Y \\
   SE\_4 & 6464 & 51.5 & 62.6 & 84.2 & 1S   & F  & Exp.  & Y \\
   SE\_5 & 6464 & 51.5 & 62.6 & 84.2 & 2SML & F  & Exp.  & Y \\
   SE\_6 & 6464 & 51.5 & 62.6 & 84.2 & 2SML & F  & Exp.  & Y \\
   \hline
   \hline
   \end{tabular}
   }
   }
\end{table*}
\end{center}

\noindent We provide supplementary material about the atomistic simulation of the water-kaolinite interface. We discuss:

\begin{itemize}
\item{The dynamical properties of the supercooled water network}
\item{The DFT calculations of kaolinite surface relaxation}
\item{The computational geometries used to model the water-kaolinite interface and how they affect the formation of ice}
\item{The formulation of the order parameter used to pinpoint ice nuclei}
\item{The ice formation on the interface S1, where structural relaxation is faster than ice nucleation}
\item{The details of the metadynamics simulations used to generate the ice nuclei for seeded molecular dynamics}
\end{itemize} 

\section{Dynamical properties of supercooled water}
\label{dynw}

\setcounter{figure}{0}
\begin{figure*}[t!]
\renewcommand\figurename{Fig.~S$\!\!$}
\centerline{\includegraphics[width=14cm]{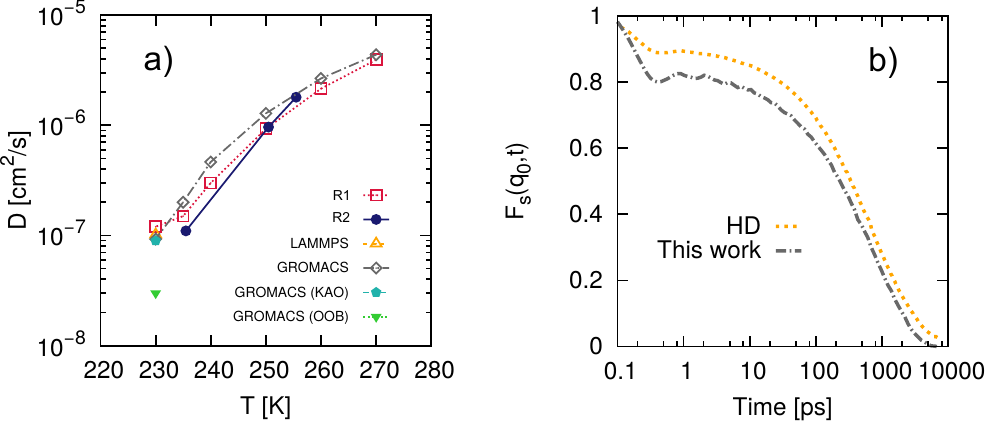}}
\caption{Dynamical properties of supercooled liquid water.  a) Diffusion coefficient of TIP4P/Ice water as a function of
temperature, calculated from NPT MD simulations of a 4096-molecule model. R1 and R2 refer to
Ref.~\citenum{weiss_kinetic_2011} and Ref.~\citenum{espinosa_homogeneous_2014} respectively.  LAMMPS refers
to a cross-check calculation performed via LAMMPS against the GROMACS data.  GROMACS(KAO) refers instead to the
computational setup we have used to model the water-KAO$_{OH}$ interface, where the diffusion coefficient has been
evaluated within the bulk-like region of the water film on top of the KAO$_{OH}$ slab via NVT MD simulations. The result
obtained by using GROMACS Out Of the BoX (GROMACS (OOB)), i.e. employing the default, basic settings is also reported.
b) Self intermediate scattering function $F_s$ as obtained by Haji-Akbari and Debenedetti~\cite{haji-akbari_direct_2015}
(HD) and in this work (This work) for a 4096-molecule model at 230 K. $q_0$=1.845 \AA$^{-1}$, corresponding to the first
peak of the structure factor S(q).} \label{FIG_S1} \end{figure*}

At the strong supercooling $\Delta T$=42 K considered in this work, the dynamics of the water network is
heterogeneous. Thus, special care must be taken to ensure the computational setup used is able to capture the
correct dynamical properties of the system. To this end, we have calculated the self-diffusion coefficient $D$ for a
4096-molecule model of TIP4P/Ice water, simulated within the NPT ensemble. The usual Einstein relation:

\begin{equation}
D=\lim_{t\to\infty}\frac{\langle ||\mathbf{r}(0)-\mathbf{r}(t)||^2\rangle}{2t\cdot \mathcal{D}}
\end{equation}

\noindent has been used, where $\langle \dots \rangle$ is the time average of the position vector $\mathbf{r}(t)$ for
all the oxygen atoms in the system and $\mathcal{D}$ is the dimensionality of the system, equal to three or two for
bulk or bi-dimensional systems respectively.  The same computational details reported in Sec. 2 of the main text have been used.  The
system has been equilibrated for 10 ns at 300 K, and subsequently quenched at the temperature of interest in 20 ns.
Production runs lasted 40 ns. We found that in this case dynamical properties are not affected by the choice of the
pressure control algorithm, the Berendsen or the Parrinello-Rahman barostats giving identical results which are reported
in Fig.~\ref{FIG_S1}a as the GROMACS dataset. Literature data (R1 and R2 in Fig.~\ref{FIG_S1}a) are also reported.  The
agreement between the different sets of data is very good at mild supercooling, but it gets worse at low temperatures
(e.g. at 235 K). The value of $D$ obtained by a cross-check calculation performed via LAMMPS, also reported, is in
excellent agreement with the result obtained via our setup (GROMACS) and the GROMACS(KAO) results in Fig.~\ref{FIG_S1}a.
These refer to the computational setup we have used to model the water-KAO$_{OH}$ interface, where the diffusion
coefficient has been evaluated within the bulk-like region of the water film on top of the KAO$_{OH}$ slab via NVT MD
simulations.  Finally, we note that using the default GROMACS settings, a
substantially smaller value of $D$ has been obtained - GROMACS(OOB) in Fig.~\ref{FIG_S1}a. We have also computed the
incoherent intermediate scattering function $F_s(q,t)$ as:

\begin{equation}
\begin{split}
F_s(q,t)=\langle\Phi_s(q,t)\rangle \\
\text{with \ \ } \Phi_s(q,t)=\frac{1}{N}\sum_{j}^N \exp{[i\mathbf{q}\cdot(\mathbf{r}_{j(0)}-\mathbf{r}_{j(t)})]} 
\end{split}
\end{equation}

\noindent where the sum runs over all the $j$ oxygen atoms having position $\mathbf{r}j(t)$ at time $t$ and $\mathbf{q}$
is a vector in reciprocal space. In an isotropic system, $F_s(q,t)$ depends only on the magnitude $q$ of the vector
$\mathbf{q}$, which selects the length scale (within the water network) probed by the scattering function.  The most
common choice is to take $q_0$ equal to the position in reciprocal space of the maximum of the structure factor S(q),
which for our model at 230 K corresponds to 1.845 \AA$^{-1}$.  $F_s(q,t)$ contains several information about the
dynamics of the system. The interested reader is referred to e.g.
Refs.~\citenum{berthier_theoretical_2011,sosso_dynamical_2014}. Here we just note that the time $\tau$ for which
$F_s(q,t)=1/e$ gives a measure of the structural relaxation time of - in this case - the water network. Our result for
bulk water at 230 K is reported in Fig.~\ref{FIG_S1} together with the $F_s(q,t)$ calculated for the same model and
slightly different settings (different MD code and possibly different values of $q_0$) in
Ref.~\citenum{haji-akbari_direct_2015}.  We obtain a structural relaxation time $\tau\sim$ 0.5 ns to be compared with
the value of $\sim$ 0.6 ns reported in Ref.~\citenum{haji-akbari_direct_2015}. This result gives us confidence in the
reliability of our MD simulations setup in describing the dynamics of the system.

\section{DFT calculations of surface relaxation}

In the main text we have discussed the surface relaxation of kaolinite. Specifically, we have shown that classical simulations using the
CLAY\_FF force field and first principles simulations using density functional theory (DFT) give very similar
result for the hydroxylated face (KAO$_{OH}$), but they disagree with respect to the relaxation of the siloxane (KAO$_{Si}$) face of kaolinite.
To date, there are no experimental indications about surface relaxation for kaolinite surfaces, so that we cannot establish whether
DFT calculations do a better job than the CLAY\_FF in reproducing the geometry of the KAO$_{Si}$ face.
As a rule of thumb, DFT calculations are likely to outperform conventional classical force fields such as CLAY\_FF.
However, it has to be said that DFT results can be particularly sensitive in this case to:

\begin{itemize}
\item{\textbf{The setup by which we model the kaolinite slab}. We used a three dimensional periodicity with
a vacuum region between slabs of $\sim$15~\AA, and the nonphysical dipole interaction across the slab (kaolinite
has a dipole orthogonal to the (001) plane) was corrected with the scheme of Neugebauer and Scheffler,~\cite{Dipole1,
Dipole2} in order to mimic a 2D system. We have performed geometry relaxations keeping the cell shape and volume fixed:
in order to demonstrate that our results are not affected by different simulation boxes, we have considered
three different starting points: (1) the cell of the experimental bulk kaolinite, adding the vacuum
on the direction orthogonal to the layer; (2) as 1, but starting from the cell of the bulk kaolinite obtained
upon full relaxation (of ions, cell shape and volume) via DFT with the PBE functional~\cite{PBE, PBE_Erratum}; (3) as 2, but using the
optPBE-vdW functional~\cite{klimes-vdW-DF}.}
\item{\textbf{The choice of the exchange-correlation (XC) functional}. In order to rule out spurious artifacts due
to the unlucky choice of a specific XC functional, we have performed geometry optimization considering several 
different commonly used  XC functionals. Specifically, two GGA functionals: PBE~\cite{PBE, PBE_Erratum} and RPBE~\cite{RPBE};
three vdW corrected PBE functionals: PBE-D2~\cite{DFT-D2}, PBE-D3~\cite{DFT-D3} and vdW(TS)~\cite{DFT-TS};
and four recently developed fully self-consistent non-local functionals: vdW-DF2~\cite{vdW-DF2}, 
revPBE-vdW~\cite{revPBE, vdW-DF}, optB86b-vdW~\cite{optB86b-vdW} and optPBE-vdW~\cite{klimes-vdW-DF}.}
\end{itemize}

Concerning the computational details: DFT calculations were performed using the plane-wave code VASP 5.4.\cite{VASP1,
VASP2, VASP3, VASP4}.  Calculations using the van der Waals density  functionals were carried out self-consistently
using the approach of Rom\'{a}n-P\'{e}rez and Soler\cite{perez&soler:implementation} as implemented in VASP by
Klime\v{s} \emph{et al.}\cite{optB86b-vdW}.  Electron-core interactions were described using the projector-augmented
wave\cite{PAW1, PAW2} (PAW) potentials supplied with  VASP.  PBE potentials for all functionals were used.  It has been
shown on a range of systems for the van der Waals functionals that this approximation with the PAW potentials does not
introduce any significant errors in the energies and structures.\cite{optB86b-vdW, JPCM:Graziano} The plane-wave energy
cut-off is 500 eV. The sampling of the reciprocal space was performed using a Monkhorst-Pack $k$-point mesh\cite{Kpoints} per simulated
supercell of $4\times2\times3$ for the bulk calculations, and $4\times2\times1$ for the slab calculations with PBE
functional. We verified that a  $2\times1\times1$ Monkhorst-Pack $k$-point mesh is already at convergence for the
geometrical properties of the slab, thus we used that for all the other XC functionals considered.

With respect to the CLAY\_FF results, we have verified that surface relaxation is not affected by different simulation boxes,
nor by the inclusion of the angular term described in Ref.~\citenum{cygan_molecular_2004}. The numbers reported in 
Table~\ref{TAB:GEO} are obtained as an average over all the angles of interest within the S3 kaolinite slab (see main text).

The results for all the performed DFT evaluations, in comparison with  those from the CLAY\_FF model, are summarized in
Table~\ref{TAB:GEO}, demonstrating that the disagreement between DFT and CLAY\_FF with respect to the surface relaxation
of the KAO$_{Si}$ face is a solid result that holds for a diverse portfolio of computational setups.

\begin{center}
\begin{table}[h!]
   \caption{
Values of the angles $\alpha$ and $\beta$ in the KAO$_{Si}$, and angle $\phi$ in the
KAO$_{OH}$ (see FIG. 1 in the main text) as obtained upon bulk or slab geometry relaxation.
Several different exchange-correlation functionals are considered, and in case of slab
calculations we also reported the methods that was used to relax the cell shape and volume (see text).
}
\label{TAB:GEO}
   {\renewcommand{\arraystretch}{1.2}
   \begin{tabular}{ l l l  c  c  c }
   \hline
   \hline
    &   &  & \multicolumn{2}{c}{KAO$_{Si}$} &  KAO$_{OH}$ \\
   System & Method & Cell  & $\alpha$ & $\beta$ & $\phi$ \\
   \hline
   \hline
   Bulk & Exp. &               & 133.9    & 104.6   & 163.4    \\
   \hline
   Bulk  & PBE  &              & 132.7    & 106.1   & 162.6    \\
   \hline
   Slab & CLAY\_FF & S3        & 118.2    & 122.4   & 165.9    \\
   \hline
   Slab & PBE & Exp.     & 144.8    & 93.3    & 168.2    \\
   \hline   Slab  & PBE & PBE    & 139.3    & 98.8    & 166.6    \\
   \hline
   Slab & PBE           & optPBE-vdW & 141.0    & 97.0    & 166.8    \\
   Slab & RPBE          & optPBE-vdW & 141.9    & 96.0    & 167.1    \\
   Slab & PBE-D2        & optPBE-vdW & 141.8    & 96.2    & 167.5    \\
   Slab & PBE-D3        & optPBE-vdW & 141.3    & 96.7    & 167.1    \\
   Slab & vdW(TS)       & optPBE-vdW & 141.3    & 96.6    & 167.0    \\
   Slab & vdW-DF2       & optPBE-vdW & 143.4    & 94.6    & 167.8    \\
   Slab & revPBE-vdW    & optPBE-vdW & 143.2    & 94.8    & 167.7    \\
   Slab & optB86b-vdW   & optPBE-vdW & 142.1    & 95.9    & 167.1    \\
   Slab & optPBE-vdW    & optPBE-vdW & 142.6    & 95.3    & 167.5    \\
   \hline
   \hline
   \end{tabular}
   }
\end{table}
\end{center}

\section{Computational geometries}

\begin{figure}[t!]
\renewcommand\figurename{Fig.~S$\!\!$}
\centerline{\includegraphics[width=8cm]{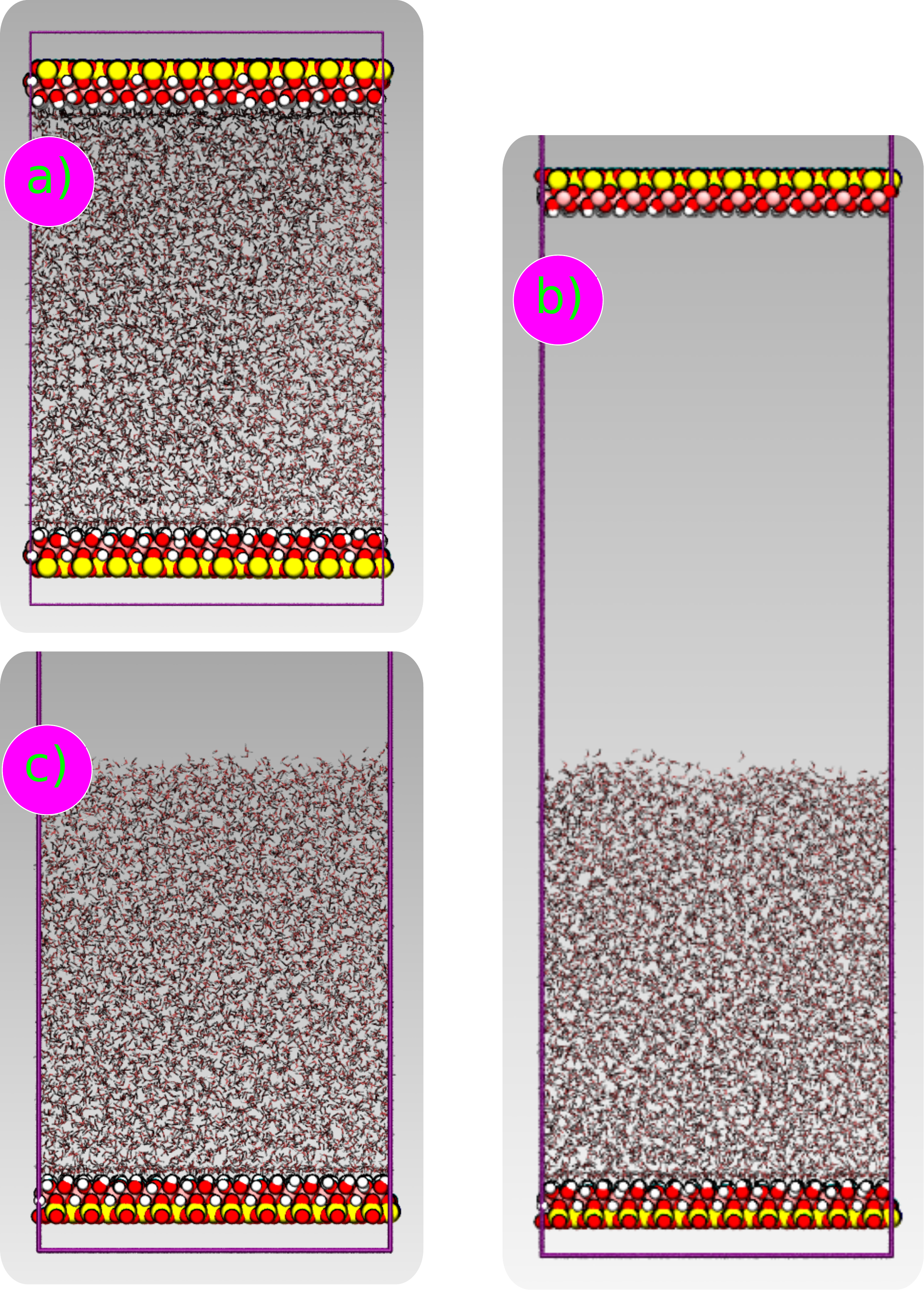}}
\caption{Representative computational geometries taken into account in this work. a) Water is sandwiched between two
mirroring kaolinite slab. Setups S1, S2 and SE\_2 employed this geometry. b) The upper kaolinite slab is moved far away
from the water film along the normal to the kaolinite surface. Setups SE\_5 and SE\_6 employed this geometry. c) The
upper kaolinite slab is removed. Setups S3, SE\_3 and SE\_4 employed this geometry. The simulation cell is shown as a
purple box.  Silicon, aluminum, oxygen and hydrogen atoms are depicted in yellow, pink, red and white respectively.
Atoms belonging to the kaolinite slab and the water film are represented by balls and sticks respectively.}
\label{FIG_S2} \end{figure}

In order to assess whether different computational geometries affect the results of our MD simulations of ice formation,
we have considered the setups described in Table II, also depicted in Fig.~\ref{FIG_S2}. \textbf{2SM} refers to mirrored two-slabs setup described
in Ref.~\citenum{zielke_simulations_2016} (Fig.~\ref{FIG_S2}a), while the setup where water is put on top of a single kaolinite slab is
labeled as \textbf{1S} (Fig.~\ref{FIG_S2}c). The \textbf{2SML} setup refers to the \textbf{2SM} setup where the upper slab has been moved
far away from the water film on top of the lower slab (Fig.~\ref{FIG_S2}b). In this way, the dipole moments of the two slabs still compensate
each other while the water network is free to relax at its natural density. Concerning restraints: \textbf{F} refers to
the situation where all the atoms in the kaolinite slab are kept frozen, avoiding the integration of the equations of
motion. Hydrogen atoms belonging to the hydroxyl groups on top of the KAO$_{OH}$ face are still allowed to move, though,
as they are bonded to the correspondent oxygen atoms via an harmonic constraint characterized by a spring constant of
2.3185$\cdot 10^3$ kJ/mol \AA$^{-2}$ acting on the O-H bond length (1.0 \AA), as required by the CLAY\_FF force
field~\cite{cygan_molecular_2004}. In the case of \textbf{N}, the positions of silicon atoms within the kaolinite slab
are restrained during the MD simulations by means of a harmonic potential characterized by a spring constant of
1.0$\cdot 10^3$ kJ/mol \AA$^{-2}$.  The O-H bonds are treated with the same harmonic constraint of \textbf{F}. All the
other kaolinite atoms are unrestrained. Note that we have chosen to restrain the silicon atoms at the bottom of the
KAO$_{OH}$ slab just in order to mimic the presence of additional kaolinite slabs below. In contrast to what has been
reported in Ref.~\citenum{zielke_simulations_2016}, we have been able to simulate a completely unrestrained slab without
observing the disruption of the system.  The \textbf{FH} case corresponds instead to the same setup as \textbf{F}, with
the difference that the dynamics of oxygen atoms of the hydroxyl groups is constrained via an harmonic potential
characterized by a spring constant of 1.0$\cdot 10^1$ kJ/mol \AA$^{-2}$, consistent with the "free-OH" setup described
in Ref.~\citenum{zielke_simulations_2016}. As we have discussed in the main text, such restraint is rather mild, and
does not prevent surface relaxation to take place.

\section{Order Parameter}
\label{sop}

In order to locate and characterize ice nuclei within our MD simulations, we have employed the following order
parameter $\lambda$: we start by
labeling as ice-like any water molecule whose oxygen atom displays a value of $lq^6>$0.45, where $lq^6$ is constructed as
follows: we first select only those oxygens which are hydrogen-bonded to four other oxygens.  For each of the $i-$th atoms of this
subset $S_{4HB}$, we calculate the local order parameter:

\begin{equation}
lq^6_i=\frac{\sum_{j=1}^{N_{S_{4HB}}} \sigma(|\mathbf{r}_{ij}|) \sum_{m=-6}^{6} q^{6*}_{i,m} \cdot q^6_{j,m} }  {\sum_{j=1}^{N_{S_{4HB}}} \sigma(|\mathbf{r}_{ij}|) }
\label{eq:L1}
\end{equation}

\noindent where $\sigma(|\mathbf{r}_{ij}|)$ is a switching function tuned so that $\sigma(|\mathbf{r}_{ij}|)$=1 when atom $j$ lies
within the first coordination shell of atom $i$ and which is zero otherwise. $q^6_{i,m}$ is the Steinhardt vector~\cite{steinhardt_bond-orientational_1983}

\begin{equation}
q^6_{i,m}=\frac { \sum_{j=1}^{N_{S_{4HB}}} \sigma(|\mathbf{r}_{ij}|) Y_{6m} (\mathbf{r}_{ij})  }    { \sum_{j=1}^{N_{S_{4HB}}} \sigma(|\mathbf{r}_{ij}|)  },
\label{eq:L2}
\end{equation}

\noindent $Y_{6m} (\mathbf{r}_{ij})$ being one of the 6th order spherical harmonics. We have used 3.2 \AA$\ $ as the cutoff for
$\sigma(|\mathbf{r}_{ij}|)$ to be consistent with Ref.~\citenum{haji-akbari_direct_2015}.  Notice that by selecting oxygen atoms within the
$S_{4HB}$ subset exclusively we ensure that the hydrogen bond network within the ice nuclei is reasonable. Having identified a set of ice-like water molecules, we pinpoint all the
connected clusters of oxygen atoms which: i) belong to the $S_{4HB}$ subset; ii) have a value of $lq^6>$0.45 and; iii)
are separated by a distance $\leq$ 3.2 \AA. We then select the largest of these clusters (i.e. the one
containing the largest number of oxygen atoms or equivalently water molecules). The final step is to find all the \textit{surface
molecules} that are connected to this cluster, as this procedure allows us to account for the diffuse interface between the solid and the liquid. 
Surface molecules are defined as the water molecules that lie within 3.2 \AA$\ $ of 
the molecules in the cluster.
The final order parameter $\lambda$ used in this work is thus the number of water molecules
within the largest ice-like cluster plus the number of surface molecules. This approach allow us to account for ice-like
atoms sitting directly on top of the kaolinite surface, which are never labeled as ice-like (and which would thus never be
included into the ice nuclei) because they are under coordinated and because they display a different symmetry to
the molecules within bulk water (which in turn leads to different values of $lq^6$).

As discussed in the main text, our seeded MD simulations allowed a very crude estimate of the critical nucleus size
which is consistent with our forward flux sampling (FFS) simulations~\cite{sosso_microscopic_2016}. The comparison between our
numbers ($\sim$ 250 and 225 $\pm$ 25 water molecules as obtained by seeded MD and FFS simulations respectively) and the
critical nucleus size for the homogeneous case at the same supercooling (and the same water model) reported in
Ref.~\citenum{haji-akbari_direct_2015} requires exactly the same order parameter to be used. In particular, the order
parameter used in Ref.~\citenum{haji-akbari_direct_2015} differs with respect to our formulation of $\lambda$ (see
Sec.~\ref{sop}) in that (i) a slightly stricter criterion has been used to label molecules as ice-like, namely
$lq^6>$0.5 to be compared with our choice of $lq^6>$0.45; and (ii) surface molecules are not included in the largest
ice-like nucleus. This means that in order to compare quantitatively our results in terms of e.g. the size of the
critical nucleus, the average number of surface molecules for a nucleus of a given size has to be added to the value of
$\lambda$ reported in Ref.~\citenum{haji-akbari_direct_2015}, resulting in an estimate of the homogeneous critical
nucleus size of $\sim$ 540 water molecules.

Finally, we note that in order to discriminate between ice-like molecules belonging to either the cubic of the hexagonal polytype,
we have employed the same approach outlined above but for the fact that we have used 3rd (instead of 6th) order spherical harmonics.
The values of the resulting order parameter $lq^3>$ can then be used~\cite{li_homogeneous_2011} to label $I_c$ and $I_h$ molecules according to
$lq^3<$ -0.85 and -0.85 $\le lq^3 \ge$ -0.70.

\section{Structural relaxation time at the water-kaolinite interface}

\begin{figure}[t!]
\begin{centering}
\centerline{\includegraphics[width=7 cm]{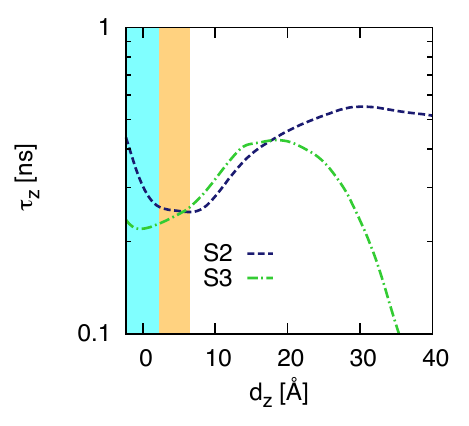}}
\par\end{centering}
\protect\caption{
Relaxation time $\tau_Z$ for S2 and S3 as a function of the distance $d_z$ along the
z axis, normal to the KAO slab, from the first peak of the water density profile ($d_z=0$).
The regions highlighted in light blue
and orange correspond to the first and second water over layer on top of the KAO slab, respectively.}
\label{FIG_S3}
\end{figure}

The dynamics of the water-KAO$_{OH}$ interface can be characterized by computing the
structural relaxation time of the water network similarly to what we have done for bulk water (see Sec.~\ref{dynw}).
However, such a calculation presents two challenges:

\begin{itemize}
\item{The water-KAO$_{OH}$ interface is a non-homogeneous system. As such, we need a different formulation for the intermediate
scattering function. To this end, we have adopted the approach described in the excellent work of Haji-Akbari and 
Debenedetti~\cite{haji-akbari_effect_2014}: the system is partitioned in slices along the normal to the substrate, and the dynamical quantities of
interest are computed within each of these regions taking into account the contributions of molecules that are found within each slice at the
beginning and at the end of the time window considered (see Eq.13 in Ref.~\citenum{haji-akbari_effect_2014}). In this way, the relaxation time discussed
in Sec.~\ref{dynw} becomes a function of the distance from the substrate $d_z$, so that we can probe the water dynamics in different
region of the water-kaolinite interface.}
\item{Dynamical quantities typically require much more statistics than structural ones to obtain converged results. In particular,
we are interested in equilibrium averages, so that we have to be sure that we are computing properties within a timescale where the
-metastable - liquid can be considered in equilibrium conditions. Thus, we here face an issue when dealing with the S1 interface: in this
case, the onset of crystallization takes place within $\sim$ 50-100 ns, and while we have verified that structural properties can be
converged within that time window, the same does not hold for quantities such as the structural relaxation time $\tau_z$}
\end{itemize}

In Fig.~\ref{FIG_S3} we report the relaxation time $\tau_Z$ of the water network as a function of the distance from the
kaolinite surface for S2 and S3.  Water dynamics is slower for S2 (and qualitatively for S1 as well, albeit we could not obtain converged results), 
as the frozen surface interacts more strongly with
the water molecules with respect to S3 at least within the first over layer (light blue region in Fig.~\ref{FIG_S3}),
consistently with early findings for Lennard-Jones interfaces~\cite{haji-akbari_effect_2014}.  The relaxation times are
very similar for S2 and S3 within the second over layer, albeit they converge to a different value within the bulk of
the water film. This is due the different computational geometry: in S2 the water film is sandwiched between two
kaolinite slab, while for S3 the presence of the water-vacuum interface affect substantially the dynamics of the system
starting from $d_z\sim$ 20\AA. Note that despite these differences, the free energy cost needed to create a templating
hexagonal patch of water molecules within the first water over layer is very similar for S2 and S3 (see main text).
Moreover, the absence of nucleation events on the same timescale for both of these two interfaces suggest the that the
dynamics of the interface is far less relevant than its structure in affecting ice formation.

\section{Nucleation or structural relaxation?}

\begin{figure}[t!]
\renewcommand\figurename{Fig.~S$\!\!$}
\centerline{\includegraphics[width=7cm]{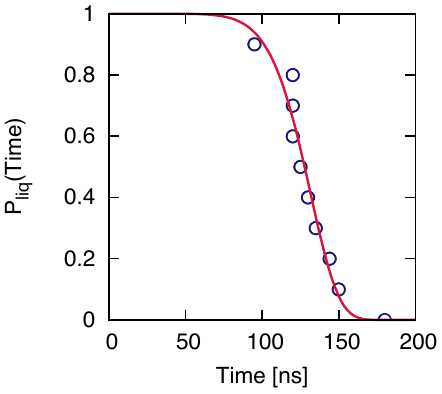}}
\caption{Survival probability $P_{liq}(t)$ (circles) obtained for ice formation at the S1 interface. The solid line represent the
fit according to Eq.~\ref{ffit} in the text.}
\label{FIG_S4}
\end{figure}

In the main text we have discussed ice formation at the S1 interface, where all the kaolinite atoms, but for the
hydrogens of the hydroxyl groups, are kept frozen during the MD simulations at the experimental positions of
bulk kaolinite. In this case, we have observed ice formation at the water-kaolinite interface within $\sim$ 100 ns
for each of the ten independent MD runs we have performed. 

We have determined the onset time $t_n$ of "nucleation" by fitting the time evolution of the order parameter $\lambda$ described in
Sec.~\ref{sop} with the following expression:

\begin{equation}
\lambda(t)=a+\frac{b}{1+\exp[c(t-t_n)]}
\end{equation}

\noindent where $a$, $b$ and $c$ are fitting parameters.
The survival probability of the liquid $P_{liq}(t)$ can then be constructed from the distribution of $t_n$ as:

\begin{equation}
P_{liq}(t)=1-\frac{1}{N_s}\sum_{i=1}^{N_s}\Theta(t-t_n^{(i)})
\end{equation}

\noindent where the sum runs over the $N_{s}$ MD simulations we have performed and $\Theta$ is a Heaviside step function.
The $P_{liq}(t)$ obtained for S1 is reported in Fig.~\ref{FIG_S4} together with the fitting of the data points with
respect to the following expression:

\begin{equation}
P_{liq}(t)=exp[-(J\cdot t)^{\gamma}]
\label{ffit}
\end{equation}

\noindent where $J$ is the nucleation rate and $\gamma$ is a parameter accounting for non-exponential kinetics. For a 
proper nucleation process, $\gamma$=1, which is consistent with a distribution of nucleation times following a Poisson
distribution due to the stochastic nature of the nucleation events. Deviation from the ideal value of $\gamma$ can be
observed in many cases even experimentally, as nicely discussed in Ref,~\citenum{sear_quantitative_2014}. However, we
have shown in Ref.~\citenum{fitzner_many_2015} that values of $\gamma\neq$1 are often observed in atomistic simulations
when nucleation times are of the same order of the relaxation time of the parent phase (in this case, the supercooled
water network). This is exactly the situation we observe in here: we obtain $\gamma$=8, spectacularly different from
the value of 1 expected, as $t_n$ are of the same order of the time scale needed for S1 to create a complete hexagonal
motif of water molecules within the first over layer, thus triggering ice formation in a non-stochastic fashion. 
As such, we argue that S1 is unstable with respect to the formation of the hexagonal templating over layer, and thus with respect to the
formation of ice itself.

\section{Metadynamics simulations}
\label{mes}

\begin{figure}[b!]
\renewcommand\figurename{Fig.~S$\!\!$}
\centerline{\includegraphics[width=9cm]{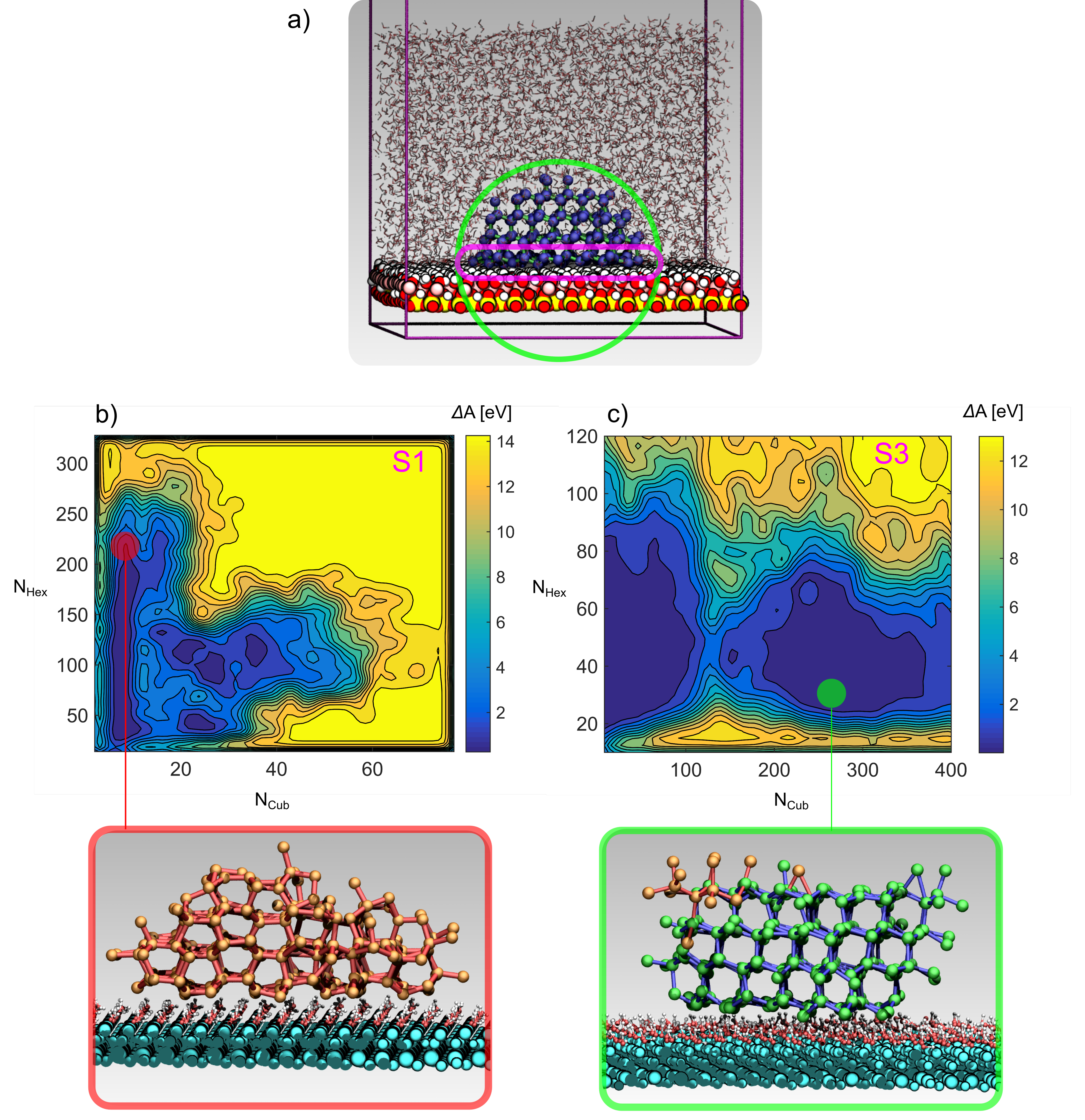}}
\caption{a) Computational setup used to perform metadynamics simulations. The algorithm acts on water molecules within the spherical region (light green circle) centered on a dummy,
fixed atom, except for those molecules within the first over layer on top of the kaolinite (purple oval).
b) and c) Metadynamics simulations of heterogeneous ice nucleation on the KAO$_{OH}$ surface. a) and b) panels refer to setups
S1 and S3
respectively. The free energy surfaces obtained are reported
as a function of the number of water molecules in the largest clusters of cubic ($N_{cub}$) and hexagonal ($N_{hex}$)
polytypes of ice.
The insets in panels b) and c) depict large
clusters of hexagonal and cubic ice respectively, the former exposing the primary prismatic face of I$_h$ (orange
spheres, red sticks)
to the KAO$_{OH}$ surface and the second one
growing on top of the basal plane of I$_c$ (green spheres, blue sticks). The KAO$_{OH}$ surface is depicted by light
blue spheres irrespective of the atomic species, but
for the -OH groups (oxygens and hydrogens as red and white spheres respectively).}
\label{FIG_S5}
\end{figure}

Metadynamics simulations have been performed using the PLUMED metadynamics plugin interfaced with the GROMACS MD package.
The best choice in terms of the collective variable (CV) 
is probably the order parameter $\lambda$ we have discussed in Sec.~\ref{sop}. This approach is now possible thanks to recent
computational advancements~\cite{tribello_dfs_2016} capable of dealing with clusters of particles as CVs.
However, this option is unfeasible here due to the very large (10$^6$) number of water molecules we have to take into account. 
In fact, the forces acting on the atoms due to the metadynamics bias are computed as the analytic derivatives
of the CV with respect to the atomic positions: unfortunately, this calculation is still exceedingly computationally expensive in the case of $\lambda$
when dealing with systems containing more than 10$^2$ particles. Thus, we have used as CV the \textit{mean} value of $lq^6_{i}$ (described in Eq.~\ref{eq:L1}), that is the value
of $lq^6_{i}$ averaged over all the water molecules of interest. 

However, considering all the water molecules in the system ($\sim$ 6000) is still 
too costly from a computational point of view. Thus, we have chosen to take into account at each metadynamics setup only those water molecules within
a spherical region or radius 16.0 \AA$\ $ centered on a dummy, fixed atom, as depicted in Fig.~\ref{FIG_S5}. In addition, in order to (i) avoid the
contribution of those water molecules on top of the kaolinite, which $lq^6_{i}$ values will be ill-defined because of the different coordination and
topology and (ii) ensure that water molecules are free to re-arrange themselves as they see fit at the water-kaolinite interface, we have excluded
from the above mentioned spherical region those water molecules within the first over layer on the surface, as illustrated in Fig.~\ref{FIG_S5}. 
By means of this strategy we reduce the number of molecules for which we have to calculate the expensive derivatives of the CV from $\sim$ 10$^3$ to to $10^2$, and
we drive explicitly nucleation events within a spherical cap on the surface, consistent with the prescription of classical nucleation theory. 
The expression for the CV we have used thus reads:

\begin{equation}
CV=\frac{1}{N_{Sph}}\sum_{j}^{N_{Sph}} lq^6_j
\end{equation}

where the sum runs over the $N_{Sph}$ atoms within the spherical cap described in Fig.~\ref{FIG_S5}. This CV is 0 and 1 for a perfectly disordered and network 
respectively. See Ref.~\citenum{li_homogeneous_2011} for the distribution of the local order parameter as obtained for supercooled water and ice.

The dynamics of the water network at the strong supercooling $\Delta T$=42 K considered is very slow. Thus, in order to observe transitions
from liquid water to ice and viceversa, as required to converge the resulting free energy surface, we have been forced to drive nucleation events quite harshly:
specifically, we have chosen Gaussians with width equal to $\sim \frac{1}{4}$ of the extent of the natural fluctuations of the CV. The initial height of the
Gaussians was 10 $k_B\cdot T$, which decays in time according to the well-tempered metadynamics framework~\cite{barducci_well-tempered_2008} with a bias factor of 200.
The bias was applied every 2000 MD steps, corresponding to 4 ps in terms of simulation time. We have also restrained the value of the CV to be $>$ 0.1, in order to
avoid non-relevant regions of the liquid basin. As a result, the free energy profiles we have obtained for S1 and S3 are converged within 0.4 eV. This level of
accuracy prevents us to draw conclusions about the thermodynamics of nucleation, but it is enough to provide a qualitative picture in terms of the different
polytypes involved, as illustrated in Fig.~\ref{FIG_S5} in the main text. There, we report free energy surfaces as a function of the number of water molecules in the largest nucleus
of $I_h$ and $I_c$. These results 
have been obtained by re-weighting the free energy profiles in terms of the original CV as detailed in Ref.~\citenum{tiwary_time-independent_2015}.

The resulting free energy surfaces in terms of the number of water molecules within the largest cluster of
hexagonal and cubic ice ($N_{Hex}$ and $N_{Cub}$) are shown in Fig.~\ref{FIG_S5}. In the case of S1, creating a nucleus
of hexagonal ice containing as many as 250 water molecules is a barrier less process, consistent with what we have
observed in the unbiased MD simulations.  In contrast, for S3 we observe the exclusive
occurrence of the cubic polytype,
which forms when the system overcomes a well-defined free energy barrier. Thus, one could think that structural
relaxation can promote the formation of different ice polytypes on the very same surface.  However, this is not the
case, as the formation of cubic ice for S3 leads to nuclei which do not proceed toward further crystallization. The
shape of the nuclei as obtained by metadynamics simulations already suggests that the basal plane of $I_c$ is not
favored to be in contact with the clay. In fact, as depicted in Fig.~\ref{FIG_S5} $I_c$ nuclei tend no to wet the
kaolinite surface, as opposite to what we have observed for $I_h$ nuclei on the S1 surface, where the prism face of
$I_h$ easily spread on the clay.

We remark that using 
a different CV (taking into account $lq^3$ instead of $lq^6$, that is, using a different angular momentum channel for the spherical harmonics) resulted in the same
outcome. Thus, it is very much possible that there is still room for improvement in terms of order parameters to drive crystal nucleation at complex interfaces.

\bibliography{Text_Sup.bib}

\begin{thebibliography}{110}%
\makeatletter
\providecommand \@ifxundefined [1]{%
 \@ifx{#1\undefined}
}%
\providecommand \@ifnum [1]{%
 \ifnum #1\expandafter \@firstoftwo
 \else \expandafter \@secondoftwo
 \fi
}%
\providecommand \@ifx [1]{%
 \ifx #1\expandafter \@firstoftwo
 \else \expandafter \@secondoftwo
 \fi
}%
\providecommand \natexlab [1]{#1}%
\providecommand \enquote  [1]{``#1''}%
\providecommand \bibnamefont  [1]{#1}%
\providecommand \bibfnamefont [1]{#1}%
\providecommand \citenamefont [1]{#1}%
\providecommand \href@noop [0]{\@secondoftwo}%
\providecommand \href [0]{\begingroup \@sanitize@url \@href}%
\providecommand \@href[1]{\@@startlink{#1}\@@href}%
\providecommand \@@href[1]{\endgroup#1\@@endlink}%
\providecommand \@sanitize@url [0]{\catcode `\\12\catcode `\$12\catcode
  `\&12\catcode `\#12\catcode `\^12\catcode `\_12\catcode `\%12\relax}%
\providecommand \@@startlink[1]{}%
\providecommand \@@endlink[0]{}%
\providecommand \url  [0]{\begingroup\@sanitize@url \@url }%
\providecommand \@url [1]{\endgroup\@href {#1}{\urlprefix }}%
\providecommand \urlprefix  [0]{URL }%
\providecommand \Eprint [0]{\href }%
\providecommand \doibase [0]{http://dx.doi.org/}%
\providecommand \selectlanguage [0]{\@gobble}%
\providecommand \bibinfo  [0]{\@secondoftwo}%
\providecommand \bibfield  [0]{\@secondoftwo}%
\providecommand \translation [1]{[#1]}%
\providecommand \BibitemOpen [0]{}%
\providecommand \bibitemStop [0]{}%
\providecommand \bibitemNoStop [0]{.\EOS\space}%
\providecommand \EOS [0]{\spacefactor3000\relax}%
\providecommand \BibitemShut  [1]{\csname bibitem#1\endcsname}%
\let\auto@bib@innerbib\@empty
\bibitem [{\citenamefont
  {Bartels-Rausch}(2013)}]{bartels-rausch_chemistry:_2013}%
  \BibitemOpen
  \bibfield  {author} {\bibinfo {author} {\bibfnamefont {T.}~\bibnamefont
  {Bartels-Rausch}},\ }\href {\doibase 10.1038/494027a} {\bibfield  {journal}
  {\bibinfo  {journal} {Nature}\ }\textbf {\bibinfo {volume} {494}},\ \bibinfo
  {pages} {27} (\bibinfo {year} {2013})}\BibitemShut {NoStop}%
\bibitem [{\citenamefont {Tang}, \citenamefont {Cziczo},\ and\ \citenamefont
  {Grassian}(2016)}]{tang_interactions_2016}%
  \BibitemOpen
  \bibfield  {author} {\bibinfo {author} {\bibfnamefont {M.}~\bibnamefont
  {Tang}}, \bibinfo {author} {\bibfnamefont {D.~J.}\ \bibnamefont {Cziczo}}, \
  and\ \bibinfo {author} {\bibfnamefont {V.~H.}\ \bibnamefont {Grassian}},\
  }\href {\doibase 10.1021/acs.chemrev.5b00529} {\bibfield  {journal} {\bibinfo
   {journal} {Chem. Rev.}\ }\textbf {\bibinfo {volume} {116}},\ \bibinfo
  {pages} {4205} (\bibinfo {year} {2016})}\BibitemShut {NoStop}%
\bibitem [{\citenamefont {Bartels-Rausch}\ \emph {et~al.}(2012)\citenamefont
  {Bartels-Rausch}, \citenamefont {Bergeron}, \citenamefont {Cartwright},
  \citenamefont {Escribano}, \citenamefont {Finney}, \citenamefont {Grothe},
  \citenamefont {Gutiérrez}, \citenamefont {Haapala}, \citenamefont {Kuhs},
  \citenamefont {Pettersson}, \citenamefont {Price}, \citenamefont
  {Sainz-Díaz}, \citenamefont {Stokes}, \citenamefont {Strazzulla},
  \citenamefont {Thomson}, \citenamefont {Trinks},\ and\ \citenamefont
  {Uras-Aytemiz}}]{bartels-rausch_ice_2012}%
  \BibitemOpen
  \bibfield  {author} {\bibinfo {author} {\bibfnamefont {T.}~\bibnamefont
  {Bartels-Rausch}}, \bibinfo {author} {\bibfnamefont {V.}~\bibnamefont
  {Bergeron}}, \bibinfo {author} {\bibfnamefont {J.~H.~E.}\ \bibnamefont
  {Cartwright}}, \bibinfo {author} {\bibfnamefont {R.}~\bibnamefont
  {Escribano}}, \bibinfo {author} {\bibfnamefont {J.~L.}\ \bibnamefont
  {Finney}}, \bibinfo {author} {\bibfnamefont {H.}~\bibnamefont {Grothe}},
  \bibinfo {author} {\bibfnamefont {P.~J.}\ \bibnamefont {Gutiérrez}},
  \bibinfo {author} {\bibfnamefont {J.}~\bibnamefont {Haapala}}, \bibinfo
  {author} {\bibfnamefont {W.~F.}\ \bibnamefont {Kuhs}}, \bibinfo {author}
  {\bibfnamefont {J.~B.~C.}\ \bibnamefont {Pettersson}}, \bibinfo {author}
  {\bibfnamefont {S.~D.}\ \bibnamefont {Price}}, \bibinfo {author}
  {\bibfnamefont {C.~I.}\ \bibnamefont {Sainz-Díaz}}, \bibinfo {author}
  {\bibfnamefont {D.~J.}\ \bibnamefont {Stokes}}, \bibinfo {author}
  {\bibfnamefont {G.}~\bibnamefont {Strazzulla}}, \bibinfo {author}
  {\bibfnamefont {E.~S.}\ \bibnamefont {Thomson}}, \bibinfo {author}
  {\bibfnamefont {H.}~\bibnamefont {Trinks}}, \ and\ \bibinfo {author}
  {\bibfnamefont {N.}~\bibnamefont {Uras-Aytemiz}},\ }\href {\doibase
  10.1103/RevModPhys.84.885} {\bibfield  {journal} {\bibinfo  {journal} {Rev.
  Mod. Phys.}\ }\textbf {\bibinfo {volume} {84}},\ \bibinfo {pages} {885}
  (\bibinfo {year} {2012})}\BibitemShut {NoStop}%
\bibitem [{\citenamefont {Tam}\ \emph {et~al.}(2009)\citenamefont {Tam},
  \citenamefont {Rowley}, \citenamefont {Petrov}, \citenamefont {Zhang},
  \citenamefont {Afagh}, \citenamefont {Woo},\ and\ \citenamefont
  {Ben}}]{tam_solution_2009}%
  \BibitemOpen
  \bibfield  {author} {\bibinfo {author} {\bibfnamefont {R.~Y.}\ \bibnamefont
  {Tam}}, \bibinfo {author} {\bibfnamefont {C.~N.}\ \bibnamefont {Rowley}},
  \bibinfo {author} {\bibfnamefont {I.}~\bibnamefont {Petrov}}, \bibinfo
  {author} {\bibfnamefont {T.}~\bibnamefont {Zhang}}, \bibinfo {author}
  {\bibfnamefont {N.~A.}\ \bibnamefont {Afagh}}, \bibinfo {author}
  {\bibfnamefont {T.~K.}\ \bibnamefont {Woo}}, \ and\ \bibinfo {author}
  {\bibfnamefont {R.~N.}\ \bibnamefont {Ben}},\ }\href {\doibase
  10.1021/ja904169a} {\bibfield  {journal} {\bibinfo  {journal} {J. Am. Chem.
  Soc.}\ }\textbf {\bibinfo {volume} {131}},\ \bibinfo {pages} {15745}
  (\bibinfo {year} {2009})}\BibitemShut {NoStop}%
\bibitem [{\citenamefont {Sellberg}\ \emph {et~al.}(2014)\citenamefont
  {Sellberg}, \citenamefont {Huang}, \citenamefont {McQueen}, \citenamefont
  {Loh}, \citenamefont {Laksmono}, \citenamefont {Schlesinger}, \citenamefont
  {Sierra}, \citenamefont {Nordlund}, \citenamefont {Hampton}, \citenamefont
  {Starodub}, \citenamefont {DePonte}, \citenamefont {Beye}, \citenamefont
  {Chen}, \citenamefont {Martin}, \citenamefont {Barty}, \citenamefont
  {Wikfeldt}, \citenamefont {Weiss}, \citenamefont {Caronna}, \citenamefont
  {Feldkamp}, \citenamefont {Skinner}, \citenamefont {Seibert}, \citenamefont
  {Messerschmidt}, \citenamefont {Williams}, \citenamefont {Boutet},
  \citenamefont {Pettersson}, \citenamefont {Bogan},\ and\ \citenamefont
  {Nilsson}}]{sellberg_ultrafast_2014}%
  \BibitemOpen
  \bibfield  {author} {\bibinfo {author} {\bibfnamefont {J.~A.}\ \bibnamefont
  {Sellberg}}, \bibinfo {author} {\bibfnamefont {C.}~\bibnamefont {Huang}},
  \bibinfo {author} {\bibfnamefont {T.~A.}\ \bibnamefont {McQueen}}, \bibinfo
  {author} {\bibfnamefont {N.~D.}\ \bibnamefont {Loh}}, \bibinfo {author}
  {\bibfnamefont {H.}~\bibnamefont {Laksmono}}, \bibinfo {author}
  {\bibfnamefont {D.}~\bibnamefont {Schlesinger}}, \bibinfo {author}
  {\bibfnamefont {R.~G.}\ \bibnamefont {Sierra}}, \bibinfo {author}
  {\bibfnamefont {D.}~\bibnamefont {Nordlund}}, \bibinfo {author}
  {\bibfnamefont {C.~Y.}\ \bibnamefont {Hampton}}, \bibinfo {author}
  {\bibfnamefont {D.}~\bibnamefont {Starodub}}, \bibinfo {author}
  {\bibfnamefont {D.~P.}\ \bibnamefont {DePonte}}, \bibinfo {author}
  {\bibfnamefont {M.}~\bibnamefont {Beye}}, \bibinfo {author} {\bibfnamefont
  {C.}~\bibnamefont {Chen}}, \bibinfo {author} {\bibfnamefont {A.~V.}\
  \bibnamefont {Martin}}, \bibinfo {author} {\bibfnamefont {A.}~\bibnamefont
  {Barty}}, \bibinfo {author} {\bibfnamefont {K.~T.}\ \bibnamefont {Wikfeldt}},
  \bibinfo {author} {\bibfnamefont {T.~M.}\ \bibnamefont {Weiss}}, \bibinfo
  {author} {\bibfnamefont {C.}~\bibnamefont {Caronna}}, \bibinfo {author}
  {\bibfnamefont {J.}~\bibnamefont {Feldkamp}}, \bibinfo {author}
  {\bibfnamefont {L.~B.}\ \bibnamefont {Skinner}}, \bibinfo {author}
  {\bibfnamefont {M.~M.}\ \bibnamefont {Seibert}}, \bibinfo {author}
  {\bibfnamefont {M.}~\bibnamefont {Messerschmidt}}, \bibinfo {author}
  {\bibfnamefont {G.~J.}\ \bibnamefont {Williams}}, \bibinfo {author}
  {\bibfnamefont {S.}~\bibnamefont {Boutet}}, \bibinfo {author} {\bibfnamefont
  {L.~G.~M.}\ \bibnamefont {Pettersson}}, \bibinfo {author} {\bibfnamefont
  {M.~J.}\ \bibnamefont {Bogan}}, \ and\ \bibinfo {author} {\bibfnamefont
  {A.}~\bibnamefont {Nilsson}},\ }\href {\doibase 10.1038/nature13266}
  {\bibfield  {journal} {\bibinfo  {journal} {Nature}\ }\textbf {\bibinfo
  {volume} {510}},\ \bibinfo {pages} {381} (\bibinfo {year}
  {2014})}\BibitemShut {NoStop}%
\bibitem [{\citenamefont {Martin}(1981)}]{martin_frazil_1981}%
  \BibitemOpen
  \bibfield  {author} {\bibinfo {author} {\bibfnamefont {S.}~\bibnamefont
  {Martin}},\ }\href {\doibase 10.1146/annurev.fl.13.010181.002115} {\bibfield
  {journal} {\bibinfo  {journal} {Ann. Rev. of Flu. Mech.}\ }\textbf {\bibinfo
  {volume} {13}},\ \bibinfo {pages} {379} (\bibinfo {year} {1981})}\BibitemShut
  {NoStop}%
\bibitem [{\citenamefont {Lepp\"{a}ranta}(2015)}]{lepparanta_2015}%
  \BibitemOpen
  \bibfield  {author} {\bibinfo {author} {\bibfnamefont {M.}~\bibnamefont
  {Lepp\"{a}ranta}},\ }\href@noop {} {\emph {\bibinfo {title} {Freezing of
  lakes and the evolution of their ice cover}}}\ (\bibinfo  {publisher}
  {Heidelberg : Springer},\ \bibinfo {year} {2015})\BibitemShut {NoStop}%
\bibitem [{\citenamefont {Welch}\ and\ \citenamefont
  {Davis}(1980)}]{welch_1980}%
  \BibitemOpen
  \bibfield  {author} {\bibinfo {author} {\bibfnamefont {S.}~\bibnamefont
  {Welch}, \bibfnamefont {R.M.;~Cox}}\ and\ \bibinfo {author} {\bibfnamefont
  {J.}~\bibnamefont {Davis}},\ }\href@noop {} {\emph {\bibinfo {title} {Solar
  Radiation and Clouds}}}\ (\bibinfo  {publisher} {Boston : American
  Meteorological Society},\ \bibinfo {year} {1980})\BibitemShut {NoStop}%
\bibitem [{\citenamefont {Yang}\ \emph {et~al.}(2014)\citenamefont {Yang},
  \citenamefont {Liou}, \citenamefont {Bi}, \citenamefont {Liu}, \citenamefont
  {Yi},\ and\ \citenamefont {Baum}}]{yang_radiative_2014}%
  \BibitemOpen
  \bibfield  {author} {\bibinfo {author} {\bibfnamefont {P.}~\bibnamefont
  {Yang}}, \bibinfo {author} {\bibfnamefont {K.-N.}\ \bibnamefont {Liou}},
  \bibinfo {author} {\bibfnamefont {L.}~\bibnamefont {Bi}}, \bibinfo {author}
  {\bibfnamefont {C.}~\bibnamefont {Liu}}, \bibinfo {author} {\bibfnamefont
  {B.}~\bibnamefont {Yi}}, \ and\ \bibinfo {author} {\bibfnamefont {B.~A.}\
  \bibnamefont {Baum}},\ }\href {\doibase 10.1007/s00376-014-0011-z} {\bibfield
   {journal} {\bibinfo  {journal} {Adv. Atmos. Sci.}\ }\textbf {\bibinfo
  {volume} {32}},\ \bibinfo {pages} {32} (\bibinfo {year} {2014})}\BibitemShut
  {NoStop}%
\bibitem [{\citenamefont {Murray}\ \emph {et~al.}(2012)\citenamefont {Murray},
  \citenamefont {O'Sullivan}, \citenamefont {Atkinson},\ and\ \citenamefont
  {Webb}}]{murray_ice_2012}%
  \BibitemOpen
  \bibfield  {author} {\bibinfo {author} {\bibfnamefont {B.~J.}\ \bibnamefont
  {Murray}}, \bibinfo {author} {\bibfnamefont {D.}~\bibnamefont {O'Sullivan}},
  \bibinfo {author} {\bibfnamefont {J.~D.}\ \bibnamefont {Atkinson}}, \ and\
  \bibinfo {author} {\bibfnamefont {M.~E.}\ \bibnamefont {Webb}},\ }\href
  {\doibase 10.1039/C2CS35200A} {\bibfield  {journal} {\bibinfo  {journal}
  {Chem. Soc. Rev.}\ }\textbf {\bibinfo {volume} {41}},\ \bibinfo {pages}
  {6519} (\bibinfo {year} {2012})}\BibitemShut {NoStop}%
\bibitem [{\citenamefont {Slater}\ \emph {et~al.}(2015)\citenamefont {Slater},
  \citenamefont {Michaelides}, \citenamefont {Salzmann},\ and\ \citenamefont
  {Lohmann}}]{slater_blue-sky_2015}%
  \BibitemOpen
  \bibfield  {author} {\bibinfo {author} {\bibfnamefont {B.}~\bibnamefont
  {Slater}}, \bibinfo {author} {\bibfnamefont {A.}~\bibnamefont {Michaelides}},
  \bibinfo {author} {\bibfnamefont {C.~G.}\ \bibnamefont {Salzmann}}, \ and\
  \bibinfo {author} {\bibfnamefont {U.}~\bibnamefont {Lohmann}},\ }\href
  {\doibase 10.1175/BAMS-D-15-00131.1} {\bibfield  {journal} {\bibinfo
  {journal} {Bull. Amer. Meteor. Soc.}\ } (\bibinfo {year} {2015}),\
  10.1175/BAMS-D-15-00131.1}\BibitemShut {NoStop}%
\bibitem [{\citenamefont {Murray}\ \emph {et~al.}(2010)\citenamefont {Murray},
  \citenamefont {Wilson}, \citenamefont {Dobbie}, \citenamefont {Cui},
  \citenamefont {Al-Jumur}, \citenamefont {Möhler}, \citenamefont {Schnaiter},
  \citenamefont {Wagner}, \citenamefont {Benz}, \citenamefont {Niemand},
  \citenamefont {Saathoff}, \citenamefont {Ebert}, \citenamefont {Wagner},\
  and\ \citenamefont {Kärcher}}]{murray_heterogeneous_2010}%
  \BibitemOpen
  \bibfield  {author} {\bibinfo {author} {\bibfnamefont {B.~J.}\ \bibnamefont
  {Murray}}, \bibinfo {author} {\bibfnamefont {T.~W.}\ \bibnamefont {Wilson}},
  \bibinfo {author} {\bibfnamefont {S.}~\bibnamefont {Dobbie}}, \bibinfo
  {author} {\bibfnamefont {Z.}~\bibnamefont {Cui}}, \bibinfo {author}
  {\bibfnamefont {S.~M. R.~K.}\ \bibnamefont {Al-Jumur}}, \bibinfo {author}
  {\bibfnamefont {O.}~\bibnamefont {Möhler}}, \bibinfo {author} {\bibfnamefont
  {M.}~\bibnamefont {Schnaiter}}, \bibinfo {author} {\bibfnamefont
  {R.}~\bibnamefont {Wagner}}, \bibinfo {author} {\bibfnamefont
  {S.}~\bibnamefont {Benz}}, \bibinfo {author} {\bibfnamefont {M.}~\bibnamefont
  {Niemand}}, \bibinfo {author} {\bibfnamefont {H.}~\bibnamefont {Saathoff}},
  \bibinfo {author} {\bibfnamefont {V.}~\bibnamefont {Ebert}}, \bibinfo
  {author} {\bibfnamefont {S.}~\bibnamefont {Wagner}}, \ and\ \bibinfo {author}
  {\bibfnamefont {B.}~\bibnamefont {Kärcher}},\ }\href {\doibase
  10.1038/ngeo817} {\bibfield  {journal} {\bibinfo  {journal} {Nat. Geosci.}\
  }\textbf {\bibinfo {volume} {3}},\ \bibinfo {pages} {233} (\bibinfo {year}
  {2010})}\BibitemShut {NoStop}%
\bibitem [{\citenamefont {Campbell}, \citenamefont {Meldrum},\ and\
  \citenamefont {Christenson}(2015)}]{campbell_is_2015}%
  \BibitemOpen
  \bibfield  {author} {\bibinfo {author} {\bibfnamefont {J.~M.}\ \bibnamefont
  {Campbell}}, \bibinfo {author} {\bibfnamefont {F.~C.}\ \bibnamefont
  {Meldrum}}, \ and\ \bibinfo {author} {\bibfnamefont {H.~K.}\ \bibnamefont
  {Christenson}},\ }\href {\doibase 10.1021/jp5113729} {\bibfield  {journal}
  {\bibinfo  {journal} {J. Phys. Chem. C}\ }\textbf {\bibinfo {volume} {119}},\
  \bibinfo {pages} {1164} (\bibinfo {year} {2015})}\BibitemShut {NoStop}%
\bibitem [{\citenamefont {Whale}\ \emph {et~al.}(2015)\citenamefont {Whale},
  \citenamefont {Rosillo-Lopez}, \citenamefont {Murray},\ and\ \citenamefont
  {Salzmann}}]{whale_ice_2015}%
  \BibitemOpen
  \bibfield  {author} {\bibinfo {author} {\bibfnamefont {T.~F.}\ \bibnamefont
  {Whale}}, \bibinfo {author} {\bibfnamefont {M.}~\bibnamefont
  {Rosillo-Lopez}}, \bibinfo {author} {\bibfnamefont {B.~J.}\ \bibnamefont
  {Murray}}, \ and\ \bibinfo {author} {\bibfnamefont {C.~G.}\ \bibnamefont
  {Salzmann}},\ }\href {\doibase 10.1021/acs.jpclett.5b01096} {\bibfield
  {journal} {\bibinfo  {journal} {J. Phys. Chem. Lett.}\ }\textbf {\bibinfo
  {volume} {6}},\ \bibinfo {pages} {3012} (\bibinfo {year} {2015})}\BibitemShut
  {NoStop}%
\bibitem [{\citenamefont {Chipot}(2007)}]{fecc}%
  \BibitemOpen
  \bibfield  {author} {\bibinfo {author} {\bibfnamefont {A.}~\bibnamefont
  {Chipot}, \bibfnamefont {C;~Pohorille}},\ }\href@noop {} {\emph {\bibinfo
  {title} {Free Energy Calculations : Theory and Applications in Chemistry and
  Biology}}}\ (\bibinfo  {publisher} {Berlin : Springer},\ \bibinfo {year}
  {2007})\BibitemShut {NoStop}%
\bibitem [{\citenamefont {Sosso}\ \emph
  {et~al.}(2016{\natexlab{a}})\citenamefont {Sosso}, \citenamefont {Chen},
  \citenamefont {Cox}, \citenamefont {Fitzner}, \citenamefont {Pedevilla},
  \citenamefont {Zen},\ and\ \citenamefont {Michaelides}}]{sosso_crystal_2016}%
  \BibitemOpen
  \bibfield  {author} {\bibinfo {author} {\bibfnamefont {G.~C.}\ \bibnamefont
  {Sosso}}, \bibinfo {author} {\bibfnamefont {J.}~\bibnamefont {Chen}},
  \bibinfo {author} {\bibfnamefont {S.~J.}\ \bibnamefont {Cox}}, \bibinfo
  {author} {\bibfnamefont {M.}~\bibnamefont {Fitzner}}, \bibinfo {author}
  {\bibfnamefont {P.}~\bibnamefont {Pedevilla}}, \bibinfo {author}
  {\bibfnamefont {A.}~\bibnamefont {Zen}}, \ and\ \bibinfo {author}
  {\bibfnamefont {A.}~\bibnamefont {Michaelides}},\ }\href {\doibase
  10.1021/acs.chemrev.5b00744} {\bibfield  {journal} {\bibinfo  {journal}
  {Chem. Rev.}\ }\textbf {\bibinfo {volume} {116}},\ \bibinfo {pages} {7078}
  (\bibinfo {year} {2016}{\natexlab{a}})}\BibitemShut {NoStop}%
\bibitem [{\citenamefont {Molinero}\ and\ \citenamefont
  {Moore}(2009)}]{molinero_water_2009}%
  \BibitemOpen
  \bibfield  {author} {\bibinfo {author} {\bibfnamefont {V.}~\bibnamefont
  {Molinero}}\ and\ \bibinfo {author} {\bibfnamefont {E.~B.}\ \bibnamefont
  {Moore}},\ }\href {\doibase 10.1021/jp805227c} {\bibfield  {journal}
  {\bibinfo  {journal} {J. Phys. Chem.}\ }\textbf {\bibinfo {volume} {113}},\
  \bibinfo {pages} {4008} (\bibinfo {year} {2009})}\BibitemShut {NoStop}%
\bibitem [{\citenamefont {Fitzner}\ \emph {et~al.}(2015)\citenamefont
  {Fitzner}, \citenamefont {Sosso}, \citenamefont {Cox},\ and\ \citenamefont
  {Michaelides}}]{fitzner_many_2015}%
  \BibitemOpen
  \bibfield  {author} {\bibinfo {author} {\bibfnamefont {M.}~\bibnamefont
  {Fitzner}}, \bibinfo {author} {\bibfnamefont {G.~C.}\ \bibnamefont {Sosso}},
  \bibinfo {author} {\bibfnamefont {S.~J.}\ \bibnamefont {Cox}}, \ and\
  \bibinfo {author} {\bibfnamefont {A.}~\bibnamefont {Michaelides}},\ }\href
  {\doibase 10.1021/jacs.5b08748} {\bibfield  {journal} {\bibinfo  {journal}
  {J. Am. Chem. Soc.}\ }\textbf {\bibinfo {volume} {137}},\ \bibinfo {pages}
  {13658} (\bibinfo {year} {2015})}\BibitemShut {NoStop}%
\bibitem [{\citenamefont {Lupi}\ and\ \citenamefont
  {Molinero}(2014)}]{lupi_does_2014}%
  \BibitemOpen
  \bibfield  {author} {\bibinfo {author} {\bibfnamefont {L.}~\bibnamefont
  {Lupi}}\ and\ \bibinfo {author} {\bibfnamefont {V.}~\bibnamefont
  {Molinero}},\ }\href {\doibase 10.1021/jp4118375} {\bibfield  {journal}
  {\bibinfo  {journal} {J. Phys. Chem. A}\ }\textbf {\bibinfo {volume} {118}},\
  \bibinfo {pages} {7330} (\bibinfo {year} {2014})}\BibitemShut {NoStop}%
\bibitem [{\citenamefont {Lupi}, \citenamefont {Hudait},\ and\ \citenamefont
  {Molinero}(2014)}]{lupi_heterogeneous_2014}%
  \BibitemOpen
  \bibfield  {author} {\bibinfo {author} {\bibfnamefont {L.}~\bibnamefont
  {Lupi}}, \bibinfo {author} {\bibfnamefont {A.}~\bibnamefont {Hudait}}, \ and\
  \bibinfo {author} {\bibfnamefont {V.}~\bibnamefont {Molinero}},\ }\href
  {\doibase 10.1021/ja411507a} {\bibfield  {journal} {\bibinfo  {journal} {J.
  Am. Chem. Soc.}\ }\textbf {\bibinfo {volume} {136}},\ \bibinfo {pages} {3156}
  (\bibinfo {year} {2014})}\BibitemShut {NoStop}%
\bibitem [{\citenamefont {Cabriolu}\ and\ \citenamefont
  {Li}(2015)}]{cabriolu_ice_2015}%
  \BibitemOpen
  \bibfield  {author} {\bibinfo {author} {\bibfnamefont {R.}~\bibnamefont
  {Cabriolu}}\ and\ \bibinfo {author} {\bibfnamefont {T.}~\bibnamefont {Li}},\
  }\href {\doibase 10.1103/PhysRevE.91.052402} {\bibfield  {journal} {\bibinfo
  {journal} {Phys. Rev. E}\ }\textbf {\bibinfo {volume} {91}},\ \bibinfo
  {pages} {052402} (\bibinfo {year} {2015})}\BibitemShut {NoStop}%
\bibitem [{\citenamefont {Lupi}, \citenamefont {Peters},\ and\ \citenamefont
  {Molinero}(2016)}]{lupi_pre-ordering_2016}%
  \BibitemOpen
  \bibfield  {author} {\bibinfo {author} {\bibfnamefont {L.}~\bibnamefont
  {Lupi}}, \bibinfo {author} {\bibfnamefont {B.}~\bibnamefont {Peters}}, \ and\
  \bibinfo {author} {\bibfnamefont {V.}~\bibnamefont {Molinero}},\ }\href
  {http://scitation.aip.org/content/aip/journal/jcp/145/21/10.1063/1.4961652}
  {\bibfield  {journal} {\bibinfo  {journal} {J. Chem. Phys.}\ }\textbf
  {\bibinfo {volume} {145}},\ \bibinfo {pages} {211910} (\bibinfo {year}
  {2016})}\BibitemShut {NoStop}%
\bibitem [{\citenamefont {Bi}, \citenamefont {Cabriolu},\ and\ \citenamefont
  {Li}(2016)}]{bi_heterogeneous_2016}%
  \BibitemOpen
  \bibfield  {author} {\bibinfo {author} {\bibfnamefont {Y.}~\bibnamefont
  {Bi}}, \bibinfo {author} {\bibfnamefont {R.}~\bibnamefont {Cabriolu}}, \ and\
  \bibinfo {author} {\bibfnamefont {T.}~\bibnamefont {Li}},\ }\href {\doibase
  10.1021/acs.jpcc.5b09740} {\bibfield  {journal} {\bibinfo  {journal} {J.
  Phys. Chem. C}\ }\textbf {\bibinfo {volume} {120}},\ \bibinfo {pages} {1507}
  (\bibinfo {year} {2016})}\BibitemShut {NoStop}%
\bibitem [{\citenamefont {Yan}\ and\ \citenamefont
  {Patey}(2011{\natexlab{a}})}]{yan_heterogeneous_2011}%
  \BibitemOpen
  \bibfield  {author} {\bibinfo {author} {\bibfnamefont {J.~Y.}\ \bibnamefont
  {Yan}}\ and\ \bibinfo {author} {\bibfnamefont {G.~N.}\ \bibnamefont
  {Patey}},\ }\href {\doibase 10.1021/jz201113m} {\bibfield  {journal}
  {\bibinfo  {journal} {J. Phys. Chem. Lett.}\ }\textbf {\bibinfo {volume}
  {2}},\ \bibinfo {pages} {2555} (\bibinfo {year}
  {2011}{\natexlab{a}})}\BibitemShut {NoStop}%
\bibitem [{\citenamefont {Yan}\ and\ \citenamefont
  {Patey}(2013{\natexlab{a}})}]{yan_ice_2013}%
  \BibitemOpen
  \bibfield  {author} {\bibinfo {author} {\bibfnamefont {J.~Y.}\ \bibnamefont
  {Yan}}\ and\ \bibinfo {author} {\bibfnamefont {G.~N.}\ \bibnamefont
  {Patey}},\ }\href {\doibase 10.1063/1.4824139} {\bibfield  {journal}
  {\bibinfo  {journal} {J. Chem. Phys.}\ }\textbf {\bibinfo {volume} {139}},\
  \bibinfo {pages} {144501} (\bibinfo {year} {2013}{\natexlab{a}})}\BibitemShut
  {NoStop}%
\bibitem [{\citenamefont {Zielke}, \citenamefont {Bertram},\ and\ \citenamefont
  {Patey}(2014)}]{zielke_molecular_2014}%
  \BibitemOpen
  \bibfield  {author} {\bibinfo {author} {\bibfnamefont {S.~A.}\ \bibnamefont
  {Zielke}}, \bibinfo {author} {\bibfnamefont {A.~K.}\ \bibnamefont {Bertram}},
  \ and\ \bibinfo {author} {\bibfnamefont {G.~N.}\ \bibnamefont {Patey}},\
  }\href {\doibase 10.1021/jp508601s} {\bibfield  {journal} {\bibinfo
  {journal} {J. Phys. Chem. B}\ }\textbf {\bibinfo {volume} {119}},\ \bibinfo
  {pages} {9049} (\bibinfo {year} {2014})}\BibitemShut {NoStop}%
\bibitem [{\citenamefont {Cox}\ \emph {et~al.}(2014)\citenamefont {Cox},
  \citenamefont {Raza}, \citenamefont {Kathmann}, \citenamefont {Slater},\ and\
  \citenamefont {Michaelides}}]{cox_microscopic_2014}%
  \BibitemOpen
  \bibfield  {author} {\bibinfo {author} {\bibfnamefont {S.~J.}\ \bibnamefont
  {Cox}}, \bibinfo {author} {\bibfnamefont {Z.}~\bibnamefont {Raza}}, \bibinfo
  {author} {\bibfnamefont {S.~M.}\ \bibnamefont {Kathmann}}, \bibinfo {author}
  {\bibfnamefont {B.}~\bibnamefont {Slater}}, \ and\ \bibinfo {author}
  {\bibfnamefont {A.}~\bibnamefont {Michaelides}},\ }\href {\doibase
  10.1039/C3FD00059A} {\bibfield  {journal} {\bibinfo  {journal} {Farad.
  Discuss.}\ }\textbf {\bibinfo {volume} {167}},\ \bibinfo {pages} {389}
  (\bibinfo {year} {2014})}\BibitemShut {NoStop}%
\bibitem [{\citenamefont {Sosso}\ \emph
  {et~al.}(2016{\natexlab{b}})\citenamefont {Sosso}, \citenamefont {Li},
  \citenamefont {Donadio}, \citenamefont {Tribello},\ and\ \citenamefont
  {Michaelides}}]{sosso_microscopic_2016}%
  \BibitemOpen
  \bibfield  {author} {\bibinfo {author} {\bibfnamefont {G.~C.}\ \bibnamefont
  {Sosso}}, \bibinfo {author} {\bibfnamefont {T.}~\bibnamefont {Li}}, \bibinfo
  {author} {\bibfnamefont {D.}~\bibnamefont {Donadio}}, \bibinfo {author}
  {\bibfnamefont {G.~A.}\ \bibnamefont {Tribello}}, \ and\ \bibinfo {author}
  {\bibfnamefont {A.}~\bibnamefont {Michaelides}},\ }\href {\doibase
  10.1021/acs.jpclett.6b01013} {\bibfield  {journal} {\bibinfo  {journal} {J.
  Phys. Chem. Lett.}\ }\textbf {\bibinfo {volume} {7}},\ \bibinfo {pages}
  {2350} (\bibinfo {year} {2016}{\natexlab{b}})}\BibitemShut {NoStop}%
\bibitem [{\citenamefont {Murray}\ \emph {et~al.}(2011)\citenamefont {Murray},
  \citenamefont {Broadley}, \citenamefont {Wilson}, \citenamefont {Atkinson},\
  and\ \citenamefont {Wills}}]{murray_heterogeneous_2011}%
  \BibitemOpen
  \bibfield  {author} {\bibinfo {author} {\bibfnamefont {B.~J.}\ \bibnamefont
  {Murray}}, \bibinfo {author} {\bibfnamefont {S.~L.}\ \bibnamefont
  {Broadley}}, \bibinfo {author} {\bibfnamefont {T.~W.}\ \bibnamefont
  {Wilson}}, \bibinfo {author} {\bibfnamefont {J.~D.}\ \bibnamefont
  {Atkinson}}, \ and\ \bibinfo {author} {\bibfnamefont {R.~H.}\ \bibnamefont
  {Wills}},\ }\href {\doibase 10.5194/acp-11-4191-2011} {\bibfield  {journal}
  {\bibinfo  {journal} {Atmos. Chem. Phys.}\ }\textbf {\bibinfo {volume}
  {11}},\ \bibinfo {pages} {4191} (\bibinfo {year} {2011})}\BibitemShut
  {NoStop}%
\bibitem [{\citenamefont {Tobo}\ \emph {et~al.}(2012)\citenamefont {Tobo},
  \citenamefont {DeMott}, \citenamefont {Raddatz}, \citenamefont {Niedermeier},
  \citenamefont {Hartmann}, \citenamefont {Kreidenweis}, \citenamefont
  {Stratmann},\ and\ \citenamefont {Wex}}]{tobo_impacts_2012}%
  \BibitemOpen
  \bibfield  {author} {\bibinfo {author} {\bibfnamefont {Y.}~\bibnamefont
  {Tobo}}, \bibinfo {author} {\bibfnamefont {P.~J.}\ \bibnamefont {DeMott}},
  \bibinfo {author} {\bibfnamefont {M.}~\bibnamefont {Raddatz}}, \bibinfo
  {author} {\bibfnamefont {D.}~\bibnamefont {Niedermeier}}, \bibinfo {author}
  {\bibfnamefont {S.}~\bibnamefont {Hartmann}}, \bibinfo {author}
  {\bibfnamefont {S.~M.}\ \bibnamefont {Kreidenweis}}, \bibinfo {author}
  {\bibfnamefont {F.}~\bibnamefont {Stratmann}}, \ and\ \bibinfo {author}
  {\bibfnamefont {H.}~\bibnamefont {Wex}},\ }\href {\doibase
  10.1029/2012GL053007} {\bibfield  {journal} {\bibinfo  {journal} {Geophys.
  Res. Lett.}\ }\textbf {\bibinfo {volume} {39}},\ \bibinfo {pages} {L19803}
  (\bibinfo {year} {2012})}\BibitemShut {NoStop}%
\bibitem [{\citenamefont {Welti}\ \emph {et~al.}(2013)\citenamefont {Welti},
  \citenamefont {Kanji}, \citenamefont {L\"{u}\"{o}d}, \citenamefont
  {Stetzer},\ and\ \citenamefont {Lohmann}}]{welti_exploring_2013}%
  \BibitemOpen
  \bibfield  {author} {\bibinfo {author} {\bibfnamefont {A.}~\bibnamefont
  {Welti}}, \bibinfo {author} {\bibfnamefont {Z.~A.}\ \bibnamefont {Kanji}},
  \bibinfo {author} {\bibfnamefont {F.}~\bibnamefont {L\"{u}\"{o}d}}, \bibinfo
  {author} {\bibfnamefont {O.}~\bibnamefont {Stetzer}}, \ and\ \bibinfo
  {author} {\bibfnamefont {U.}~\bibnamefont {Lohmann}},\ }\href {\doibase
  10.1175/JAS-D-12-0252.1} {\bibfield  {journal} {\bibinfo  {journal} {J.
  Atmos. Sci.}\ }\textbf {\bibinfo {volume} {71}},\ \bibinfo {pages} {16}
  (\bibinfo {year} {2013})}\BibitemShut {NoStop}%
\bibitem [{\citenamefont {Wex}\ \emph {et~al.}(2014)\citenamefont {Wex},
  \citenamefont {DeMott}, \citenamefont {Tobo}, \citenamefont {Hartmann},
  \citenamefont {R\"{o}sch}, \citenamefont {Clauss}, \citenamefont {Tomsche},
  \citenamefont {Niedermeier},\ and\ \citenamefont
  {Stratmann}}]{wex_kaolinite_2014}%
  \BibitemOpen
  \bibfield  {author} {\bibinfo {author} {\bibfnamefont {H.}~\bibnamefont
  {Wex}}, \bibinfo {author} {\bibfnamefont {P.~J.}\ \bibnamefont {DeMott}},
  \bibinfo {author} {\bibfnamefont {Y.}~\bibnamefont {Tobo}}, \bibinfo {author}
  {\bibfnamefont {S.}~\bibnamefont {Hartmann}}, \bibinfo {author}
  {\bibfnamefont {M.}~\bibnamefont {R\"{o}sch}}, \bibinfo {author}
  {\bibfnamefont {T.}~\bibnamefont {Clauss}}, \bibinfo {author} {\bibfnamefont
  {L.}~\bibnamefont {Tomsche}}, \bibinfo {author} {\bibfnamefont
  {D.}~\bibnamefont {Niedermeier}}, \ and\ \bibinfo {author} {\bibfnamefont
  {F.}~\bibnamefont {Stratmann}},\ }\href {\doibase 10.5194/acp-14-5529-2014}
  {\bibfield  {journal} {\bibinfo  {journal} {Atmos. Chem. Phys.}\ }\textbf
  {\bibinfo {volume} {14}},\ \bibinfo {pages} {5529} (\bibinfo {year}
  {2014})}\BibitemShut {NoStop}%
\bibitem [{\citenamefont {Hu}\ and\ \citenamefont
  {Michaelides}(2010)}]{hu_kaolinite_2010}%
  \BibitemOpen
  \bibfield  {author} {\bibinfo {author} {\bibfnamefont {X.~L.}\ \bibnamefont
  {Hu}}\ and\ \bibinfo {author} {\bibfnamefont {A.}~\bibnamefont
  {Michaelides}},\ }\href {\doibase 10.1016/j.susc.2009.10.026} {\bibfield
  {journal} {\bibinfo  {journal} {Surf. Sci.}\ }\textbf {\bibinfo {volume}
  {604}},\ \bibinfo {pages} {111} (\bibinfo {year} {2010})}\BibitemShut
  {NoStop}%
\bibitem [{\citenamefont {Tunega}, \citenamefont {Gerzabek},\ and\
  \citenamefont {Lischka}(2004)}]{tunega_ab_2004}%
  \BibitemOpen
  \bibfield  {author} {\bibinfo {author} {\bibfnamefont {D.}~\bibnamefont
  {Tunega}}, \bibinfo {author} {\bibfnamefont {M.~H.}\ \bibnamefont
  {Gerzabek}}, \ and\ \bibinfo {author} {\bibfnamefont {H.}~\bibnamefont
  {Lischka}},\ }\href {\doibase 10.1021/jp037121g} {\bibfield  {journal}
  {\bibinfo  {journal} {J. Phys. Chem. B}\ }\textbf {\bibinfo {volume} {108}},\
  \bibinfo {pages} {5930} (\bibinfo {year} {2004})}\BibitemShut {NoStop}%
\bibitem [{\citenamefont {Zielke}, \citenamefont {Bertram},\ and\ \citenamefont
  {Patey}(2016)}]{zielke_simulations_2016}%
  \BibitemOpen
  \bibfield  {author} {\bibinfo {author} {\bibfnamefont {S.~A.}\ \bibnamefont
  {Zielke}}, \bibinfo {author} {\bibfnamefont {A.~K.}\ \bibnamefont {Bertram}},
  \ and\ \bibinfo {author} {\bibfnamefont {G.~N.}\ \bibnamefont {Patey}},\
  }\href@noop {} {\bibfield  {journal} {\bibinfo  {journal} {J. Phys. Chem. B}\
  }\textbf {\bibinfo {volume} {120}},\ \bibinfo {pages} {1726} (\bibinfo {year}
  {2016})}\BibitemShut {NoStop}%
\bibitem [{\citenamefont {Allen}, \citenamefont {Valeriani},\ and\
  \citenamefont {Rein~ten Wolde}(2009)}]{allen_forward_2009}%
  \BibitemOpen
  \bibfield  {author} {\bibinfo {author} {\bibfnamefont {R.~J.}\ \bibnamefont
  {Allen}}, \bibinfo {author} {\bibfnamefont {C.}~\bibnamefont {Valeriani}}, \
  and\ \bibinfo {author} {\bibfnamefont {P.}~\bibnamefont {Rein~ten Wolde}},\
  }\href {\doibase 10.1088/0953-8984/21/46/463102} {\bibfield  {journal}
  {\bibinfo  {journal} {J. Phys. Cond. Matt.}\ }\textbf {\bibinfo {volume}
  {21}},\ \bibinfo {pages} {463102} (\bibinfo {year} {2009})}\BibitemShut
  {NoStop}%
\bibitem [{\citenamefont {Li}, \citenamefont {Donadio},\ and\ \citenamefont
  {Galli}(2013)}]{li_ice_2013}%
  \BibitemOpen
  \bibfield  {author} {\bibinfo {author} {\bibfnamefont {T.}~\bibnamefont
  {Li}}, \bibinfo {author} {\bibfnamefont {D.}~\bibnamefont {Donadio}}, \ and\
  \bibinfo {author} {\bibfnamefont {G.}~\bibnamefont {Galli}},\ }\href
  {\doibase 10.1038/ncomms2918} {\bibfield  {journal} {\bibinfo  {journal}
  {Nat. Comm.}\ }\textbf {\bibinfo {volume} {4}},\ \bibinfo {pages} {1887}
  (\bibinfo {year} {2013})}\BibitemShut {NoStop}%
\bibitem [{\citenamefont {Bi}\ and\ \citenamefont
  {Li}(2014)}]{bi_probing_2014}%
  \BibitemOpen
  \bibfield  {author} {\bibinfo {author} {\bibfnamefont {Y.}~\bibnamefont
  {Bi}}\ and\ \bibinfo {author} {\bibfnamefont {T.}~\bibnamefont {Li}},\ }\href
  {\doibase 10.1021/jp503000u} {\bibfield  {journal} {\bibinfo  {journal} {J.
  Phys. Chem. B}\ }\textbf {\bibinfo {volume} {118}},\ \bibinfo {pages} {13324}
  (\bibinfo {year} {2014})}\BibitemShut {NoStop}%
\bibitem [{\citenamefont {Gianetti}\ \emph {et~al.}(2016)\citenamefont
  {Gianetti}, \citenamefont {Haji-Akbari}, \citenamefont {Longinotti},\ and\
  \citenamefont {Debenedetti}}]{gianetti_computational_2016}%
  \BibitemOpen
  \bibfield  {author} {\bibinfo {author} {\bibfnamefont {M.~M.}\ \bibnamefont
  {Gianetti}}, \bibinfo {author} {\bibfnamefont {A.}~\bibnamefont
  {Haji-Akbari}}, \bibinfo {author} {\bibfnamefont {M.~P.}\ \bibnamefont
  {Longinotti}}, \ and\ \bibinfo {author} {\bibfnamefont {P.~G.}\ \bibnamefont
  {Debenedetti}},\ }\href {\doibase 10.1039/C5CP06535F} {\bibfield  {journal}
  {\bibinfo  {journal} {Phys. Chem. Chem. Phys.}\ }\textbf {\bibinfo {volume}
  {18}},\ \bibinfo {pages} {4102} (\bibinfo {year} {2016})}\BibitemShut
  {NoStop}%
\bibitem [{\citenamefont {Van Der~Spoel}\ \emph {et~al.}(2005)\citenamefont
  {Van Der~Spoel}, \citenamefont {Lindahl}, \citenamefont {Hess}, \citenamefont
  {Groenhof}, \citenamefont {Mark},\ and\ \citenamefont
  {Berendsen}}]{van_der_spoel_gromacs:_2005}%
  \BibitemOpen
  \bibfield  {author} {\bibinfo {author} {\bibfnamefont {D.}~\bibnamefont {Van
  Der~Spoel}}, \bibinfo {author} {\bibfnamefont {E.}~\bibnamefont {Lindahl}},
  \bibinfo {author} {\bibfnamefont {B.}~\bibnamefont {Hess}}, \bibinfo {author}
  {\bibfnamefont {G.}~\bibnamefont {Groenhof}}, \bibinfo {author}
  {\bibfnamefont {A.~E.}\ \bibnamefont {Mark}}, \ and\ \bibinfo {author}
  {\bibfnamefont {H.~J.~C.}\ \bibnamefont {Berendsen}},\ }\href {\doibase
  10.1002/jcc.20291} {\bibfield  {journal} {\bibinfo  {journal} {J. Comp.
  Chem.}\ }\textbf {\bibinfo {volume} {26}},\ \bibinfo {pages} {1701} (\bibinfo
  {year} {2005})}\BibitemShut {NoStop}%
\bibitem [{\citenamefont {Cygan}, \citenamefont {Liang},\ and\ \citenamefont
  {Kalinichev}(2004)}]{cygan_molecular_2004}%
  \BibitemOpen
  \bibfield  {author} {\bibinfo {author} {\bibfnamefont {R.~T.}\ \bibnamefont
  {Cygan}}, \bibinfo {author} {\bibfnamefont {J.-J.}\ \bibnamefont {Liang}}, \
  and\ \bibinfo {author} {\bibfnamefont {A.~G.}\ \bibnamefont {Kalinichev}},\
  }\href {\doibase 10.1021/jp0363287} {\bibfield  {journal} {\bibinfo
  {journal} {J. Phys. Chem. B}\ }\textbf {\bibinfo {volume} {108}},\ \bibinfo
  {pages} {1255} (\bibinfo {year} {2004})}\BibitemShut {NoStop}%
\bibitem [{\citenamefont {Abascal}\ \emph {et~al.}(2005)\citenamefont
  {Abascal}, \citenamefont {Sanz}, \citenamefont {Fernandez},\ and\
  \citenamefont {Vega}}]{abascal_potential_2005}%
  \BibitemOpen
  \bibfield  {author} {\bibinfo {author} {\bibfnamefont {J.~L.~F.}\
  \bibnamefont {Abascal}}, \bibinfo {author} {\bibfnamefont {E.}~\bibnamefont
  {Sanz}}, \bibinfo {author} {\bibfnamefont {R.~G.}\ \bibnamefont {Fernandez}},
  \ and\ \bibinfo {author} {\bibfnamefont {C.}~\bibnamefont {Vega}},\ }\href
  {\doibase 10.1063/1.1931662} {\bibfield  {journal} {\bibinfo  {journal} {J.
  Chem. Phys.}\ }\textbf {\bibinfo {volume} {122}},\ \bibinfo {pages} {234511}
  (\bibinfo {year} {2005})}\BibitemShut {NoStop}%
\bibitem [{\citenamefont {Lorentz}(1881)}]{lorentz_ueber_1881}%
  \BibitemOpen
  \bibfield  {author} {\bibinfo {author} {\bibfnamefont {H.~A.}\ \bibnamefont
  {Lorentz}},\ }\href {\doibase 10.1002/andp.18812480110} {\bibfield  {journal}
  {\bibinfo  {journal} {Ann. der Physik}\ }\textbf {\bibinfo {volume} {248}},\
  \bibinfo {pages} {127} (\bibinfo {year} {1881})}\BibitemShut {NoStop}%
\bibitem [{\citenamefont {Berthelot}\ and\ \citenamefont
  {Hebd}(1898)}]{ahthefrench}%
  \BibitemOpen
  \bibfield  {author} {\bibinfo {author} {\bibfnamefont {D.}~\bibnamefont
  {Berthelot}}\ and\ \bibinfo {author} {\bibfnamefont {C.~R.}\ \bibnamefont
  {Hebd}},\ }\href@noop {} {\bibfield  {journal} {\bibinfo  {journal} {Seanc.
  Acad. Sci.}\ }\textbf {\bibinfo {volume} {126}},\ \bibinfo {pages} {1703}
  (\bibinfo {year} {1898})}\BibitemShut {NoStop}%
\bibitem [{Note1()}]{Note1}%
  \BibitemOpen
  \bibinfo {note} {Specifically, the adsorption energy $E_{ads}$ of a single
  water molecule on top of the hydroxylated (001) surface of kaolinite
  (calculated as $E_{ads}=E_{KAO+W}-E_{KAO}-E_{W}$ where $E_{KAO+W}$, $E_{KAO}$
  and $E_{W}$ are the total energy of a kaolinite slab with a water molecule on
  top, the energy of the slab alone and the energy of the water molecule in
  vacuum respectively) on its most favorable adsorption site is -646 $\pm $ 60
  meV, in excellent agreement with the quantum Monte Carlo result of -648 $\pm
  $ 18 meV~\cite {ANDREA_PREPRINT}}\BibitemShut {NoStop}%
\bibitem [{\citenamefont {Bussi}, \citenamefont {Donadio},\ and\ \citenamefont
  {Parrinello}(2007)}]{bussi_canonical_2007}%
  \BibitemOpen
  \bibfield  {author} {\bibinfo {author} {\bibfnamefont {G.}~\bibnamefont
  {Bussi}}, \bibinfo {author} {\bibfnamefont {D.}~\bibnamefont {Donadio}}, \
  and\ \bibinfo {author} {\bibfnamefont {M.}~\bibnamefont {Parrinello}},\
  }\href {\doibase 10.1063/1.2408420} {\bibfield  {journal} {\bibinfo
  {journal} {J. Chem. Phys.}\ }\textbf {\bibinfo {volume} {126}},\ \bibinfo
  {pages} {014101} (\bibinfo {year} {2007})}\BibitemShut {NoStop}%
\bibitem [{\citenamefont {Miyamoto}\ and\ \citenamefont
  {Kollman}(1992)}]{miyamoto_settle:_1992}%
  \BibitemOpen
  \bibfield  {author} {\bibinfo {author} {\bibfnamefont {S.}~\bibnamefont
  {Miyamoto}}\ and\ \bibinfo {author} {\bibfnamefont {P.~A.}\ \bibnamefont
  {Kollman}},\ }\href {\doibase 10.1002/jcc.540130805} {\bibfield  {journal}
  {\bibinfo  {journal} {J. Comp. Chem.}\ }\textbf {\bibinfo {volume} {13}},\
  \bibinfo {pages} {952} (\bibinfo {year} {1992})}\BibitemShut {NoStop}%
\bibitem [{\citenamefont {Hess}\ \emph {et~al.}(1997)\citenamefont {Hess},
  \citenamefont {Bekker}, \citenamefont {Berendsen},\ and\ \citenamefont
  {Fraaije}}]{hess_lincs:_1997}%
  \BibitemOpen
  \bibfield  {author} {\bibinfo {author} {\bibfnamefont {B.}~\bibnamefont
  {Hess}}, \bibinfo {author} {\bibfnamefont {H.}~\bibnamefont {Bekker}},
  \bibinfo {author} {\bibfnamefont {H.~J.~C.}\ \bibnamefont {Berendsen}}, \
  and\ \bibinfo {author} {\bibfnamefont {J.~G. E.~M.}\ \bibnamefont
  {Fraaije}},\ }\href {\doibase
  10.1002/(SICI)1096-987X(199709)18:12<1463::AID-JCC4>3.0.CO;2-H} {\bibfield
  {journal} {\bibinfo  {journal} {J. Comp. Chem.}\ }\textbf {\bibinfo {volume}
  {18}},\ \bibinfo {pages} {1463} (\bibinfo {year} {1997})}\BibitemShut
  {NoStop}%
\bibitem [{\citenamefont {Moore}\ and\ \citenamefont
  {Molinero}(2011)}]{moore_is_2011}%
  \BibitemOpen
  \bibfield  {author} {\bibinfo {author} {\bibfnamefont {E.~B.}\ \bibnamefont
  {Moore}}\ and\ \bibinfo {author} {\bibfnamefont {V.}~\bibnamefont
  {Molinero}},\ }\href {\doibase 10.1039/c1cp22022e} {\bibfield  {journal}
  {\bibinfo  {journal} {Phys. Chem. Chem. Phys.}\ }\textbf {\bibinfo {volume}
  {13}},\ \bibinfo {pages} {20008} (\bibinfo {year} {2011})}\BibitemShut
  {NoStop}%
\bibitem [{\citenamefont {Hansen}\ \emph {et~al.}(2008)\citenamefont {Hansen},
  \citenamefont {Koza}, \citenamefont {Lindner},\ and\ \citenamefont
  {Kuhs}}]{hansen_formation_2008}%
  \BibitemOpen
  \bibfield  {author} {\bibinfo {author} {\bibfnamefont {T.~C.}\ \bibnamefont
  {Hansen}}, \bibinfo {author} {\bibfnamefont {M.~M.}\ \bibnamefont {Koza}},
  \bibinfo {author} {\bibfnamefont {P.}~\bibnamefont {Lindner}}, \ and\
  \bibinfo {author} {\bibfnamefont {W.~F.}\ \bibnamefont {Kuhs}},\ }\href
  {\doibase 10.1088/0953-8984/20/28/285105} {\bibfield  {journal} {\bibinfo
  {journal} {J. Phys. Cond. Matt.}\ }\textbf {\bibinfo {volume} {20}},\
  \bibinfo {pages} {285105} (\bibinfo {year} {2008})}\BibitemShut {NoStop}%
\bibitem [{\citenamefont {Malkin}\ \emph {et~al.}(2014)\citenamefont {Malkin},
  \citenamefont {Murray}, \citenamefont {Salzmann}, \citenamefont {Molinero},
  \citenamefont {Pickering},\ and\ \citenamefont
  {Whale}}]{malkin_stacking_2014}%
  \BibitemOpen
  \bibfield  {author} {\bibinfo {author} {\bibfnamefont {T.~L.}\ \bibnamefont
  {Malkin}}, \bibinfo {author} {\bibfnamefont {B.~J.}\ \bibnamefont {Murray}},
  \bibinfo {author} {\bibfnamefont {C.~G.}\ \bibnamefont {Salzmann}}, \bibinfo
  {author} {\bibfnamefont {V.}~\bibnamefont {Molinero}}, \bibinfo {author}
  {\bibfnamefont {S.~J.}\ \bibnamefont {Pickering}}, \ and\ \bibinfo {author}
  {\bibfnamefont {T.~F.}\ \bibnamefont {Whale}},\ }\href {\doibase
  10.1039/C4CP02893G} {\bibfield  {journal} {\bibinfo  {journal} {Phys. Chem.
  Chem. Phys.}\ }\textbf {\bibinfo {volume} {17}},\ \bibinfo {pages} {60}
  (\bibinfo {year} {2014})}\BibitemShut {NoStop}%
\bibitem [{\citenamefont {Slater}\ and\ \citenamefont
  {Quigley}(2014)}]{slater_crystal_2014}%
  \BibitemOpen
  \bibfield  {author} {\bibinfo {author} {\bibfnamefont {B.}~\bibnamefont
  {Slater}}\ and\ \bibinfo {author} {\bibfnamefont {D.}~\bibnamefont
  {Quigley}},\ }\href {\doibase 10.1038/nmat4017} {\bibfield  {journal}
  {\bibinfo  {journal} {Nat. Mat.}\ }\textbf {\bibinfo {volume} {13}},\
  \bibinfo {pages} {670} (\bibinfo {year} {2014})}\BibitemShut {NoStop}%
\bibitem [{\citenamefont {Quigley}(2014)}]{quigley_communication:_2014}%
  \BibitemOpen
  \bibfield  {author} {\bibinfo {author} {\bibfnamefont {D.}~\bibnamefont
  {Quigley}},\ }\href {\doibase 10.1063/1.4896376} {\bibfield  {journal}
  {\bibinfo  {journal} {J. Chem. Phys.}\ }\textbf {\bibinfo {volume} {141}},\
  \bibinfo {pages} {121101} (\bibinfo {year} {2014})}\BibitemShut {NoStop}%
\bibitem [{\citenamefont {Sanz}\ \emph {et~al.}(2013)\citenamefont {Sanz},
  \citenamefont {Vega}, \citenamefont {Espinosa}, \citenamefont
  {Caballero-Bernal}, \citenamefont {Abascal},\ and\ \citenamefont
  {Valeriani}}]{sanz_homogeneous_2013}%
  \BibitemOpen
  \bibfield  {author} {\bibinfo {author} {\bibfnamefont {E.}~\bibnamefont
  {Sanz}}, \bibinfo {author} {\bibfnamefont {C.}~\bibnamefont {Vega}}, \bibinfo
  {author} {\bibfnamefont {J.~R.}\ \bibnamefont {Espinosa}}, \bibinfo {author}
  {\bibfnamefont {R.}~\bibnamefont {Caballero-Bernal}}, \bibinfo {author}
  {\bibfnamefont {J.~L.~F.}\ \bibnamefont {Abascal}}, \ and\ \bibinfo {author}
  {\bibfnamefont {C.}~\bibnamefont {Valeriani}},\ }\href {\doibase
  10.1021/ja4028814} {\bibfield  {journal} {\bibinfo  {journal} {J. Am. Chem.
  Soc.}\ }\textbf {\bibinfo {volume} {135}},\ \bibinfo {pages} {15008}
  (\bibinfo {year} {2013})}\BibitemShut {NoStop}%
\bibitem [{\citenamefont {Espinosa}\ \emph {et~al.}(2014)\citenamefont
  {Espinosa}, \citenamefont {Sanz}, \citenamefont {Valeriani},\ and\
  \citenamefont {Vega}}]{espinosa_homogeneous_2014}%
  \BibitemOpen
  \bibfield  {author} {\bibinfo {author} {\bibfnamefont {J.~R.}\ \bibnamefont
  {Espinosa}}, \bibinfo {author} {\bibfnamefont {E.}~\bibnamefont {Sanz}},
  \bibinfo {author} {\bibfnamefont {C.}~\bibnamefont {Valeriani}}, \ and\
  \bibinfo {author} {\bibfnamefont {C.}~\bibnamefont {Vega}},\ }\href {\doibase
  10.1063/1.4897524} {\bibfield  {journal} {\bibinfo  {journal} {J. Chem.
  Phys.}\ }\textbf {\bibinfo {volume} {141}},\ \bibinfo {pages} {18C529}
  (\bibinfo {year} {2014})}\BibitemShut {NoStop}%
\bibitem [{\citenamefont {Espinosa}\ \emph {et~al.}(2016)\citenamefont
  {Espinosa}, \citenamefont {Vega}, \citenamefont {Valeriani},\ and\
  \citenamefont {Sanz}}]{espinosa_seeding_2016}%
  \BibitemOpen
  \bibfield  {author} {\bibinfo {author} {\bibfnamefont {J.~R.}\ \bibnamefont
  {Espinosa}}, \bibinfo {author} {\bibfnamefont {C.}~\bibnamefont {Vega}},
  \bibinfo {author} {\bibfnamefont {C.}~\bibnamefont {Valeriani}}, \ and\
  \bibinfo {author} {\bibfnamefont {E.}~\bibnamefont {Sanz}},\ }\href {\doibase
  10.1063/1.4939641} {\bibfield  {journal} {\bibinfo  {journal} {J. Chem.
  Phys.}\ }\textbf {\bibinfo {volume} {144}},\ \bibinfo {pages} {034501}
  (\bibinfo {year} {2016})}\BibitemShut {NoStop}%
\bibitem [{\citenamefont {Laio}\ and\ \citenamefont
  {Parrinello}(2002)}]{laio_escaping_2002}%
  \BibitemOpen
  \bibfield  {author} {\bibinfo {author} {\bibfnamefont {A.}~\bibnamefont
  {Laio}}\ and\ \bibinfo {author} {\bibfnamefont {M.}~\bibnamefont
  {Parrinello}},\ }\href {\doibase 10.1073/pnas.202427399} {\bibfield
  {journal} {\bibinfo  {journal} {Proc. Natl. Acad. Sci. U.S.A.}\ }\textbf
  {\bibinfo {volume} {99}},\ \bibinfo {pages} {12562} (\bibinfo {year}
  {2002})}\BibitemShut {NoStop}%
\bibitem [{\citenamefont {Laio}\ and\ \citenamefont
  {Gervasio}(2008)}]{laio_metadynamics:_2008}%
  \BibitemOpen
  \bibfield  {author} {\bibinfo {author} {\bibfnamefont {A.}~\bibnamefont
  {Laio}}\ and\ \bibinfo {author} {\bibfnamefont {F.~L.}\ \bibnamefont
  {Gervasio}},\ }\href {\doibase 10.1088/0034-4885/71/12/126601} {\bibfield
  {journal} {\bibinfo  {journal} {Rep. Prog. Phys.}\ }\textbf {\bibinfo
  {volume} {71}},\ \bibinfo {pages} {126601} (\bibinfo {year}
  {2008})}\BibitemShut {NoStop}%
\bibitem [{\citenamefont {Steinhardt}, \citenamefont {Nelson},\ and\
  \citenamefont {Ronchetti}(1983)}]{steinhardt_bond-orientational_1983}%
  \BibitemOpen
  \bibfield  {author} {\bibinfo {author} {\bibfnamefont {P.~J.}\ \bibnamefont
  {Steinhardt}}, \bibinfo {author} {\bibfnamefont {D.~R.}\ \bibnamefont
  {Nelson}}, \ and\ \bibinfo {author} {\bibfnamefont {M.}~\bibnamefont
  {Ronchetti}},\ }\href {\doibase 10.1103/PhysRevB.28.784} {\bibfield
  {journal} {\bibinfo  {journal} {Phys. Rev. B}\ }\textbf {\bibinfo {volume}
  {28}},\ \bibinfo {pages} {784} (\bibinfo {year} {1983})}\BibitemShut
  {NoStop}%
\bibitem [{\citenamefont {Haji-Akbari}\ and\ \citenamefont
  {Debenedetti}(2015)}]{haji-akbari_direct_2015}%
  \BibitemOpen
  \bibfield  {author} {\bibinfo {author} {\bibfnamefont {A.}~\bibnamefont
  {Haji-Akbari}}\ and\ \bibinfo {author} {\bibfnamefont {P.~G.}\ \bibnamefont
  {Debenedetti}},\ }\href {\doibase 10.1073/pnas.1509267112} {\bibfield
  {journal} {\bibinfo  {journal} {Proc. Natl. Acad. Sci. U.S.A.}\ }\textbf
  {\bibinfo {volume} {112}},\ \bibinfo {pages} {10582} (\bibinfo {year}
  {2015})}\BibitemShut {NoStop}%
\bibitem [{mos()}]{mostly}%
  \BibitemOpen
  \href@noop {} {}\bibinfo {note} {Water molecules within each nucleus have
  been labelled as $I_h$ or $I_c$ according to the order parameter illustrated
  in the SM. We consider a nucleus to be made \textit{predominantely} of e.g.
  $I_h$ if the fraction of $I_h$ molecules is at least 1.5 times larger than
  that of $I_c$. We have also verified that by using a topological
  criterion~\cite{haji-akbari_direct_2015} for identifying the building blocks
  of $I_h$ and $I_c$ the probability densities reported in Fig. 3 remain
  basically unchanged.}\BibitemShut {Stop}%
\bibitem [{\citenamefont {Klime\v{s}}, \citenamefont {Bowler},\ and\
  \citenamefont {Michaelides}(2010)}]{klimes-vdW-DF}%
  \BibitemOpen
  \bibfield  {author} {\bibinfo {author} {\bibfnamefont {J.}~\bibnamefont
  {Klime\v{s}}}, \bibinfo {author} {\bibfnamefont {D.~R.}\ \bibnamefont
  {Bowler}}, \ and\ \bibinfo {author} {\bibfnamefont {A.}~\bibnamefont
  {Michaelides}},\ }\href@noop {} {\bibfield  {journal} {\bibinfo  {journal}
  {J. Phys. Cond. Mat.}\ }\textbf {\bibinfo {volume} {22}},\ \bibinfo {pages}
  {022201} (\bibinfo {year} {2010})}\BibitemShut {NoStop}%
\bibitem [{\citenamefont {Pruppacher}\ and\ \citenamefont
  {Klett}(1997)}]{PK97}%
  \BibitemOpen
  \bibfield  {author} {\bibinfo {author} {\bibfnamefont {H.~R.}\ \bibnamefont
  {Pruppacher}}\ and\ \bibinfo {author} {\bibfnamefont {J.~D.}\ \bibnamefont
  {Klett}},\ }\href@noop {} {\emph {\bibinfo {title} {Microphysics Of Clouds
  And Precipitation - Second Revised And Enlarged Edition with an Introduction
  to Cloud Chemistry and Cloud Electricity}}}\ (\bibinfo  {publisher} {Kluwer
  Academic Publishers},\ \bibinfo {address} {Dordrecht, The Netherlands},\
  \bibinfo {year} {1997})\BibitemShut {NoStop}%
\bibitem [{\citenamefont {Le~Roux}\ and\ \citenamefont
  {Jund}(2010)}]{le_roux_ring_2010}%
  \BibitemOpen
  \bibfield  {author} {\bibinfo {author} {\bibfnamefont {S.}~\bibnamefont
  {Le~Roux}}\ and\ \bibinfo {author} {\bibfnamefont {P.}~\bibnamefont {Jund}},\
  }\href {\doibase 10.1016/j.commatsci.2010.04.023} {\bibfield  {journal}
  {\bibinfo  {journal} {Comp. Mat. Sci.}\ }\textbf {\bibinfo {volume} {49}},\
  \bibinfo {pages} {70} (\bibinfo {year} {2010})}\BibitemShut {NoStop}%
\bibitem [{\citenamefont {King}(1967)}]{king_ring_1967}%
  \BibitemOpen
  \bibfield  {author} {\bibinfo {author} {\bibfnamefont {S.~V.}\ \bibnamefont
  {King}},\ }\href {\doibase 10.1038/2131112a0} {\bibfield  {journal} {\bibinfo
   {journal} {Nature}\ }\textbf {\bibinfo {volume} {213}},\ \bibinfo {pages}
  {1112} (\bibinfo {year} {1967})}\BibitemShut {NoStop}%
\bibitem [{\citenamefont {Franzblau}(1991)}]{franzblau_computation_1991}%
  \BibitemOpen
  \bibfield  {author} {\bibinfo {author} {\bibfnamefont {D.~S.}\ \bibnamefont
  {Franzblau}},\ }\href {\doibase 10.1103/PhysRevB.44.4925} {\bibfield
  {journal} {\bibinfo  {journal} {Phys. Rev. B}\ }\textbf {\bibinfo {volume}
  {44}},\ \bibinfo {pages} {4925} (\bibinfo {year} {1991})}\BibitemShut
  {NoStop}%
\bibitem [{equ()}]{equill}%
  \BibitemOpen
  \href@noop {} {}\bibinfo {note} {The system is obviously in a metastable
  state. In fact, S1 is even unstable with respect to ice formation on the 100
  ns timescale. However, we here seek to compare the different interfaces S1 S2
  S3 before the onset of ice formation, where the system can be considered in
  equilibrium with respect to the computation of the free energy profiles
  reported in Fig. 4a. This is why we have considered 10 independent
  simulations for each interface, each 10 ns long, in order to ensure a
  meaningful equilibrium statistics.}\BibitemShut {Stop}%
\bibitem [{\citenamefont {Haji-Akbari}\ and\ \citenamefont
  {Debenedetti}(2014)}]{haji-akbari_effect_2014}%
  \BibitemOpen
  \bibfield  {author} {\bibinfo {author} {\bibfnamefont {A.}~\bibnamefont
  {Haji-Akbari}}\ and\ \bibinfo {author} {\bibfnamefont {P.~G.}\ \bibnamefont
  {Debenedetti}},\ }\href {\doibase 10.1063/1.4885365} {\bibfield  {journal}
  {\bibinfo  {journal} {J. Chem. Phys.}\ }\textbf {\bibinfo {volume} {141}},\
  \bibinfo {pages} {024506} (\bibinfo {year} {2014})}\BibitemShut {NoStop}%
\bibitem [{\citenamefont {Yan}\ and\ \citenamefont
  {Patey}(2011{\natexlab{b}})}]{patey2011}%
  \BibitemOpen
  \bibfield  {author} {\bibinfo {author} {\bibfnamefont {J.~Y.}\ \bibnamefont
  {Yan}}\ and\ \bibinfo {author} {\bibfnamefont {G.~N.}\ \bibnamefont
  {Patey}},\ }\href@noop {} {\bibfield  {journal} {\bibinfo  {journal} {J.
  Phys. Chem. Lett.}\ }\textbf {\bibinfo {volume} {2}},\ \bibinfo {pages}
  {2555} (\bibinfo {year} {2011}{\natexlab{b}})}\BibitemShut {NoStop}%
\bibitem [{\citenamefont {Yan}\ and\ \citenamefont
  {Patey}(2013{\natexlab{b}})}]{patey2013}%
  \BibitemOpen
  \bibfield  {author} {\bibinfo {author} {\bibfnamefont {J.~Y.}\ \bibnamefont
  {Yan}}\ and\ \bibinfo {author} {\bibfnamefont {G.~N.}\ \bibnamefont
  {Patey}},\ }\href {\doibase http://dx.doi.org/10.1063/1.4824139} {\bibfield
  {journal} {\bibinfo  {journal} {J. Chem. Phys.}\ }\textbf {\bibinfo {volume}
  {139}},\ \bibinfo {pages} {144501} (\bibinfo {year}
  {2013}{\natexlab{b}})}\BibitemShut {NoStop}%
\bibitem [{\citenamefont {Vega}, \citenamefont {Sanz},\ and\ \citenamefont
  {Abascal}(2005)}]{vega_melting_2005}%
  \BibitemOpen
  \bibfield  {author} {\bibinfo {author} {\bibfnamefont {C.}~\bibnamefont
  {Vega}}, \bibinfo {author} {\bibfnamefont {E.}~\bibnamefont {Sanz}}, \ and\
  \bibinfo {author} {\bibfnamefont {J.~L.~F.}\ \bibnamefont {Abascal}},\ }\href
  {\doibase 10.1063/1.1862245} {\bibfield  {journal} {\bibinfo  {journal} {J.
  Chem. Phys.}\ }\textbf {\bibinfo {volume} {122}},\ \bibinfo {pages} {114507}
  (\bibinfo {year} {2005})}\BibitemShut {NoStop}%
\bibitem [{\citenamefont {Orsi}(2014)}]{orsi_comparative_2014}%
  \BibitemOpen
  \bibfield  {author} {\bibinfo {author} {\bibfnamefont {M.}~\bibnamefont
  {Orsi}},\ }\href {\doibase 10.1080/00268976.2013.844373} {\bibfield
  {journal} {\bibinfo  {journal} {Mol. Phys.}\ }\textbf {\bibinfo {volume}
  {112}},\ \bibinfo {pages} {1566} (\bibinfo {year} {2014})}\BibitemShut
  {NoStop}%
\bibitem [{\citenamefont {Haji-Akbari}\ \emph {et~al.}(2014)\citenamefont
  {Haji-Akbari}, \citenamefont {DeFever}, \citenamefont {Sarupria},\ and\
  \citenamefont {Debenedetti}}]{haji-akbari_suppression_2014}%
  \BibitemOpen
  \bibfield  {author} {\bibinfo {author} {\bibfnamefont {A.}~\bibnamefont
  {Haji-Akbari}}, \bibinfo {author} {\bibfnamefont {R.~S.}\ \bibnamefont
  {DeFever}}, \bibinfo {author} {\bibfnamefont {S.}~\bibnamefont {Sarupria}}, \
  and\ \bibinfo {author} {\bibfnamefont {P.~G.}\ \bibnamefont {Debenedetti}},\
  }\href {\doibase 10.1039/C4CP03948C} {\bibfield  {journal} {\bibinfo
  {journal} {Phys. Chem. Chem. Phys.}\ }\textbf {\bibinfo {volume} {16}},\
  \bibinfo {pages} {25916} (\bibinfo {year} {2014})}\BibitemShut {NoStop}%
\bibitem [{\citenamefont {Mazza}\ \emph {et~al.}(2006)\citenamefont {Mazza},
  \citenamefont {Giovambattista}, \citenamefont {Starr},\ and\ \citenamefont
  {Stanley}}]{mazza_relation_2006}%
  \BibitemOpen
  \bibfield  {author} {\bibinfo {author} {\bibfnamefont {M.~G.}\ \bibnamefont
  {Mazza}}, \bibinfo {author} {\bibfnamefont {N.}~\bibnamefont
  {Giovambattista}}, \bibinfo {author} {\bibfnamefont {F.~W.}\ \bibnamefont
  {Starr}}, \ and\ \bibinfo {author} {\bibfnamefont {H.~E.}\ \bibnamefont
  {Stanley}},\ }\href {\doibase 10.1103/PhysRevLett.96.057803} {\bibfield
  {journal} {\bibinfo  {journal} {Phys. Rev. Lett.}\ }\textbf {\bibinfo
  {volume} {96}},\ \bibinfo {pages} {057803} (\bibinfo {year}
  {2006})}\BibitemShut {NoStop}%
\bibitem [{\citenamefont {Kumar}\ \emph {et~al.}(2007)\citenamefont {Kumar},
  \citenamefont {Buldyrev}, \citenamefont {Becker}, \citenamefont {Poole},
  \citenamefont {Starr},\ and\ \citenamefont {Stanley}}]{kumar_relation_2007}%
  \BibitemOpen
  \bibfield  {author} {\bibinfo {author} {\bibfnamefont {P.}~\bibnamefont
  {Kumar}}, \bibinfo {author} {\bibfnamefont {S.~V.}\ \bibnamefont {Buldyrev}},
  \bibinfo {author} {\bibfnamefont {S.~R.}\ \bibnamefont {Becker}}, \bibinfo
  {author} {\bibfnamefont {P.~H.}\ \bibnamefont {Poole}}, \bibinfo {author}
  {\bibfnamefont {F.~W.}\ \bibnamefont {Starr}}, \ and\ \bibinfo {author}
  {\bibfnamefont {H.~E.}\ \bibnamefont {Stanley}},\ }\href {\doibase
  10.1073/pnas.0702608104} {\bibfield  {journal} {\bibinfo  {journal} {Proc.
  Nat. Acad. Sci.}\ }\textbf {\bibinfo {volume} {104}},\ \bibinfo {pages}
  {9575} (\bibinfo {year} {2007})}\BibitemShut {NoStop}%
\bibitem [{\citenamefont {Liu}\ and\ \citenamefont
  {Lu}(2006)}]{liu_thermodynamic_2006}%
  \BibitemOpen
  \bibfield  {author} {\bibinfo {author} {\bibfnamefont {X.-D.}\ \bibnamefont
  {Liu}}\ and\ \bibinfo {author} {\bibfnamefont {X.-C.}\ \bibnamefont {Lu}},\
  }\href {\doibase 10.1002/anie.200601740} {\bibfield  {journal} {\bibinfo
  {journal} {Angew. Chem. Int. Ed.}\ }\textbf {\bibinfo {volume} {45}},\
  \bibinfo {pages} {6300} (\bibinfo {year} {2006})}\BibitemShut {NoStop}%
\bibitem [{\citenamefont {Haria}, \citenamefont {Grest},\ and\ \citenamefont
  {Lorenz}(2013)}]{haria_viscosity_2013}%
  \BibitemOpen
  \bibfield  {author} {\bibinfo {author} {\bibfnamefont {N.~R.}\ \bibnamefont
  {Haria}}, \bibinfo {author} {\bibfnamefont {G.~S.}\ \bibnamefont {Grest}}, \
  and\ \bibinfo {author} {\bibfnamefont {C.~D.}\ \bibnamefont {Lorenz}},\
  }\href {\doibase 10.1021/jp312181u} {\bibfield  {journal} {\bibinfo
  {journal} {J. Phys. Chem. C}\ }\textbf {\bibinfo {volume} {117}},\ \bibinfo
  {pages} {6096} (\bibinfo {year} {2013})}\BibitemShut {NoStop}%
\bibitem [{\citenamefont {Hu}\ and\ \citenamefont
  {Michaelides}(2007)}]{hu_ice_2007}%
  \BibitemOpen
  \bibfield  {author} {\bibinfo {author} {\bibfnamefont {X.~L.}\ \bibnamefont
  {Hu}}\ and\ \bibinfo {author} {\bibfnamefont {A.}~\bibnamefont
  {Michaelides}},\ }\href {\doibase 10.1016/j.susc.2007.09.012} {\bibfield
  {journal} {\bibinfo  {journal} {Surf. Sci.}\ }\textbf {\bibinfo {volume}
  {601}},\ \bibinfo {pages} {5378} (\bibinfo {year} {2007})}\BibitemShut
  {NoStop}%
\bibitem [{\citenamefont {Wang}\ \emph {et~al.}(2005)\citenamefont {Wang},
  \citenamefont {Kalinichev}, \citenamefont {Kirkpatrick},\ and\ \citenamefont
  {Cygan}}]{wang_structure_2005}%
  \BibitemOpen
  \bibfield  {author} {\bibinfo {author} {\bibfnamefont {J.}~\bibnamefont
  {Wang}}, \bibinfo {author} {\bibfnamefont {A.~G.}\ \bibnamefont
  {Kalinichev}}, \bibinfo {author} {\bibfnamefont {R.~J.}\ \bibnamefont
  {Kirkpatrick}}, \ and\ \bibinfo {author} {\bibfnamefont {R.~T.}\ \bibnamefont
  {Cygan}},\ }\href {\doibase 10.1021/jp045299c} {\bibfield  {journal}
  {\bibinfo  {journal} {J. Phys. Chem. B}\ }\textbf {\bibinfo {volume} {109}},\
  \bibinfo {pages} {15893} (\bibinfo {year} {2005})}\BibitemShut {NoStop}%
\bibitem [{\citenamefont {Odelius}, \citenamefont {Bernasconi},\ and\
  \citenamefont {Parrinello}(1997)}]{odelius_two_1997}%
  \BibitemOpen
  \bibfield  {author} {\bibinfo {author} {\bibfnamefont {M.}~\bibnamefont
  {Odelius}}, \bibinfo {author} {\bibfnamefont {M.}~\bibnamefont {Bernasconi}},
  \ and\ \bibinfo {author} {\bibfnamefont {M.}~\bibnamefont {Parrinello}},\
  }\href {\doibase 10.1103/PhysRevLett.78.2855} {\bibfield  {journal} {\bibinfo
   {journal} {Phys. Rev. Lett.}\ }\textbf {\bibinfo {volume} {78}},\ \bibinfo
  {pages} {2855} (\bibinfo {year} {1997})}\BibitemShut {NoStop}%
\bibitem [{\citenamefont {Weiss}\ \emph {et~al.}(2011)\citenamefont {Weiss},
  \citenamefont {Rullich}, \citenamefont {Köhler},\ and\ \citenamefont
  {Frauenheim}}]{weiss_kinetic_2011}%
  \BibitemOpen
  \bibfield  {author} {\bibinfo {author} {\bibfnamefont {V.~C.}\ \bibnamefont
  {Weiss}}, \bibinfo {author} {\bibfnamefont {M.}~\bibnamefont {Rullich}},
  \bibinfo {author} {\bibfnamefont {C.}~\bibnamefont {Köhler}}, \ and\
  \bibinfo {author} {\bibfnamefont {T.}~\bibnamefont {Frauenheim}},\ }\href
  {\doibase 10.1063/1.3609768} {\bibfield  {journal} {\bibinfo  {journal} {J.
  Chem. Phys.}\ }\textbf {\bibinfo {volume} {135}},\ \bibinfo {pages} {034701}
  (\bibinfo {year} {2011})}\BibitemShut {NoStop}%
\bibitem [{\citenamefont {Berthier}\ and\ \citenamefont
  {Biroli}(2011)}]{berthier_theoretical_2011}%
  \BibitemOpen
  \bibfield  {author} {\bibinfo {author} {\bibfnamefont {L.}~\bibnamefont
  {Berthier}}\ and\ \bibinfo {author} {\bibfnamefont {G.}~\bibnamefont
  {Biroli}},\ }\href {\doibase 10.1103/RevModPhys.83.587} {\bibfield  {journal}
  {\bibinfo  {journal} {Rev. Mod. Phys.}\ }\textbf {\bibinfo {volume} {83}},\
  \bibinfo {pages} {587} (\bibinfo {year} {2011})}\BibitemShut {NoStop}%
\bibitem [{\citenamefont {Sosso}\ \emph {et~al.}(2014)\citenamefont {Sosso},
  \citenamefont {Colombo}, \citenamefont {Behler}, \citenamefont {Del~Gado},\
  and\ \citenamefont {Bernasconi}}]{sosso_dynamical_2014}%
  \BibitemOpen
  \bibfield  {author} {\bibinfo {author} {\bibfnamefont {G.~C.}\ \bibnamefont
  {Sosso}}, \bibinfo {author} {\bibfnamefont {J.}~\bibnamefont {Colombo}},
  \bibinfo {author} {\bibfnamefont {J.}~\bibnamefont {Behler}}, \bibinfo
  {author} {\bibfnamefont {E.}~\bibnamefont {Del~Gado}}, \ and\ \bibinfo
  {author} {\bibfnamefont {M.}~\bibnamefont {Bernasconi}},\ }\href {\doibase
  10.1021/jp507361f} {\bibfield  {journal} {\bibinfo  {journal} {J. Phys. Chem.
  B}\ }\textbf {\bibinfo {volume} {118}},\ \bibinfo {pages} {13621} (\bibinfo
  {year} {2014})}\BibitemShut {NoStop}%
\bibitem [{\citenamefont {Makov}\ and\ \citenamefont {Payne}(1995)}]{Dipole1}%
  \BibitemOpen
  \bibfield  {author} {\bibinfo {author} {\bibfnamefont {G.}~\bibnamefont
  {Makov}}\ and\ \bibinfo {author} {\bibfnamefont {M.~C.}\ \bibnamefont
  {Payne}},\ }\href@noop {} {\bibfield  {journal} {\bibinfo  {journal} {Phys.
  Rev. B}\ }\textbf {\bibinfo {volume} {51}},\ \bibinfo {pages} {4014}
  (\bibinfo {year} {1995})}\BibitemShut {NoStop}%
\bibitem [{\citenamefont {Neugebauer}\ and\ \citenamefont
  {Scheffler}(1992)}]{Dipole2}%
  \BibitemOpen
  \bibfield  {author} {\bibinfo {author} {\bibfnamefont {J.}~\bibnamefont
  {Neugebauer}}\ and\ \bibinfo {author} {\bibfnamefont {M.}~\bibnamefont
  {Scheffler}},\ }\href@noop {} {\bibfield  {journal} {\bibinfo  {journal}
  {Phys. Rev. B}\ }\textbf {\bibinfo {volume} {42}},\ \bibinfo {pages} {16967}
  (\bibinfo {year} {1992})}\BibitemShut {NoStop}%
\bibitem [{\citenamefont {Perdew}, \citenamefont {Burke},\ and\ \citenamefont
  {Ernzerhof}(1996)}]{PBE}%
  \BibitemOpen
  \bibfield  {author} {\bibinfo {author} {\bibfnamefont {J.~P.}\ \bibnamefont
  {Perdew}}, \bibinfo {author} {\bibfnamefont {K.}~\bibnamefont {Burke}}, \
  and\ \bibinfo {author} {\bibfnamefont {M.}~\bibnamefont {Ernzerhof}},\
  }\href@noop {} {\bibfield  {journal} {\bibinfo  {journal} {Phys. Rev. Lett.}\
  }\textbf {\bibinfo {volume} {77}},\ \bibinfo {pages} {3865} (\bibinfo {year}
  {1996})}\BibitemShut {NoStop}%
\bibitem [{\citenamefont {Perdew}, \citenamefont {Burke},\ and\ \citenamefont
  {Ernzerhof}(1997)}]{PBE_Erratum}%
  \BibitemOpen
  \bibfield  {author} {\bibinfo {author} {\bibfnamefont {J.~P.}\ \bibnamefont
  {Perdew}}, \bibinfo {author} {\bibfnamefont {K.}~\bibnamefont {Burke}}, \
  and\ \bibinfo {author} {\bibfnamefont {M.}~\bibnamefont {Ernzerhof}},\
  }\href@noop {} {\bibfield  {journal} {\bibinfo  {journal} {Phys. Rev. Lett.}\
  }\textbf {\bibinfo {volume} {78}},\ \bibinfo {pages} {1396} (\bibinfo {year}
  {1997})}\BibitemShut {NoStop}%
\bibitem [{\citenamefont {Hammer}, \citenamefont {Hansen},\ and\ \citenamefont
  {N{\o}rskov}(1999)}]{RPBE}%
  \BibitemOpen
  \bibfield  {author} {\bibinfo {author} {\bibfnamefont {B.}~\bibnamefont
  {Hammer}}, \bibinfo {author} {\bibfnamefont {L.~B.}\ \bibnamefont {Hansen}},
  \ and\ \bibinfo {author} {\bibfnamefont {J.~K.}\ \bibnamefont {N{\o}rskov}},\
  }\href@noop {} {\bibfield  {journal} {\bibinfo  {journal} {Phys. Rev. B}\
  }\textbf {\bibinfo {volume} {59}},\ \bibinfo {pages} {7413} (\bibinfo {year}
  {1999})}\BibitemShut {NoStop}%
\bibitem [{\citenamefont {Grimme}(2006)}]{DFT-D2}%
  \BibitemOpen
  \bibfield  {author} {\bibinfo {author} {\bibfnamefont {S.}~\bibnamefont
  {Grimme}},\ }\href@noop {} {\bibfield  {journal} {\bibinfo  {journal} {J.
  Comp. Chem.}\ }\textbf {\bibinfo {volume} {27}},\ \bibinfo {pages} {1787}
  (\bibinfo {year} {2006})}\BibitemShut {NoStop}%
\bibitem [{\citenamefont {Grimme}\ \emph {et~al.}(2010)\citenamefont {Grimme},
  \citenamefont {Antony}, \citenamefont {Ehrlich},\ and\ \citenamefont
  {Krieg}}]{DFT-D3}%
  \BibitemOpen
  \bibfield  {author} {\bibinfo {author} {\bibfnamefont {S.}~\bibnamefont
  {Grimme}}, \bibinfo {author} {\bibfnamefont {J.}~\bibnamefont {Antony}},
  \bibinfo {author} {\bibfnamefont {S.}~\bibnamefont {Ehrlich}}, \ and\
  \bibinfo {author} {\bibfnamefont {H.}~\bibnamefont {Krieg}},\ }\href
  {\doibase 10.1063/1.3382344} {\bibfield  {journal} {\bibinfo  {journal} {J.
  Chem. Phys.}\ }\textbf {\bibinfo {volume} {132}},\ \bibinfo {pages} {154104}
  (\bibinfo {year} {2010})}\BibitemShut {NoStop}%
\bibitem [{\citenamefont {Tkatchenko}\ and\ \citenamefont
  {Scheffler}(2009)}]{DFT-TS}%
  \BibitemOpen
  \bibfield  {author} {\bibinfo {author} {\bibfnamefont {A.}~\bibnamefont
  {Tkatchenko}}\ and\ \bibinfo {author} {\bibfnamefont {M.}~\bibnamefont
  {Scheffler}},\ }\href@noop {} {\bibfield  {journal} {\bibinfo  {journal}
  {Phys. Rev. Lett.}\ }\textbf {\bibinfo {volume} {102}},\ \bibinfo {pages}
  {073005} (\bibinfo {year} {2009})}\BibitemShut {NoStop}%
\bibitem [{\citenamefont {Lee}\ \emph {et~al.}(2010)\citenamefont {Lee},
  \citenamefont {Murray}, \citenamefont {Kong}, \citenamefont {Lundqvist},\
  and\ \citenamefont {Langreth}}]{vdW-DF2}%
  \BibitemOpen
  \bibfield  {author} {\bibinfo {author} {\bibfnamefont {K.}~\bibnamefont
  {Lee}}, \bibinfo {author} {\bibfnamefont {E.~D.}\ \bibnamefont {Murray}},
  \bibinfo {author} {\bibfnamefont {L.}~\bibnamefont {Kong}}, \bibinfo {author}
  {\bibfnamefont {B.~I.}\ \bibnamefont {Lundqvist}}, \ and\ \bibinfo {author}
  {\bibfnamefont {D.~C.}\ \bibnamefont {Langreth}},\ }\href@noop {} {\bibfield
  {journal} {\bibinfo  {journal} {Phys. Rev. B}\ }\textbf {\bibinfo {volume}
  {82}},\ \bibinfo {pages} {081101} (\bibinfo {year} {2010})}\BibitemShut
  {NoStop}%
\bibitem [{\citenamefont {Zhang}\ and\ \citenamefont {Yang}(1998)}]{revPBE}%
  \BibitemOpen
  \bibfield  {author} {\bibinfo {author} {\bibfnamefont {Y.}~\bibnamefont
  {Zhang}}\ and\ \bibinfo {author} {\bibfnamefont {W.}~\bibnamefont {Yang}},\
  }\href@noop {} {\bibfield  {journal} {\bibinfo  {journal} {Phys. Rev. Lett.}\
  }\textbf {\bibinfo {volume} {80}},\ \bibinfo {pages} {890} (\bibinfo {year}
  {1998})}\BibitemShut {NoStop}%
\bibitem [{\citenamefont {Dion}\ \emph {et~al.}(2004)\citenamefont {Dion},
  \citenamefont {Rydberg}, \citenamefont {Schroder}, \citenamefont {Langreth},\
  and\ \citenamefont {Lundqvist}}]{vdW-DF}%
  \BibitemOpen
  \bibfield  {author} {\bibinfo {author} {\bibfnamefont {M.}~\bibnamefont
  {Dion}}, \bibinfo {author} {\bibfnamefont {H.}~\bibnamefont {Rydberg}},
  \bibinfo {author} {\bibfnamefont {E.}~\bibnamefont {Schroder}}, \bibinfo
  {author} {\bibfnamefont {D.~C.}\ \bibnamefont {Langreth}}, \ and\ \bibinfo
  {author} {\bibfnamefont {B.~I.}\ \bibnamefont {Lundqvist}},\ }\href@noop {}
  {\bibfield  {journal} {\bibinfo  {journal} {Phys. Rev. Lett.}\ }\textbf
  {\bibinfo {volume} {92}},\ \bibinfo {pages} {246401} (\bibinfo {year}
  {2004})}\BibitemShut {NoStop}%
\bibitem [{\citenamefont {Klime\v{s}}, \citenamefont {Bowler},\ and\
  \citenamefont {Michaelides}(2011)}]{optB86b-vdW}%
  \BibitemOpen
  \bibfield  {author} {\bibinfo {author} {\bibfnamefont {J.}~\bibnamefont
  {Klime\v{s}}}, \bibinfo {author} {\bibfnamefont {D.~R.}\ \bibnamefont
  {Bowler}}, \ and\ \bibinfo {author} {\bibfnamefont {A.}~\bibnamefont
  {Michaelides}},\ }\href@noop {} {\bibfield  {journal} {\bibinfo  {journal}
  {Phys. Rev. B}\ }\textbf {\bibinfo {volume} {83}},\ \bibinfo {pages} {195131}
  (\bibinfo {year} {2011})}\BibitemShut {NoStop}%
\bibitem [{\citenamefont {Kresse}\ and\ \citenamefont {Hafner}(1993)}]{VASP1}%
  \BibitemOpen
  \bibfield  {author} {\bibinfo {author} {\bibfnamefont {G.}~\bibnamefont
  {Kresse}}\ and\ \bibinfo {author} {\bibfnamefont {J.}~\bibnamefont
  {Hafner}},\ }\href@noop {} {\bibfield  {journal} {\bibinfo  {journal} {Phys.
  Rev. B}\ }\textbf {\bibinfo {volume} {558}},\ \bibinfo {pages} {47} (\bibinfo
  {year} {1993})}\BibitemShut {NoStop}%
\bibitem [{\citenamefont {Kresse}\ and\ \citenamefont {Hafner}(1994)}]{VASP2}%
  \BibitemOpen
  \bibfield  {author} {\bibinfo {author} {\bibfnamefont {G.}~\bibnamefont
  {Kresse}}\ and\ \bibinfo {author} {\bibfnamefont {J.}~\bibnamefont
  {Hafner}},\ }\href@noop {} {\bibfield  {journal} {\bibinfo  {journal} {Phys.
  Rev. B}\ }\textbf {\bibinfo {volume} {49}},\ \bibinfo {pages} {14251}
  (\bibinfo {year} {1994})}\BibitemShut {NoStop}%
\bibitem [{\citenamefont {Kresse}\ and\ \citenamefont
  {Furthmuller}(1996{\natexlab{a}})}]{VASP3}%
  \BibitemOpen
  \bibfield  {author} {\bibinfo {author} {\bibfnamefont {G.}~\bibnamefont
  {Kresse}}\ and\ \bibinfo {author} {\bibfnamefont {J.}~\bibnamefont
  {Furthmuller}},\ }\href@noop {} {\bibfield  {journal} {\bibinfo  {journal}
  {Comput. Mat. Sci.}\ }\textbf {\bibinfo {volume} {6}},\ \bibinfo {pages} {15}
  (\bibinfo {year} {1996}{\natexlab{a}})}\BibitemShut {NoStop}%
\bibitem [{\citenamefont {Kresse}\ and\ \citenamefont
  {Furthmuller}(1996{\natexlab{b}})}]{VASP4}%
  \BibitemOpen
  \bibfield  {author} {\bibinfo {author} {\bibfnamefont {G.}~\bibnamefont
  {Kresse}}\ and\ \bibinfo {author} {\bibfnamefont {J.}~\bibnamefont
  {Furthmuller}},\ }\href@noop {} {\bibfield  {journal} {\bibinfo  {journal}
  {Phys. Rev. B}\ }\textbf {\bibinfo {volume} {54}},\ \bibinfo {pages} {11169}
  (\bibinfo {year} {1996}{\natexlab{b}})}\BibitemShut {NoStop}%
\bibitem [{\citenamefont {Rom\'an-P\'erez}\ and\ \citenamefont
  {Soler}(2009)}]{perez&soler:implementation}%
  \BibitemOpen
  \bibfield  {author} {\bibinfo {author} {\bibfnamefont {G.}~\bibnamefont
  {Rom\'an-P\'erez}}\ and\ \bibinfo {author} {\bibfnamefont {J.~M.}\
  \bibnamefont {Soler}},\ }\href
  {http://link.aps.org/doi/10.1103/PhysRevLett.103.096102} {\bibfield
  {journal} {\bibinfo  {journal} {Phys. Rev. Lett.}\ }\textbf {\bibinfo
  {volume} {103}},\ \bibinfo {pages} {096102} (\bibinfo {year}
  {2009})}\BibitemShut {NoStop}%
\bibitem [{\citenamefont {Blochl}(1994)}]{PAW1}%
  \BibitemOpen
  \bibfield  {author} {\bibinfo {author} {\bibfnamefont {P.~E.}\ \bibnamefont
  {Blochl}},\ }\href@noop {} {\bibfield  {journal} {\bibinfo  {journal} {Phys.
  Rev. B}\ }\textbf {\bibinfo {volume} {50}},\ \bibinfo {pages} {17953}
  (\bibinfo {year} {1994})}\BibitemShut {NoStop}%
\bibitem [{\citenamefont {Kresse}\ and\ \citenamefont {Joubert}(1999)}]{PAW2}%
  \BibitemOpen
  \bibfield  {author} {\bibinfo {author} {\bibfnamefont {G.}~\bibnamefont
  {Kresse}}\ and\ \bibinfo {author} {\bibfnamefont {D.}~\bibnamefont
  {Joubert}},\ }\href@noop {} {\bibfield  {journal} {\bibinfo  {journal} {Phys.
  Rev. B}\ }\textbf {\bibinfo {volume} {59}},\ \bibinfo {pages} {1758}
  (\bibinfo {year} {1999})}\BibitemShut {NoStop}%
\bibitem [{\citenamefont {Graziano}\ \emph {et~al.}(2012)\citenamefont
  {Graziano}, \citenamefont {Klime\v{s}}, \citenamefont {Fernandez-Alonso},\
  and\ \citenamefont {Michaelides}}]{JPCM:Graziano}%
  \BibitemOpen
  \bibfield  {author} {\bibinfo {author} {\bibfnamefont {G.}~\bibnamefont
  {Graziano}}, \bibinfo {author} {\bibfnamefont {J.}~\bibnamefont
  {Klime\v{s}}}, \bibinfo {author} {\bibfnamefont {F.}~\bibnamefont
  {Fernandez-Alonso}}, \ and\ \bibinfo {author} {\bibfnamefont
  {A.}~\bibnamefont {Michaelides}},\ }\href@noop {} {\bibfield  {journal}
  {\bibinfo  {journal} {J. Phys. Condens. Matter}\ }\textbf {\bibinfo {volume}
  {24}},\ \bibinfo {pages} {424216} (\bibinfo {year} {2012})}\BibitemShut
  {NoStop}%
\bibitem [{\citenamefont {Monkhorst}\ and\ \citenamefont
  {Pack}(1976)}]{Kpoints}%
  \BibitemOpen
  \bibfield  {author} {\bibinfo {author} {\bibfnamefont {H.~J.}\ \bibnamefont
  {Monkhorst}}\ and\ \bibinfo {author} {\bibfnamefont {J.~D.}\ \bibnamefont
  {Pack}},\ }\href@noop {} {\bibfield  {journal} {\bibinfo  {journal} {Phys.
  Rev. B}\ }\textbf {\bibinfo {volume} {13}},\ \bibinfo {pages} {5188}
  (\bibinfo {year} {1976})}\BibitemShut {NoStop}%
\bibitem [{\citenamefont {Li}\ \emph {et~al.}(2011)\citenamefont {Li},
  \citenamefont {Donadio}, \citenamefont {Russo},\ and\ \citenamefont
  {Galli}}]{li_homogeneous_2011}%
  \BibitemOpen
  \bibfield  {author} {\bibinfo {author} {\bibfnamefont {T.}~\bibnamefont
  {Li}}, \bibinfo {author} {\bibfnamefont {D.}~\bibnamefont {Donadio}},
  \bibinfo {author} {\bibfnamefont {G.}~\bibnamefont {Russo}}, \ and\ \bibinfo
  {author} {\bibfnamefont {G.}~\bibnamefont {Galli}},\ }\href {\doibase
  10.1039/C1CP22167A} {\bibfield  {journal} {\bibinfo  {journal} {Phys. Chem.
  Chem. Phys.}\ }\textbf {\bibinfo {volume} {13}},\ \bibinfo {pages} {19807}
  (\bibinfo {year} {2011})}\BibitemShut {NoStop}%
\bibitem [{\citenamefont {Sear}(2014)}]{sear_quantitative_2014}%
  \BibitemOpen
  \bibfield  {author} {\bibinfo {author} {\bibfnamefont {R.~P.}\ \bibnamefont
  {Sear}},\ }\href {\doibase 10.1039/C4CE00344F} {\bibfield  {journal}
  {\bibinfo  {journal} {Cryst. Eng. Comm.}\ }\textbf {\bibinfo {volume} {16}},\
  \bibinfo {pages} {6506} (\bibinfo {year} {2014})}\BibitemShut {NoStop}%
\bibitem [{tri()}]{tribello_dfs_2016}%
  \BibitemOpen
  \href@noop {} {}\bibinfo {note} {G.A. Tribello, F. Giberti, G.C. Sosso, M.
  Salvalaglio and M. Parrinello: "Analyzing and Driving Cluster Formation in
  Atomistic Simulations", submitted}\BibitemShut {NoStop}%
\bibitem [{\citenamefont {Barducci}, \citenamefont {Bussi},\ and\ \citenamefont
  {Parrinello}(2008)}]{barducci_well-tempered_2008}%
  \BibitemOpen
  \bibfield  {author} {\bibinfo {author} {\bibfnamefont {A.}~\bibnamefont
  {Barducci}}, \bibinfo {author} {\bibfnamefont {G.}~\bibnamefont {Bussi}}, \
  and\ \bibinfo {author} {\bibfnamefont {M.}~\bibnamefont {Parrinello}},\
  }\href {\doibase 10.1103/PhysRevLett.100.020603} {\bibfield  {journal}
  {\bibinfo  {journal} {Phys. Rev. Lett.}\ }\textbf {\bibinfo {volume} {100}},\
  \bibinfo {pages} {020603} (\bibinfo {year} {2008})}\BibitemShut {NoStop}%
\bibitem [{\citenamefont {Tiwary}\ and\ \citenamefont
  {Parrinello}(2015)}]{tiwary_time-independent_2015}%
  \BibitemOpen
  \bibfield  {author} {\bibinfo {author} {\bibfnamefont {P.}~\bibnamefont
  {Tiwary}}\ and\ \bibinfo {author} {\bibfnamefont {M.}~\bibnamefont
  {Parrinello}},\ }\href {\doibase 10.1021/jp504920s} {\bibfield  {journal}
  {\bibinfo  {journal} {J. Phys. Chem. B}\ }\textbf {\bibinfo {volume} {119}},\
  \bibinfo {pages} {736} (\bibinfo {year} {2015})}\BibitemShut {NoStop}%
\bibitem [{AND()}]{ANDREA_PREPRINT}%
  \BibitemOpen
  \href@noop {} {}\bibinfo {note} {A. Zen, L.M. Roch, S.J. Cox, X.L. Hu, S.
  Sorella, D. Alf\`{e} and A. Michaelides. "Toward Accurate Adsorption
  Energetics on Clay Surfaces". (submitted)}\BibitemShut {NoStop}%
\end{thebibliography}%

\end{document}